\documentclass[11pt,twocolumn,english]{article}
\usepackage[T1]{fontenc}
\usepackage[latin9]{inputenc}
\usepackage[landscape,a4paper]{geometry}
\usepackage{color}
\definecolor{note_fontcolor}{rgb}{0.800781, 0.800781, 0.800781}
\definecolor{rossos}{cmyk}{0,1,1,0.55}
\definecolor{bluscuro}{rgb}{0.15, 0.2, .85}
\definecolor{grigioaltocontrasto}{rgb}{0.37, 0.37, 0.37}
\definecolor{bluchiaro}{cmyk}{1,.3,0.,0.1}
\definecolor{viola}{cmyk}{0.65,0.70,0,0}
\definecolor{arancio}{cmyk}{0.10,0.60,1,0}
\usepackage[usenames,dvipsnames]{xcolor}
\usepackage[font=small,labelfont=bf]{caption}
\usepackage{babel}
\usepackage{array}
\usepackage{multirow}
\usepackage{amstext}
\usepackage{amssymb}
\usepackage{dsfont}
\usepackage{graphicx}
\usepackage[unicode=true,pdfusetitle,
 bookmarks=true,bookmarksnumbered=false,bookmarksopen=false,
 breaklinks=false,pdfborder={0 0 1},
 colorlinks=true,linktocpage]
 {hyperref}

\makeatother

\newcommand{\noun}[1]{\textsc{#1}}
\providecommand{\tabularnewline}{\\}

\usepackage{fancyhdr}
\usepackage{makecell}
\usepackage{ifthen}
\newcommand{\beq}{\begin{equation}}
\newcommand{\eeq}{\end{equation}}
\newcommand{\bea}{\begin{eqnarray}}
\newcommand{\eea}{\end{eqnarray}}
\newcommand{\ltth}[1]{\ifthenelse{#1>0.003}{#1}{0}}

\newcommand{\gev}{\textrm{GeV}}
\newcommand{\mellinO}{\mathcal{M}_{\mathcal{O}}}
\newcommand{\DeltaMO}{\Delta_{m_{t}}^{(\mellinO)}}
\newcommand{\DeltaMOtilde}{\tilde{\Delta}_{\theta}^{(\mathcal{M_{O}})}}
\newcommand{\cov}{\bold{cov}}
\newcommand{\DeltaBinO}{\Delta_{\theta}^{(b_{j,\mathcal{O}})}}
\makeatother
\begin{document}

\title{\bf{Fragmentation Uncertainties in Hadronic Observables for Top-quark Mass Measurements}}

\author{Gennaro Corcella$^{a}$\thanks{gennaro.corcella@lnf.infn.it}, Roberto Franceschini$^{b}$\thanks{roberto.franceschini@uniroma3.it}, and Doojin Kim$^{c}$\thanks{doojin.kim@cern.ch}\\
$^{a}${\small{}INFN, Laboratori Nazionali di Frascati
Via E. Fermi 40, 00044 Frascati (RM), Italy}\\
$^{b}${\small{}Dipartimento di Matematica e Fisica,
Universit\'a  degli Studi Roma Tre and INFN, sezione di Roma Tre, I-00146
Rome, Italy}\\
$^{c}${\small{}Theoretical Physics Department, CERN, CH-1211 Geneva 23, Switzerland}
}
  \maketitle
  \thispagestyle{fancy}
{\scriptsize
\tableofcontents{}}
{\scriptsize
\subsubsection*{{\bf ~ \hfill Abstract \hfill ~ }}
We study the Monte Carlo uncertainties due to modeling of 
hadronization and showering in the extraction of the top-quark mass from
observables that use exclusive hadronic final states in top decays,
such as $t\to\text{anything+J/\ensuremath{\psi}}$ or $t\to\text{anything}+(B\to\text{charged tracks})$, where $B$ is a $B$-hadron.
To this end, we investigate the sensitivity of the top-quark mass,
determined by means of a few observables already proposed in
the literature as well as some new proposals, to the 
relevant parameters of event generators, such as HERWIG 6 and PYTHIA 8.
We find that constraining those parameters at $\mathcal{O}(1\%-10\%)$ is required to avoid a Monte Carlo uncertainty on $m_t$ greater than 500 MeV.
For the sake of achieving the needed accuracy on such parameters, we examine the sensitivity of the top-quark mass measured from spectral features, such as peaks, endpoints and distributions of $E_{B}$, $m_{B\ell}$, and some $m_{T2}$-like variables. We find that restricting oneself
to regions sufficiently close to the
endpoints enables one to substantially decrease the dependence on the Monte Carlo parameters, but at the price of inflating significantly the statistical uncertainties. To ameliorate this situation we study 
how well the data on top-quark production and decay at the LHC
can be utilized to constrain the showering and hadronization
variables. We find that a global exploration of several calibration observables, sensitive to the Monte Carlo parameters
but very mildly to $m_{t}$, can offer useful constraints on the parameters, as long as such quantities are measured with a 1\% precision.
}

\section{Introduction}
The top quark mass ($m_t$) is a fundamental parameter of the
Standard Model which, together with the $W$ mass,
constrained the Higgs mass even before its discovery.
It plays a crucial role in the determination of 
the Standard Model vacuum life-time, which has been recently found to be at the border
between stability and metastability~\cite{degrassi,andreassen}.
It is therefore of paramount importance to measure $m_t$ 
with the highest possible accuracy and provide a reliable
determination of the error on $m_t$.
Furthermore, much work has been lately carried out
in order to estimate the uncertainty on $m_t$, once it
is interpreted in terms of well-defined theoretical quantities,
such as the pole mass \cite{corcella,hoang,hoang1}.

In fact, standard measurements, based on the template, matrix-element
and ideogram methods (see, $e.g.$, the analyses
in \cite{CMS-Collaboration:2016nr,Aad:2012bh,Aad:2015sf}),
rely on parton shower generators
such as HERWIG \cite{Corcella:2001fp} or PYTHIA \cite{Sjostrand:2014rr}.
They simulate the hard scattering at leading order (LO) and
multiple radiation in the soft or collinear approximation,
with phenomenological models for hadronization and underlying events.
By contrast, so-called alternative methods use other observables,
{\it e.g.,} total cross sections or kinematic endpoints, 
which, besides Monte Carlo simulations,
can be compared directly with fixed-order and
possibly resummed QCD calculations, thus allowing a
straightforward theoretical interpretation of the extracted mass
(see, for example, Ref.~\cite{sigmaatl} on the pole mass extraction
from the total $t\bar t$ production cross section).

More recent NLO+shower programs such as aMC@NLO \cite{mcnlo} and POWHEG \cite{powheg} implement NLO hard-scattering amplitudes, but still rely on 
HERWIG and PYTHIA for parton cascades and non-perturbative phenomena. 
NLO corrections to top quark production have been available in both aMC@NLO and POWHEG for some time, while much effort has been later devoted to improve the treatment of top quark decays. 
In the aMC@NLO code, NLO top decays are implemented for single-top events \cite{rikk}; in $t\bar t$ production, the decays are still on shell, but
spin correlations and part of the off-shell contributions are included via MadSpin \cite{madspin}.
In the POWHEG framework, NLO corrections to top decays were implemented
first in Ref.~\cite{pow}, with an approximate treatment of
top-width effects, and more recently, in Ref.~\cite{bb4l}, the interference between top production and decay, as well as non-resonant contributions have been included.
Furthermore, the SHERPA code \cite{sherpa} implements $t\bar t$ production
in conjunction with up to three jets at NLO, merged to parton showers
along the lines of Ref.~\cite{sherpa1}. Top decays in SHERPA include
spin correlations and are accounted for in the LO approximation
\cite{sherpa2}.

At present, since most standard $m_t$ determinations rely on
the reconstruction of $b$-flavored jets in top decays ($t\to bW$), 
one of the main uncertainties 
is from the $b$-jet energy scale (bJES), which amounts to about
250 MeV of the overall 700 MeV in the world average determination \cite{wave}.
Therefore,  
several attempts have been made so far to overcome the difficulties
of such standard methods, in particular 
to calibrate  the jet energies to very high accuracy. Some strategies
use the $W$-boson mass as a constraint to calibrate \emph{in-situ}
the jet reconstruction~\cite{Aad:2012bh}:
this method, however, cannot account for the differences between $b$-jets
and light-flavored jets, as the $W$ decays mostly into light
or charm quarks.
In other attempts the jet energy scale has
been effectively constrained by exploiting the anti-correlation of the $b$-jet
energy and angular variables~\cite{Aad:2015sf} imposed by the $V-A$
matrix element that in the Standard Model describes top-quark decay. 

Despite all these efforts, the jet energy scale keeps being a bottleneck
for the improvement of standard mass measurements; therefore,
new methods have been conceived to go around this uncertainty.
Among the earliest attempts, we recall those
based on the exclusive fragmentation of the $b$-quark to $J/\psi$~\cite{CMS-Collaboration:2016nr,Kharchilava:2000yk}  and studies of the decay length of
$B$ hadrons \cite{Hill:2005dq,CMS-PAS-TOP-12-030}. In the $J/\psi$
method, leptonic decays of the $J/\psi$ are used to identify the resonance.
The distribution in the invariant mass formed by the two leptons from the $J/\psi$ and the lepton from $W$ decay, $m_{\ell\bar{\ell}\ell'}$,
is the key observable which is then compared with a set of Monte Carlo simulation samples generated with different input top quark masses.
The overall strategy does {\it not} directly involve any jet-energy measurement in the determination of the quantity sensitive to the top quark
mass, and therefore it is free from jet-energy calibration
issues.\footnote{The relevant selection criteria are usually imposed on events accompanying jets as well as the leptonic $J/\psi$. 
Nevertheless, the effect of jet calibration
is not \emph{direct} because the mis-measurement
of energy affects only the phase space of the events, not the observable itself. The Monte Carlo simulations, which are used to extract
$m_t$ from the observed distribution, implement
the nominal phase space which is not identical to that for the events recorded in the experiment. This difference in practice
induces very marginal effects on $m_t$, as one can
check them by making a slight change for the phase space in the
event simulation.} 
Similarly, the determination of the $B$ hadron decay length 
does not involve directly any jet-energy measurement, as it is based
on the identification of the secondary vertex in the event through
tracking: therefore, this 
measurement is clean from jet-calibration issues.

For all methods involving the tagging of exclusive hadronic final
states, \emph{e.g.,} the di-lepton $J/\psi$ final state mentioned
above, we expect 
the following number of recorded events at 14 TeV LHC:
\begin{equation}
  N_{\text{ev}}\sim10^{5}\frac{\mathcal{L}}{1\text{ab}^{-1}}\frac{\epsilon_{\text{tagging}}}{10^{-4}}\cdot\epsilon_{\text{cuts}}\cdot BR(t\bar{t})\,, \label{eq:Nevents-Bhadron-tag}
\end{equation}
where $\mathcal{L}$, $\epsilon_{\text{tagging}}$, $\epsilon_{\text{cuts}}$, and $BR(t\bar{t})$ denote the integrated luminosity in ab$^{-1}$, the tagging efficiency for hadronic objects of interest, the fraction of events satisfying the associated selection cuts, and the branching ratio of $t\bar{t}$ into a given final state, respectively. 

For the $J/\psi$ clean di-lepton tagging, $\epsilon_{\text{tagging}}$
will be ${\cal O}(10^{-5})$, just taking into account branching fractions.
Adding up all decay modes of $B$ hadrons that can be reconstructed
from just charged tracks (\emph{e.g.,} $B^{+}\to K^{+}\pi^{+}\pi^{-}$),
we may increase the number of events by a factor ${\cal O}$(1-10). Therefore we adopt $\epsilon_{\text{tagging}}\sim10^{-4}$ as baseline in
eq.~(\ref{eq:Nevents-Bhadron-tag}).
Clearly these methods face the challenge of collecting the sufficient number of events
in order to fulfill precision
mass measurements; for $\epsilon_{\text{cuts}}\cdot BR(t\bar{t})\lesssim1$
and an integrated luminosity of several ab$^{-1}$,
the desired statistics
may nevertheless be reached \cite{CMS-PAS-FTR-13-017}.
Although not being affected by uncertainties from jet measurements, in
 the $J/\psi$ analysis a crucial role is nevertheless played by 
 the hadronization of $b$ quarks into $B$-hadrons and the associated
 Monte Carlo uncertainty.
Reference~\cite{CMS-PAS-FTR-13-017} estimates that the contribution
due to bottom-quark fragmentation is
about 300 MeV, for a total theoretical error of 900 MeV

In case of less exclusive requirements on $B$-hadron decays, it
is possible to obtain larger data samples. This would be the case,
for instance, of an analysis based on $B\to\ell+X$, which may allow to extract
$m_t$ from the $m_{\ell\ell}$ distribution obtained by pairing up the \emph{soft }lepton from $B$ decay and the
lepton from the $W$ boson in the same top-quark decay. This method
has been employed by CDF~\cite{Aaltonen:2009zl} and might be used
in LHC data as well.

Exactly as it is the case for mass measurement from the $J/\psi$ method, 
in a mass determination that uses soft leptons from  
$B$ hadrons, we should expect a sensitivity
to the showering and hadronization description of the event and corresponding systematics.
A similar conclusion holds for
methods that use the decay length of $B$ hadrons~\cite{Hill:2005dq,CMS-PAS-TOP-12-030}
and those that exploit the mass of the secondary vertex from charged
particles \cite{CMS-Collaboration:2016eu}. 
It is the purpose of this work to study the hadronization
uncertainties
 that affect all these methods.

Understanding
the transition of $b$ quarks into $B$ hadrons is
a crucial step to correctly predict the observables involved in the above-described mass measurements.
This aspect of the physics of quarks and hadrons is ruled by non-perturbative
QCD and it is clearly a challenge to describe it accurately starting
from first principles. One possibility to fill in this gap
is to advocate a factorization of long-distance
effects, namely the hadronization mechanism, from the hard-scattering
events. In this way it is possible to measure from data a
non-perturbative \emph{fragmentation
  function}  $D_{H,q}(z)$, accounting for the 
transition of a quark $q$ into a hadron $H$, carrying
an energy fraction $z$. Most
precise measurements of these fragmentation functions have been obtained
at LEP~\cite{Abdallah:2011az,Aleph:2001sy} and SLD~\cite{Abe:2002fc}
experiments from which
it is possible to obtain
$D_{H,q}(z)$ at an energy scale $Q^{2}$ of the order of the inelasticity
of the reaction. For processes
$e^{+}e^{-}\to b\bar{b}$ at the LEP, it is $Q^{2}\simeq m_{Z}^{2}$:
once known at some $Q^{2}$, the fragmentation functions can be evolved
to different $Q^{2}$ by means of perturbative evolution equations,
analogous to those used for parton distribution functions.

Bottom-quark fragmentation is usually described  using the
perturbative-fragmentation formalism \cite{mele}:
up to power corrections ${\cal O}(m_b^2/Q^2)$, the heavy-quark spectrum
can be factorized as a massless coefficient function and a
process-independent
perturbative fragmentation function, associated with the transition
of a massless parton into a heavy quark.
Such an approach allows a straightforward resummation of
the large mass $\ln(m_b^2/Q^2)$ and threshold contributions, $\sim[1/(1-z)]_+$
and $\sim [\ln(1-z)/(1-z)]_+$, which become large for $z\to 1$,
corresponding to soft- or collinear-gluon radiation.
In the perturbative-fragmentation approach, the state of the art
is next-to-leading-order (NLO) accuracy in
$\alpha_S$, with next-to-leading logarithmic
(NLL) resummation: Refs.~\cite{cno,ccm,corc} apply this formalism
for bottom fragmentation to $e^+e^--$ annihilation, top and Higgs
decays, respectively. Heavy-hadron spectra are then obtained
by convoluting the heavy-quark energy distribution with
non-perturbative fragmentation functions, such as the Kartvelishvili
\cite{kart} or Peterson \cite{peter} models, containing
few parameters which must be tuned to experimental data.

More recently, the approach based on Soft Collinear Effective
Theory (SCET) and Heavy Quark Effective Theory (HQET) was used
in the NNLO+NNNLL approximation in order to determine the $b$-quark
fragmentation function from $e^+e^-$ annihilation at the $Z$ pole
\cite{Fickinger:2016fk}. The authors found a general good agreement
with the data, although some discrepancy is still present for
large values of the energy fraction, where one is mostly sensitive
to non-perturbative effects.

Although having lately become very accurate, calculations based on the
fragmentation-function formalism are nonetheless too inclusive
for a complete description of the final states and
in general they need to be performed independently for each observable
and hard-scattering reaction.
Moreover, this approach is based on factorization, so it 
is not obvious whether the non-perturbative fragmentation function measured 
in a color-neutral environment, such as $e^{+}e^{-}$ machines,
could be straightforwardly extended to
colored parton scatterings at hadron colliders.

As an alternative to the fragmentation function approach, there have been attempts to describe the formation of hadrons with phenomenological
models inspired
by some features that can be derived from first principles, albeit
depending on a few parameters;
namely the string model~\cite{Andersson:1983iaa,Bowler:1981sb}
or the cluster model~\cite{cluster}.
This sort of approach is most common in shower Monte
Carlo event generators,
which are usually shipped with a ``tuning'' of the phenomenological
parameters in such a way that they are tailored
to reproduce a large set of previously
available data like $e^{+}e^{-}\to {\rm hadrons}$.
This approach is based on the universality of the hadronization
transition, which, in principle would allow one to use
models tuned at $e^+e^-$ machines even at hadron colliders. However,
as discussed above, it is not guaranteed that they will still
be reliable in a colored environment, wherein even initial-state
radiation and underlying event will play a role.
Furthermore, the number of parameters involved is usually quite large,
especially in the string model, and there is no clear factorization
between perturbative and non-perturbative physics: 
every time one tries to tune a specific sample of experimental data,
one should be warned that the possible agreement with other data may be
spoiled.

Moreover, when using 
fragmentation functions and hadronization models, it 
is hard to make a reliable estimate of theory uncertainties. In general,
there are pitfalls in carrying the properties of 
$e^{+}e^{-}$ measurements to hadronic
machines;
although it is assumed that they will be accurate enough to capture correctly
effects of order $\Lambda_{QCD}$, the evaluation of the systematics due
to this assumption is difficult to assess.
For this reason
it would be desirable to determine the
hadronization parameters directly in hadronic collisions, so to remove
all the issues tied with the use of $e^{+}e^{-}$ data. 

Extracting accurate fragmentation functions at hadronic colliders
might bring about
a formidable task. In fact, one needs to overcome the
difficulty to obtain precise data and match
them with sufficiently accurate theory calculations.
For the purpose of top-quark phenomenology, this will likely need
an NNLO accuracy in order not to spoil a sub-GeV accuracy on the
top-quark mass
determination.\footnote{This seems to apply to
  the extraction of $m_t$ from the first Mellin moment
of the $B$-lepton invariant mass $\langle m_{B\ell}\rangle $,
as for this observable Ref. \cite{Biswas:2010vq} reports around 1.5-2.0
GeV error using an NLO prediction. Other observables might need even
higher-order calculations. In fact, attempts with NLO accurate fragmentation
functions have shown that the error on the top quark mass might be
up to few GeV \cite{Agashe:2016xq}.}
Furthermore, the complexity of reconstructing the top-quark rest frame
impedes the direct measurement of energy-fraction
variables such as the $x_{B}=2\cdot E_{B}/\sqrt{s}$ measured at LEP
and SLD experiments, where the $Z$ rest frame coincides with the
laboratory frame.

The use of hadronization models for precision physics at hadron colliders
is challenging because simple models have had hard time
to reproduce accurate experimental results
\cite{Gieseke:2004sh,Bellm:2015rm,Karneyeu:2014uq}
and elaborate models with many parameters are needed to describe $e^{+}e^{-}\to hadrons$ data.
The large number of phenomenological parameters needed to describe
present data makes it difficult to find a unique favorite set that best reproduces them. Even if such ``best fit'' point is found
(or hand-picked), it is often the case that the determination of each
parameter is tied to that of the others contained in the model.
Such a non-trivial correlation makes it hard
to evaluate the impact on physical observables of the change of each
parameter, hence to define a theory uncertainty due to the modeling
of hadronization. 

A further complication arises from the fact that any hadronization
model is strictly related to the 
the parton cascade, hence even
the non-perturbative parameters are closely entangled with
those characterizing the shower. All of this shows that it
is necessary to carry out a coordinated study to constrain both
hadronization and shower parameters from
data, including possible $pp\to t\bar{t}$ samples, in order to
feel confident with the fits and their possible use in other
environments.
This was the purpose
of the tunes of the PYTHIA 8 generator,
such as the Monash tune~\cite{Skands:2014fj},
which we will take as baseline for our study, and of the efforts in the ATLAS
collaboration~\cite{ATL-PHYS-PUB-2015-007} towards adding $t\bar{t}$
data in the fits.

Color reconnection is another source of non-perturbative uncertainty in $m_t$, accounting for about 300 MeV in the estimate of the error in the
world-average analysis \cite{wave}.
Although color reconnection can even take place, {\it e.g.,} in $e^+e^-\to WW \to 4$~jets processes at $e^+e^-$ colliders, events where
bottom quarks in top quark decays are color-connected to initial-state antiquarks do not have their counterpart in $e^+e^-\to b\bar b$ annihilation. Therefore, color reconnection in $t\bar t$ events should deserve further investigation and may even need Monte Carlo retuning at hadron colliders.
Studies on the impact of color reconnection
on $m_t$ were undertaken in Refs.~\cite{spyros,corc1},
in the frameworks of PYTHIA and HERWIG, respectively.
Ref.~\cite{spyros} investigated
the description of color reconnection within the PYTHIA
string model, finding that, even restricting to the
most sophisticated models, it can have an impact up 
to 500 MeV on the top mass.
On the other hand, Ref.~\cite{corc1} addressed this
issue by simulating fictitious top-flavored hadrons in HERWIG
and comparing final-state distributions, such as the $BW$ invariant
mass, with standard $t\bar t$ events. In fact,
assuming that top-flavored hadrons decay according to the
spectator model, the $b$ quark
is forced to connect with the spectator or with antiquarks in its 
own shower, so that color reconnection with the initial state
or with $\bar t$-decay products is suppressed.

In this work we study the impact of the hadronization model parameters
on precision measurements of the top-quark mass, for which a target
accuracy around 0.5\% is set. The aimed accuracy is about the uncertainty 
where current most precise measurements of $m_t$ tend
to accumulate, mainly because of the limited knowledge
of the jet energy. As
the study of $B$-hadrons is motivated by going beyond the present
limitations, this is clearly the precision level
that hadronic observables need to reach in order to be useful. This
accuracy corresponds to an absolute error $\delta m_{t}\simeq\Lambda_{\rm QCD}$,
which for hadronic observables implies the necessity to carefully
model hadronization effects, whose impact on the single 
$B$-hadron kinematics
is naturally of the same order of magnitude as $\Lambda_{\rm QCD}$. 

In the following, we study several observables that have been discussed
in the literature to measure the top mass, and assess the error
on $m_{t}$ that would stem from a variation of hadronization
and related shower parameters. As some of these observables have been
proposed in studies of $b$-jets, we will define analogous variables
for single $B$-hadrons: for example, a measurement
of the invariant mass $m_{bl}$ of the $b$-tagged jet and a charged
lepton has been modified to the invariant mass formed by the $B$-hadron and
a charged lepton. For each observable, we derive the sensitivity
of the top quark mass to the parameters of hadronization and showering.
We then collect results in the tables of the following sections, expressing them in terms of logarithmic derivatives like:
\begin{equation}
\Delta_{\theta}^{(O)}\equiv\frac{\bar{\theta}}{\bar{O}}\frac{\partial O}{\partial\theta},\label{eq:DeltaGenericDef}
\end{equation}
where $O$ is a generic observable in $t\bar t$ events, 
the bar in $\bar{O}$ and $\bar{\theta}$ indicates an average value in a given range,
and $\theta$ denotes a generic Monte Carlo hadronization
or showering parameter. While studying these sensitivities,
we shall compute the
precision in the knowledge of the value of these parameters 
that is necessary to achieve in order to warrant a sub-GeV
theoretical error on the $m_t$ determination from hadronic
observables. 

Furthermore, we carry out similar studies to quantify
the sensitivity to showering and hadronization parameters of observables
that are \emph{not} sensitive to the top-quark mass.\footnote{An example in the literature is the $B$-hadron energy
fraction in the top-quark rest frame \cite{mescia}.}
We henceforth name these quantities 
``calibration observables'' 
which are useful to gauge the suitability of the tunes obtained elsewhere
to describe hadronization in the top-quark specific context.
As the calibration observables do not (or very mildly) depend on $m_t$, the
measurements of these observables can be used to constrain the showering
and hadronization parameters that have the greatest impact on the
observables sensitive to $m_t$, thus resulting in an improvement
of the accuracy and the robustness of the mass measurement.
A combined determination from data of both $m_t$ and 
Monte Carlo variables could be called an ``in situ'' calibration
of the nuisance theoretical parameters that affect the top-quark mass determination with hadronic observables. In a way this could be thought as a strategy similar to the in-situ calibration of the jet-energy correction obtained from the reconstruction of the hadronic $W$ boson peak in top quark events of Ref.~\cite{CMS-PAS-TOP-14-001}. Here, of course, we would be calibrating parameters of the phenomenological hadronization models and of the parton shower. The use of $N+1$ observables, where $N$ are mostly sensitive to Monte Carlo parameters and one is mostly sensitive to $m_t$, could be considered as an ultimate strategy for the top quark mass determination or for the cross-check of theoretical uncertainties in the global picture of top quark mass determinations at the Large Hadron Collider. In the following, combining the results on calibration observables and mass sensitive ones, we will outline the ingredients and the inherent difficulties of such analysis. 

  To deliver our ideas efficiently, the paper is organized as follows. In Section~\ref{sec:parameters}, we begin with identifying relevant hadronization and shower parameters in PYTHIA 8 and HERWIG 6,
  together with their variation ranges to compute associated
  sensitivities. We then discuss kinematic observables to be used throughout this paper and the reconstruction schemes for the variables carrying some subtlety in Section~\ref{sec:observables}. Section~\ref{sec:calibration} is devoted to reporting our results on the calibration observables, while we present 
  our analysis on observables sensitive to $m_t$ in Section~\ref{sec:topmass}. Our conclusions and outlook will be presented in Section~\ref{sec:conclusions}.
  Finally, two appendices are reserved for discussing some details of our
  computation.

\section{Parameters and calculations \label{sec:parameters}}

Hereafter we shall study the process
\begin{equation}
pp\to t\bar{t}\to \ell^{-}\ell^{+}b\bar{b}\bar{\nu}\nu,\label{eq:process}
\end{equation}
using PYTHIA~8.2~\cite{Sjostrand:2014rr}
or HERWIG~6~\cite{Corcella:2001fp}, namely LO hard scattering,
parton showers and hadronization according to the string
\cite{Andersson:1983iaa,Bowler:1981sb} or cluster
\cite{cluster} models, respectively.
In the simulation of $t\bar t$ events, standard
HERWIG and PYTHIA factorize production
and decay phases and the top-quark width is neglected.
Such effects are included in
the new $b\bar b4\ell$ generator \cite{bb4l}, implemented in the framework
of the POWHEG-BOX \cite{powheg}, which
simulates the full $pp\to b\bar b\ell^+\ell^-\nu_\ell
\bar\nu_\ell$ at NLO, including the interference between top production and
decay, as well as non-resonant contributions.
Investigations of width effects and full NLO corrections to top
decays are nevertheless beyond the scopes of the present paper
and we defer it to future explorations.
We point out, however, that both HERWIG and PYTHIA implement
matrix-element corrections to parton showers in
top decays, along the lines of
\cite{corsey} and \cite{norb}, respectively, and therefore hard-
and large-angle gluon radiation in $t\to bW$ processes will be
taken into account in the results which we will present hereafter.

\subsection{Variation of PYTHIA parameters}

In order to reproduce the data, a number 
of hadronization parameters can be tuned; some of them
are specific to heavy flavors and the formation
of $B$ hadrons, while others are meant to describe hadronization of 
light flavors. In our study we concentrate on heavy-flavor parameters
and study variations of the Lund string model
$a$ and $b$, by considering deviations from the best-fit values which are
used to describe light-flavored hadrons. Furthermore, we will also vary
the $r_{B}$ parameter, the Bowler modification of the 
fragmentation function for massive quarks,
which reads \cite{Bowler:1981sb}:
\begin{equation}
f_B(z)\sim{1\over{z^{1+b\,r_Bm^2_q}}}(1-z)^a\exp\left(-\frac{b\,m_T^2}{z}\right).
\label{bow}
\end{equation}
Here the $z$ variable implies the fraction of $E+p_z$ taken
by the $B$ hadron under the assumption that a $b$ quark of energy $E$ moves along the $z$-axis, while $m_T$ symbolizes the associated $B$-hadron transverse mass. 

As discussed in the introduction,
hadronization models are specific to the showering to which they are
attached, and therefore it is important to study the sensitivity of
top-mass observables to variations of the showering quantities.
Parameters for which a sensitivity is expected include $p_{T,\text{min}}$ which is 
the scale at which the parton shower is stopped and hadronization
models take over, the value of the strong coupling constant at the
$Z$-boson mass that is used in the final-state parton shower,
and the value of the $b$ quark mass. In addition, 
we also check the effect induced by discrete choices in using the
$b$-quark or the $W$ boson as a recoiler to impose momentum conservation
when we have splittings from the $b$ quark in the shower.\footnote{The choice between $b$ quark and $W$ boson as a recoiler is a discrete choice, hence it cannot be treated in the same way as we treat other continuous parameters. The calibration or tuning procedure described below can be 
regarded as a way either to find which recoil scheme should be adopted, or to make the recoil parameter become a continuous one and choose
the recoiler randomly in such a way that the $b$ and the $W$ act as a recoiler in a fraction of events suggested by the tuning.}

It is important to remark that the chosen range of variation of these
parameters has little importance for the calculation of the sensitivity defined in eq.~(\ref{eq:DeltaGenericDef}). In fact, the derivative
of the observables with respect to the Monte Carlo parameters is typically
stable across large ranges, and as will be shown later on, even 
the sensitivity $\Delta_\theta^{(O)}$ is largely independent of it.
What mainly leads to the choice of given ranges is to
have sufficiently large differences between spectra for different
parametrizations, so that the derivatives can be computed accurately,
without being overwhelmed by the statistical errors due to the finite
size of the Monte Carlo sample.

At the same time, we avoid taking values 
too far from the default, since this might generate
unforeseen changes in the Monte Carlo predictions.
For these reasons, we shall vary parameters up and down by at most 20\% of their central values and
evaluate numerically the derivatives in all
available data points. For a summary of the PYTHIA parameters
and the ranges of variation, we refer to Table~\ref{tab:varied}. 
{In principle, other parameters could be varied and tuned as they affect the kinematics of top quark events. For instance the initial state $\alpha_{s}$ parameter of \noun{Pythia8}, \noun{SpaceShower:alphaSvalue},
  could be considered.
  We choose to not investigate them, as they can be fixed from specific  measurements such as the jet multiplicity used in  \cite{CMS-Collaboration:af}.}

\begin{table}
\begin{centering}
\scriptsize{
	\begin{tabular}{|c|c|c|c|}
	\hline 
	 & \noun{Pythia8} parameter & range & Monash default\tabularnewline
	\hline 
	\hline 
	$p_{T,\text{min}}$ & \noun{TimeShower:pTmin } & 0.25-1.00 GeV & 0.5\tabularnewline
	\hline 
	$\alpha_{s,\text{FSR}}$ & \noun{TimeShower:alphaSvalue } & 0.1092 - 0.1638 & 0.1365\tabularnewline
	\hline 
	recoil  & \noun{TimeShower:recoilToColoured} & \emph{on} and \emph{off} & \emph{on}\tabularnewline
	\hline 
	$b$ quark mass & \noun{5:m0} & 3.8-5.8 GeV & 4.8 GeV\tabularnewline
	\hline 
	Bowler's $r_{B}$  & \noun{StringZ:rFactB } & 0.713-0.813 & 0.855\tabularnewline
	\hline 
	string model $a$ & \noun{StringZ:aNonstandardB } & 0.54-0.82 & 0.68\tabularnewline
	\hline 
	string model $b$ & \noun{StringZ:bNonstandardB } & 0.78-1.18 & 0.98\tabularnewline
	\hline 
	\end{tabular}
}
\par\end{centering}
\caption{\label{tab:varied}
Ranges and central values of the parameters that
we varied. Note that some values are not varied around the default
values of the Monash tuning. For instance we run
$r_{B}$ around the mid-point between \noun{Pythia6.4}
and \noun{Pythia8-Monash} values.}
\end{table}

\subsection{Variation of HERWIG parameters}

For the sake of comparison, we also investigate the impact of the HERWIG shower and hadronization parameters on the
top-quark mass measurement.
In fact, HERWIG and PYTHIA generators differ in several aspects:
for example, the ordering variables of the showers are not the same,
matrix-element corrections are implemented according to different strategies
and, above all, models for hadronization and underlying events are different.

As far as HERWIG is concerned, hadronization occurs according to the
cluster model, which is strictly related to the angular ordering of the
parton shower, yielding color pre-confinement
even before the hadronization transition.
In the following, we shall use the HERWIG 6 event generator,
written in Fortran language. In fact, 
the object-oriented
code HERWIG 7 \cite{Bellm:2017aa}
presents a number of improvements, especially when using the
new dipole shower model with the modified kinematics
for massive quarks \cite{Bellm:2017idv}.
However, although the latest version HERWIG 7.1 
exhibits remarkable improvements for the purpose of
bottom-quark fragmentation, 
the comparison with the $e^+e^-\to b\bar b$ data is still not optimal.

As for HERWIG 6, in
Ref.~\cite{Corcella:2005tg} the authors tried to tune it to
LEP and SLD data, finding that the comparison could be much improved
with respect to the default parametrization, although some 
discrepancy still persists and the prediction is only
marginally consistent with the data.
As a whole, we decided to follow \cite{Corcella:2005tg}
and stick to using HERWIG 6 in the present paper.
As done in the case of PYTHIA,
we identify the relevant shower and hadronization parameters
and vary them by at most 20\% around their respective default values.

The relevant parameters of the cluster model are the following:
 CLMSR(2) 
which controls the Gaussian smearing of a $B$-hadron with respect
to the $b$-quark direction,  PSPLT(2) 
which governs the mass distribution
of the decays of $b$-flavored clusters, and CLMAX 
and CLPOW 
which determine the highest allowed cluster masses.
Their default values and variation ranges are summarized in Table~\ref{tab:varied-1}.
Furthermore, unlike Ref.~\cite{Corcella:2005tg} which just accounted
for cluster-hadronization parameters, we shall also explore
the dependence of top-quark mass observables on the following parameters: 
RMASS(5) and RMASS(13), the bottom and gluon effective masses,
and the virtuality cutoffs, VQCUT for quarks and VGCUT for gluons, which are added to the parton masses in the shower (see Table~\ref{tab:varied-1}).
We also investigate the impact of changing
QCDLAM 
as it plays the role of an effective $\Lambda_{\rm QCD}$
in the so-called Catani--Marchesini--Webber (CMW) 
definition of the strong coupling constant
$\alpha_S$ in the parton shower
(see Ref.~\cite{Catani:1990rr} for the discussion on its relation with respect to the standard
$\Lambda_{\rm QCD}$ in the $\overline{\rm MS}$ scheme).
As well as for other paramters,  we vary QCDLAM around its default value, the  range of variation  is tabulated in Table~\ref{tab:varied-1}. 
\begin{table}
\begin{centering}
\scriptsize{%
	\begin{tabular*}{1\linewidth}{@{\extracolsep{\fill}}|>{\centering}p{0.49\linewidth}|c|c|c|}
	\hline 
	 & parameter & range & default\tabularnewline
	\hline 
	\hline 
	Cluster spectrum parameter & PSPLT(2) & 0.9 - 1 & 1\tabularnewline
	\hline 
	Power in maximum cluster mass  & CLPOW & 1.8 - 2.2 & 2\tabularnewline
	\hline 
	Maximum cluster mass  & CLMAX & 3.0 - 3.7 & 3.35\tabularnewline
	\hline 
	CMW $\Lambda_{QCD}$ & QCDLAM & 0.16 - 2 & 0.18\tabularnewline
	\hline 
	Smearing width of $B$-hadron direction &  CLMSR(2) & 0.1 - 0.2 & 0\tabularnewline
	\hline 
	Quark shower cutoff & VQCUT & 0.4 - 0.55 & 0.48\tabularnewline
	\hline 
	Gluon shower cutoff & VGCUT & 0.05 - 0.15 & 0.1\tabularnewline
	\hline 
	Gluon effective mass & RMASS(13) & 0.65 - 0.85 & 0.75\tabularnewline
	\hline 
	Bottom-quark mass & RMASS(5) & 4.6 - 5.3 & 4.95\tabularnewline
	\hline 
	\end{tabular*}
}
\par\end{centering}
\caption{\label{tab:varied-1}  \noun{Herwig}~6 parameters under consideration and ranges of their variation.}

\end{table}

\section{Observables \label{sec:observables}}

In this section, we identify the observables that we employ to study their sensitivities to parameter variations. We first discuss the observables relevant to top quark mass measurements, followed by our proposed variables for the tuning or the calibration of Monte Carlo parameters. For the formulation of how to compute these observables, we will follow the naming scheme for $t\bar{t}$ final states that we illustrate in Figure~\ref{fig:Schematic-view-of-event}.

\begin{center}
\begin{figure}
\begin{centering}
\includegraphics[width=0.45\paperwidth]{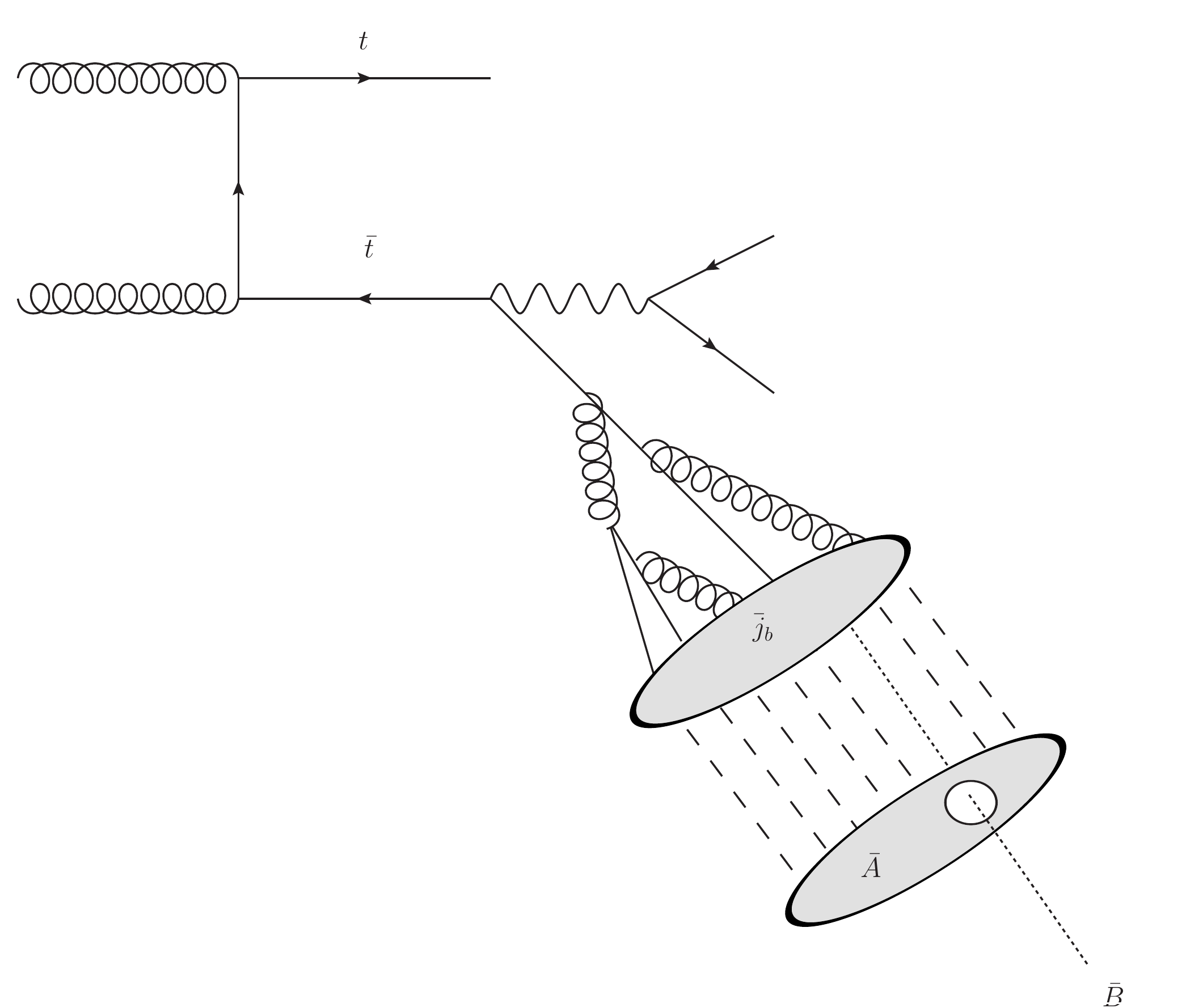}
\par\end{centering}
\caption{\label{fig:Schematic-view-of-event}Schematic view of the event, including
  the transformation of a $\bar{b}$ quark into hadrons, namely
  into a $\bar{B}$
and other light-flavored hadrons denoted collectively as $\bar{A}$,
which together form a $b$-tagged jet $j_{\bar{b}}$. Similarly, a $b$
quark coming from the $t$ gives rise to jets and hadrons denoted
by the same symbols without a bar.}
\end{figure}
\par\end{center}

\subsection{$m_{t}$-determination observables}

We first list up the observables that we consider for the top-quark mass measurement.
\begin{itemize}
\item $E_{B}$: energy of each tagged $B$-hadron;
\item $p_{T,B}$: $B$-hadron transverse momentum;
\item $E_{B}+E_{\bar B}$ and $p_{T,B}+p_{T,\bar B}$: sum of the energies
  and transverse momenta of $B$ and $\bar B$ 
  (in events where both $B$ hadrons are tagged);
\item $m_{B\ell}$: invariant mass of the $B$-hadron and one
lepton from $W$ decay 
(the prescription for combinatorial ambiguity arising in forming $m_{B\ell}$  will be discussed shortly);
\item $m_{T2}$~\cite{Lester:1999tx} and $m_{T2,\perp}$~\cite{Matchev:2009ad} of the $B$ and the $B\ell$ subsystems
defined below;
\item $m_{BB\ell\ell}$, the total mass of the system constructed by the two leptons and the two tagged $B$-hadrons in the event.
\end{itemize}
We choose these observables because some of them have already been
discussed in the literature to carry out measurements of $m_{t}$
from hadronic exclusive final states: for instance, 
$m_{B\ell}$ \cite{Biswas:2010vq}, $m_{T2}$ and its variants
\cite{CMS-PAS-TOP-11-027,CMS-PAS-TOP-15-008} and $E_{B}$ \cite{Agashe:2016xq}.
Furthermore, we consider observables that have been discussed for
$b$-jets, and can naturally extend them by
replacing $b$-jets with $B$ hadrons. For instance,
the sum of $B$-hadron energies $E_{B}+E_{\bar B}$ is inspired by
the analogous  $b$-jet quantity  $E_{j_{b}}+E_{j_{\bar{b}}}$
defined in \cite{Biswas:2010vq}.
To the best of our knowledge, the variable $m_{BB\ell\ell}$ has not been considered before. It is meant to be sensitive to $m_{t}$ as it probes the the total mass of the $t\bar{t}$ system. Therefore, 
we expect the bulk of its distribution
to be sensitive to the top mass in a way similar to the bulk of the $m_{B\ell}$ distribution. In the lower tail, the variable $m_{BB\ell\ell}$ is anticipated to have sensitivity to the top quark mass because of a threshold effect similar to the one in the proposal of Ref.~\cite{Alioli:2013mz}.

It is noteworthy
that the observables $E_{B}+E_{\bar B}$, $p_{T,B}+p_{T,\bar B}$, $m_{BB\ell\ell}$, and all the $m_{T2}$ variables require two tagged $B$-hadrons in each event, and therefore they may be hardly measurable with enough
statistics at the LHC. However, it is equally interesting to study them
to see how much information on the top quark mass could be present
in the kinematics of the whole event, rather than in a single
top-quark decay, probed by other observables such as $m_{B\ell}$,
$p_{T,B}$, and $E_{B}$. 
While it may be challenging to secure a sufficient
  number of events with fully-reconstructed $B$-hadrons,
  we also remark that for the measurements which do not actually demand tagged $B$-hadron(s), {\it e.g.},  
those based on generic tracks from the secondary vertex, 
it can be easier to achieve enough statistics.
Therefore, our study of these observables with $B$ hadrons is useful to see what can be gained using both sides of the $t\bar{t}$ events in these measurements.

\subsubsection{Reconstruction schemes}

For observables involving particle pairings, 
we need to specify a prescription: for example, in order to compute
$m_{B\ell}$, we need to choose which lepton should be paired with the tagged
$B$ in the event.

Speaking of $m_{B\ell}$ first, we define two variants.
In the first one, we pick the lepton that gives the smaller $m_{B\ell}$ for
each lepton charge, that is to say, for each event we obtain two
values, $\text{min}(m_{B\ell^{+}},m_{B\ell^{-}})$ and
$\text{min}(m_{\bar{B}\ell^{+}},m_{\bar{B}\ell^{-}})$,
and put both in relevant histograms. 
This selection scheme enables us to avoid leptons causing an endpoint violation of the true distribution
arising from the $B$ and $\ell$ in the same top decay. We
call such a computed invariant mass 
$m_{B\ell,\text{min}}$.
We remark that, in the 
situation in which only one $B$-hadron
is identified per event,
this simply corresponds to just picking $\text{min}(m_{B\ell^{+}},m_{B\ell^{-}})$.
In the second variant of $m_{B\ell,\text{true }}$ 
we use the event records to pair the $B$-hadron with the lepton
that comes from the same top quark as the $B$.

For more complex observables, such as $m_{T2}$, we have to specify
further how we deal with constituent particles in the event. The $b$-jet $j_{b}$
is made of the $B$ hadron and some other hadrons that we call $A$,
as shown in Figure~\ref{fig:Schematic-view-of-event}. In a computation
of $m_{T2}$ from $b$-jets, we need to calculate the transverse mass
in combination with $j_{b}$ and some vector in the transverse plane that stands for one source of missing transverse energy (mET), associated
with the neutrino in $W$ decay.
When dealing with the $m_{T2}$ of the $B$-hadron,
we are naturally led to replace $j_{b}$ by $B$, but this leaves
an ambiguity as of what to do with the rest of the jet. The hadrons
$A$ in $j_{b}$ are detectable particles, hence not
invisible, but at the same time they are extraneous to the
$B$-related transverse mass. For this reason, 
we are allowed to treat them
as either extra radiation or invisible particles. In
the latter case, they would be added to the mET of the event,
as if they were not measured; in the former case, they
would be regarded as extra radiation, as some
kind of upstream momentum in top-quark decay. We shall call these two
variants $m_{T2}^{\text{(mET)}}$ and $m_{T2}^{\text{(ISR)}}$,
respectively. For
a generic reaction 
\[
pp\to t\bar{t}\to \ell^{-}\ell^{+}+B+A+\bar{B}+\bar{A}+\bar{\nu}\nu+X,
\]
the two options would give rise to mET and ISR computed as in
Table~\ref{tab:Variants-of-reco}. In our study, we
shall also consider the $m_{T2,\perp}$ variable~\cite{Matchev:2009ad}, which was considered
at the jet level by CMS in \cite{CMS-PAS-TOP-11-027}.
The variable is defined in a similar fashion to the standard $m_{T2}$, but in the plane perpendicular to the total upstream transverse momentum, so that it is insensitive to ISR, which affects the dynamical details of the $m_{T2}$ spectra.
With the $m_{T2,\perp}$ variables, we therefore study the
relevant sensitivities in the context of decay kinematics of top quarks, excluding any  top quark production-related dynamics.\footnote{Of course, this statement assumes precise measurement of the upstream momentum. In this sense, production-related information may enter the variables indirectly.}

\begin{table}
\begin{centering}
\begin{tabular}{|c|c|c|}
\hline 
 & mET definition & ISR\tabularnewline
\hline 
\hline 
$m_{T2}^{\text{(mET)}}$ & $\text{mET=}-\sum_{i=\{B,\ell,X\}}\bar{p}_{T,i}$ & $X$\tabularnewline
\hline 
$m_{T2}^{\text{(ISR)}}$ & $\text{mET}=-\sum_{i=\{A,B,\ell,X\}}\bar{p}_{T,i}$ & $X+A+\bar{A}$\tabularnewline
\hline 
$m_{T2}^{\text{(true)}}$ & $\text{mET}=-\left(\bar{p}_{T,\nu}+\bar{p}_{T,\bar{\nu}}\right)$ & \makecell{$\sum_{x}$ for x not coming \\ form either $t$ or $\bar{t}$} \tabularnewline
\hline 
\end{tabular}
\par\end{centering}
\caption{\label{tab:Variants-of-reco}Variants of the reconstruction scheme
for $m_{T2}$.}
\end{table}

\subsubsection{Definition of the sub-system observables}

For the computation of the $m_{T2}$ variables it is possible to consider
different subsets of top-decay products as visible particles~\cite{Burns:2008va}.
Barring the light hadrons $A$, each top quark gives rise to a $B$-hadron
and a charged lepton $\ell$, with two possible B-hadron-accompanying options for their
treatment. We can opt to form a compound system of these two and
compute the transverse masses using $p_{B}+p_{\ell}$ as visible momentum
or ignore $\ell$ and calculate the transverse
mass using the $B$-hadrons and the mET,
including the charged leptons as if they were invisible. These
two options are labeled:
\begin{itemize}
\item ``$B\ell$-subsystem'' $m_{T2,B\ell}$: the transverse mass is computed
from $p_{B}+p_{\ell}$ for some pairing of $B$ and $\ell$ with only the
neutrino considered as invisible and the relevant test invisible mass is set to the neutrino mass;
\item ``$B$-subsystem'' $m_{T2,B}$: the transverse mass is computed
  from $p_{B}$ with the $W$ treated as invisible and the relevant
  test invisible mass is set to the true $W$ mass.
\end{itemize}
Both options for $m_{T2}$ yield distributions with a kinematic
endpoint at $m_{t}$, hence they can be used to measure the top mass. 
The ``$B$-subsystem'' mass $m_{T2,B}$ has the virtue
that it does not require any hypothesis on which lepton should be paired with
each $B$. The ``$B\ell$-subsystem'' $m_{T2,B\ell}$,
on the contrary, faces the issue of pairing leptons and hadrons,
but generally it has the greatest sensitivity to
$m_t$, especially in the bulk of the distribution, where $m_{T2,B}$ instead
exhibits a very mild dependence 
 on $m_{t}$.
Overall, the two options
offer different pros and cons, making a comparison of
their performances rather interesting.

\subsection{Calibration observables}

Even in presence of a perfect fit to precise $e^{+}e^{-}$ data, there
is no guarantee that the Monte Carlo tuning will be a good enough
description of LHC data (fragmentation properties of light
flavors have proven this to be the case \cite{dEnterria:2014mq}).
It is therefore crucial to undertake checks of these tunings
specific for light- and heavy-quark physics. 

With this motivation in mind, we study  ``calibration observables'', 
i.e. quantities that we expect to
have no or little sensitivity to $m_t$ but still depend
on showering and hadronization in $t\bar{t}$ events.
Exploiting this feature, 
one can use these observables
to check the accuracy of the tunings or to constrain the Monte Carlo
parameters ``in-situ'' for the mass measurement. In the latter
case, the top-quark sample would be used to constrain simultaneously
a number of showering and hadronization quantities from the data,
as well as $m_t$, in the same manner as
the jet energy scale \emph{in-situ} calibration carried
out in \cite{CMS-PAS-TOP-14-001}. 
We highlight the fact that in this in-situ calibration procedure  one may end up with a point of
Monte Carlo parameter space that may or may not be close to the point used of the standard tunings, $e.g.,$
the Monash one. The best-fit calibration point, in fact, is very specific to 
top quark events and in principle even to the selection cuts used to isolate top quarks from background. Therefore, it has a rather 
different meaning than usual tunings, and we do not expect it to be a Monte Carlo set-up usable outside the domain of 
the top quark mass measurement.

Furthermore, we point out
that our calibration is intended only for
  the purpose of reducing
the impact of fragmentation uncertainties on the $m_t$ measurements.
For this reason, one might even think of studying the response
of the observables to changes in the hadronization model in dijet
events with two $b$-tags and better constraining the Monte Carlo
parameters: however, this would go in the direction of producing a tuning,
rather than an \emph{in-situ} calibration. In the following, we shall
not discuss
the use of dijet and other heavy-flavor production processes, but we
just leave them as a possible validation of the calibration procedure.
Clearly, if a stark deviation from the process-universality of the
Monte Carlo
parameters had to be observed, this should raise questions on 
the appropriateness of
the theory description employed to model $t\bar{t}$ events
and other heavy-flavor production.

Studies to identify observables useful for a calibration specific
to top quark have been carried out by ATLAS in the tuning work 
 \cite{ATL-PHYS-PUB-2015-007}, as well as in 
Ref.~\cite{tancrediTopLHCWG14}. These investigations
consider the observables: 
\begin{itemize}
\item $p_{T,B}/p_{T,j_{b}}$: the ratio of the $B$-hadron transverse momentum {magnitude}
over that of the $b$-jet;
\item $\rho(r)=\frac{1}{\Delta r}\frac{1}{E_{j}}\sum_{\textrm{track}}E(\textrm{track})\theta\left(\left|r-\Delta R_{j,\textrm{track}}\right|<\delta r\right)$:
  the radial jet-energy density, defined and measured
  as in \cite{ATLAS-Collaboration:2013tx}.
\end{itemize}
We remark that these observables are sensitive to the presence of
the heavy hadron and to the energy distribution in
the jet. Hence, they are suitable to probe the dynamics of the conversion
of a single parton into a hadron, but feel only indirectly the effects
of other partons (hadrons) in the event.
However, for the precision we are aiming
at, it is important to test the possible cross-talk among partons
in the event: in fact, in the
hadronization transition, each parton is necessarily 
connected with the others, 
due to the need to form color-singlet hadrons. To probe these global
effects in the formation of hadrons, we
study the following variables:
\begin{itemize}
  \item $\chi_{B}=2E_{B}/X_{B}$,
   where possible
   options for $X_B$ are 
   $m_{j_{b}{j}_{\bar{b}}}$, $\sqrt{s_{\min}}$,
   \ $\left|p_{T,j_{b}}\right|+\left|p_{T,{j}_{\bar{b}}}\right|$ and
   $E_{j_b}+E_{\bar j_{\bar b}}$.
\end{itemize}
These variables are sensitive (in different manners) to the existence
of a $b\bar b$ system, hence they probe hadronization in a more global
way than the single $b$-jet-associated observables examined in
\cite{tancrediTopLHCWG14} and \cite{ATL-PHYS-PUB-2015-007}. All
the options for $X_{B}$ 
tend to $\sqrt{s}$ in $e^{+}e^{-}$ collisions or
any other fixed partonic
center-of-mass energy. In this context, this property is useful
as it allows a more direct comparison to $e^{+}e^{-}$ data, if necessary.

\begin{figure}
\begin{centering}
\includegraphics[width=1\linewidth]{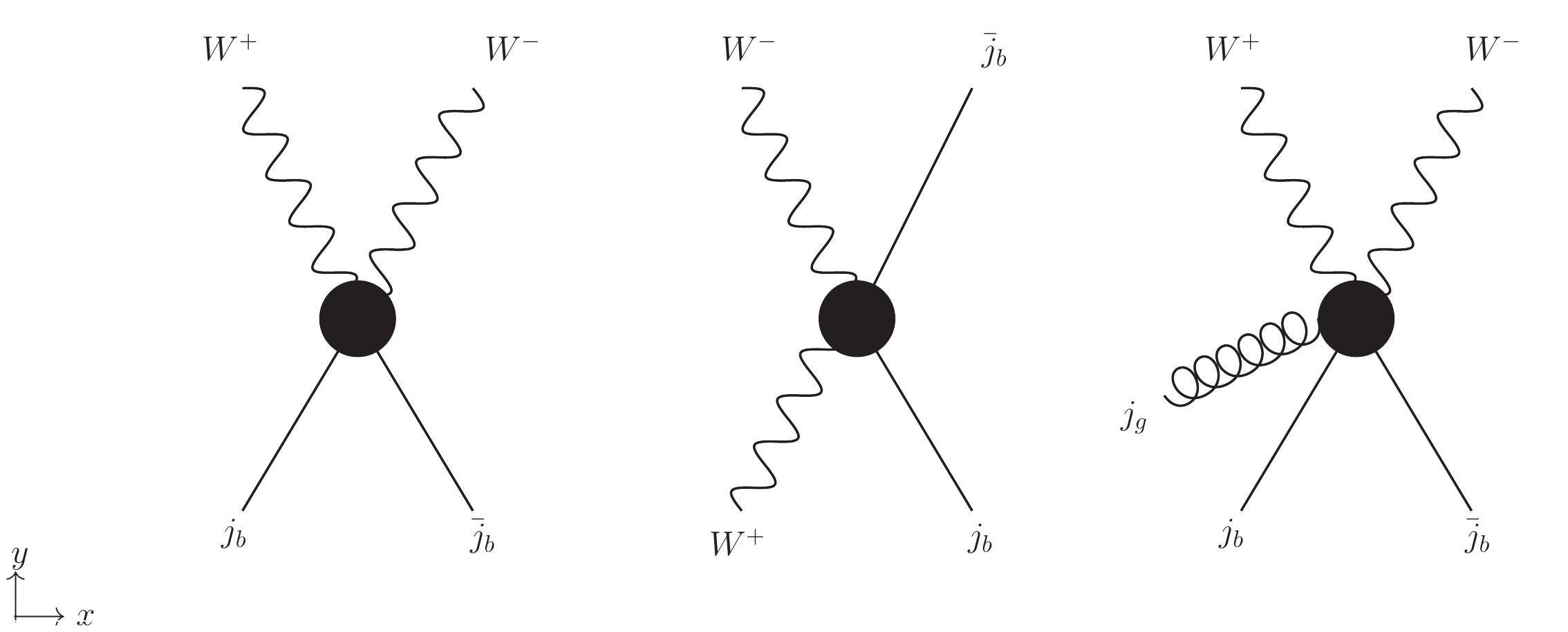}
\par\end{centering}
\caption{\label{fig:Three-kinematical-configurations} Spacial representation of three kinematic configurations of the $W$ bosons and jets in the transverse plane. The three configurations depicted can be distinguished looking at several  $\chi_{B}(X)$ variables. The first two can have the same
$\left|p_{T,j_{b}}\right|+\left|p_{T,{j}_{\bar{b}}}\right|$, but different
$m_{j_b j_{\bar{b}}}$, whereas the first and the third differ for $\sqrt{s_{\min}}$,
despite having the same $m_{j_b j_{\bar{b}}}$
and $\left|p_{T,j_{b}}\right|+\left|p_{T,{j}_{\bar{b}}}\right|$.}
\end{figure}

The options for $X_{B}$ 
are sensitive to different aspects of the $pp\to bW^{+}\bar{b}W^{-}+X$
kinematics, such as the relation
between the $b$ quarks with
the initial state, which, being colored, can influence the hadronization.
For example, in Figure~\ref{fig:Three-kinematical-configurations}
we display two configurations (first and second event sketches) that
can have similar total hardness, {\it e.g.,}
measured by $\left|p_{T,j_{b}}\right|+\left|p_{T,{j}_{\bar{b}}}\right|$,
but might have considerably different $m_{j_b j_{\bar{b}}}$.
For small $m_{j_b j_{\bar{b}}}$,
the two $b$ quarks will tend to be more collinear, hence they have a greater
chance to interfere with each other when hadronizing. On the other hand,
the first and third event sketches might have similar
$\left|p_{T,j_{b}}\right|+\left|p_{T,{j}_{\bar{b}}}\right|$
and $m_{j_b j_{\bar{b}}}$, but the hadronization
may still differ because of the different center-of-mass energy,
which is  larger in the third type
of events. The notion of center-of-mass energy of the initial partons
in presence of invisible particles can be captured by the variable
$\sqrt{s_{\text{min}}}$ proposed in \cite{Konar:2011sf}, which
would be a discriminant between the third and the first type of event.

In addition, if one wishes to probe the whole kinematic phase-space
of the $b\bar{b}$ system and its hadrons, one can think of further
observables. For a probe of emissions from the $b$ quarks we can
use the relative azimuthal angle 
$\Delta\phi(j_{b}j_{\bar{b}})$, that is affected by $b\to bg$
splittings, hence it can be meaningful to constrain parton shower parameters. A
similar angular observable for $B$-hadrons $\Delta\phi(B\bar B)$ can
also be employed thanks to its sensitivity to the angular features of
the transition of quark into hadrons. Attempting to isolate the fragmentation
effect of these angular observables, in the following we will also
consider the difference between hadron and jet angular separation,
that is $|\Delta\phi(j_{b}j_{\bar{b}})-\Delta\phi(B\bar B)|$. Besides employing
the azimuth $\phi$, we use the complementary information of 
polar angles in the distributions of 
$\Delta R(j_{b}j_{\bar{b}})$, $\Delta R(B\bar B)$,
and $|\Delta R(j_{b}j_{\bar{b}})-\Delta R(B\bar B)|$.\footnote{In general,
  given two particles at azimuthal angles $\phi_1$ and $\phi_2$ with
  rapidities $\eta_1$ and $\eta_2$, it is
  $\Delta R=\sqrt{(\phi_1-\phi_2)^2+(\eta_1-\eta_2)^2}$.}
Moreover, we consider, as possible
candidate calibration observables, the mass of the $b$-jet, $m_{j_{b}}$,
which might help to constrain the shower parameters, and the ratio of
the invariant
mass of two $B$-hadrons over the mass of the two jets,
namely $m_{B\bar B}/m_{j_bj_{\bar b}}$. 

A caution must be taken: some of the observables might suffer from
experimental effects that hinder the use of their precise measurement
for a calibration. For instance, the $b$-jet mass or the
$X_{B}$ variables used to normalize $\chi_{B}$ are, in principle,
sensitive to jet energy scale uncertainties, which may spoil
the accuracy on the calibration observables.
We shall therefore consider two scenarios for the achievable
precision on the calibration observables: an optimistic scenario in which
the jet energy scale uncertainties do not inflate the total uncertainty,
and another one where the accuracy is
limited by these uncertainties at around 1\%.

As a further exploration,
we also study variables where jets are avoided altogether. One of them is the jet radial energy density $\rho(r)$ defined above, which needs jets only to identify the hadrons that enter the calculation.  
In addition  we also consider other observables such as the ratios:
\begin{itemize} 
\item{$\frac{E_{B}}{E_{\ell}}$} 
\item{$\frac{E_{B}}{E_{\ell}+E_{\bar{\ell}}}$,} 
\end{itemize}
in which we use the lepton energy to make a
dimensionless quantity of the energy out of a $B$-hadron. 
For these variables there is no use of jets  so that a measurement with systematic uncertainties far below the percent level is achievable.

\subsubsection{Variants of the calibration observables}

Due to the complexity of the $t\bar{t}$ decay final state, it is possible
to conceive variants of the above defined $\chi_{B}$. For instance,
the inclusion of leptons in the ``visible'' system used for the
computation of $\sqrt{s_{\text{min}}}$ is somewhat arbitrary. Furthermore,
if the calibration observables were to be studied in a high-$p_T$ sample
of $pp\to b\bar{b}$ events, there would be no leptons at all so that a
comparison of resulting outputs to $t\bar{t}$ results would not be straightforward.
Therefore, in addition to the standard definition of $\sqrt{s_{\text{min}}}$
in Ref.~\cite{Konar:2011sf}, which is written in terms of the ``visible'' transverse momentum
\begin{equation}
\sqrt{s_{\text{min}}}(M_{\text{inv}})=\sqrt{E_{\text{vis}}{}^{2}-P_{\text{z,vis}}^{2}}+\sqrt{\text{mET}^{2}+(\Sigma_{\text{inv}}M_{\text{inv}})^{2}} \,,\label{eq:smin}
\end{equation}
where $M_{\text{inv}}$ are hypothetical masses of invisible particles,
we also employ a modified version in which the corresponding quantity is computed
as if the charged leptons were invisible, hence part of the 
mET. 
We denote this definition as $\sqrt{s_{\text{min,}b\bar b}}$, where
$E_{\text{vis}}$ and $P_{z,\text{vis}}$ are expressed in terms of the
$b\bar{b}$ system only.
Although $M_{\text{inv}}$ can carry whatever values, in the case of $\sqrt{s_{\min}}$, the masses of the invisible particles are simply set to be the same as
the neutrino masses for LHC purposes.
By contrast, in the case that the whole $W$ boson is regarded as invisible, one may conceive to use the physical $W$-boson mass
for $M_{\text{inv}}$ in eq.~(\ref{eq:smin}). 
We have explicitly checked the effect of not taking the physical mass of the
$W$ boson on our results and observed little dependence on this arbitrariness;
therefore, throughout our analysis, we shall put $M_{\text{inv}}=0$
for $\sqrt{s_{\text{min,}b\bar b}}$.

\section{Results on the calibration observables \label{sec:calibration}}

\begin{figure}
\centering
\includegraphics[width=0.49\linewidth]{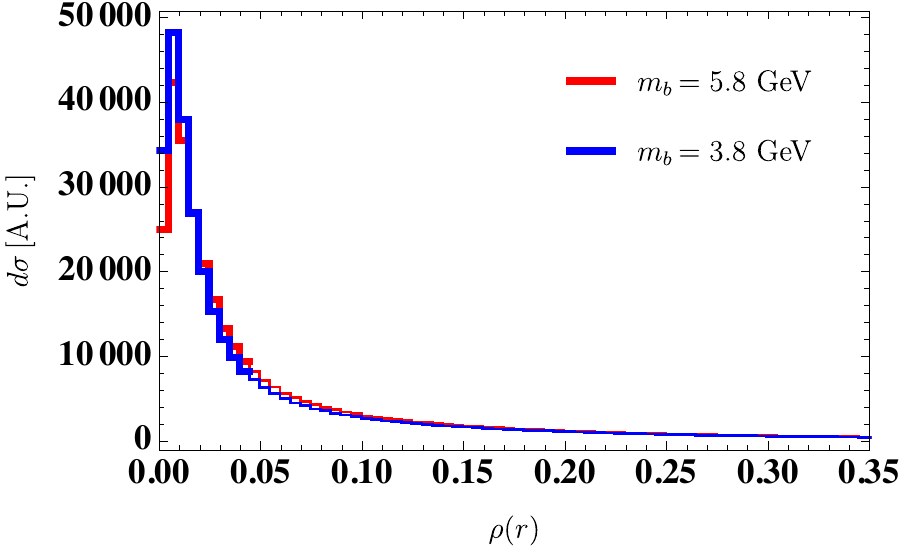}\includegraphics[width=0.49\linewidth]{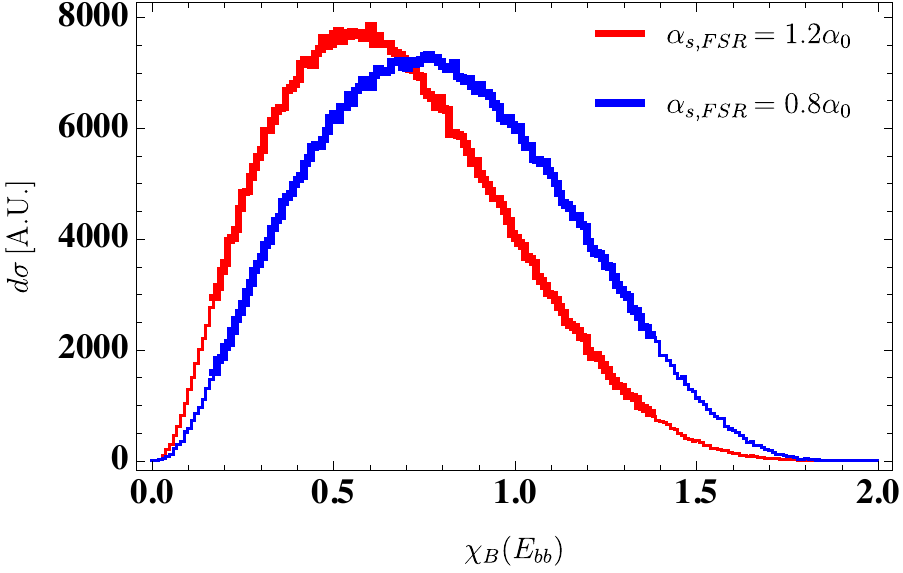}\\
\includegraphics[width=0.49\linewidth]{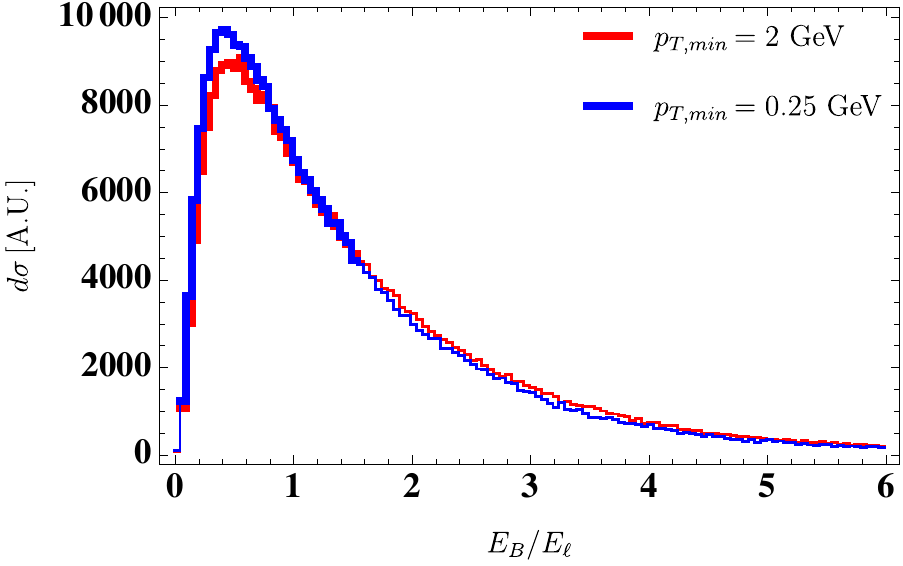}\includegraphics[width=0.49\linewidth]{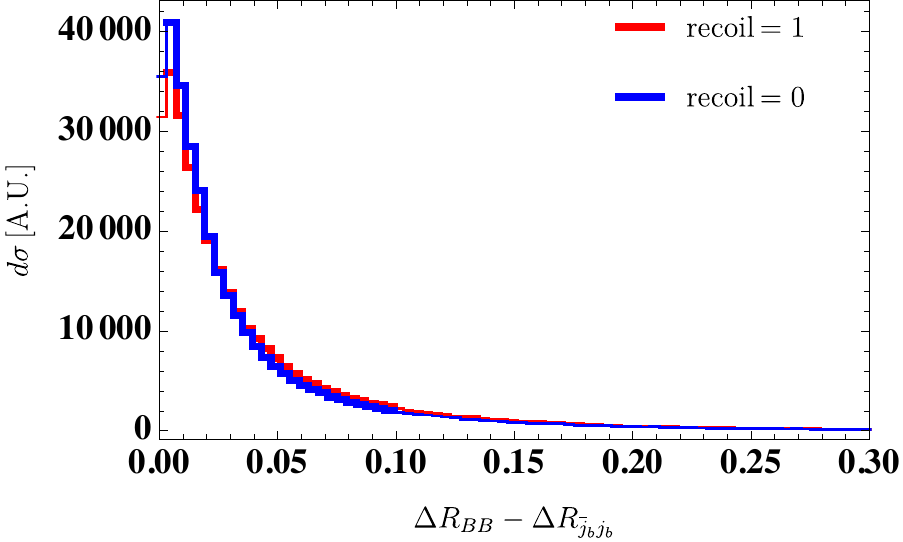}\\
\includegraphics[width=0.49\linewidth]{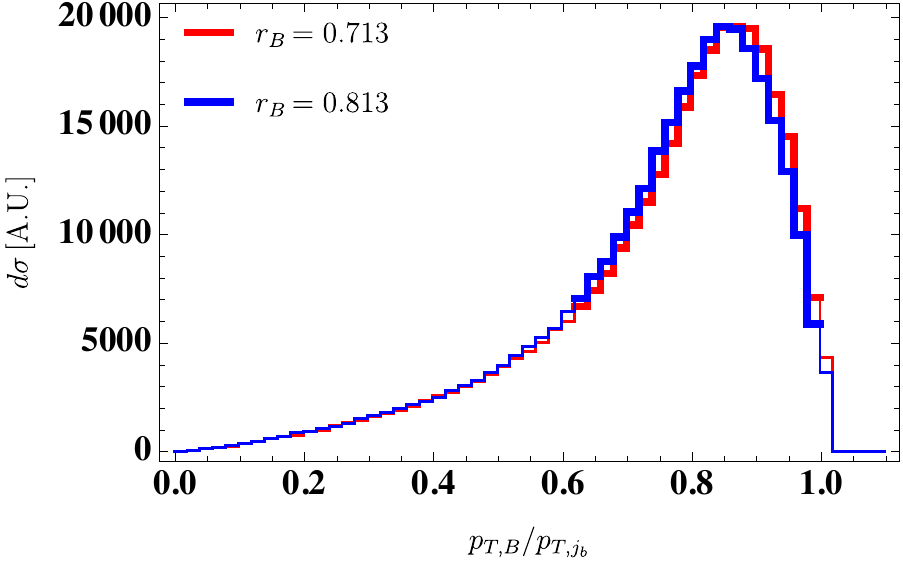}\includegraphics[width=0.49\linewidth]{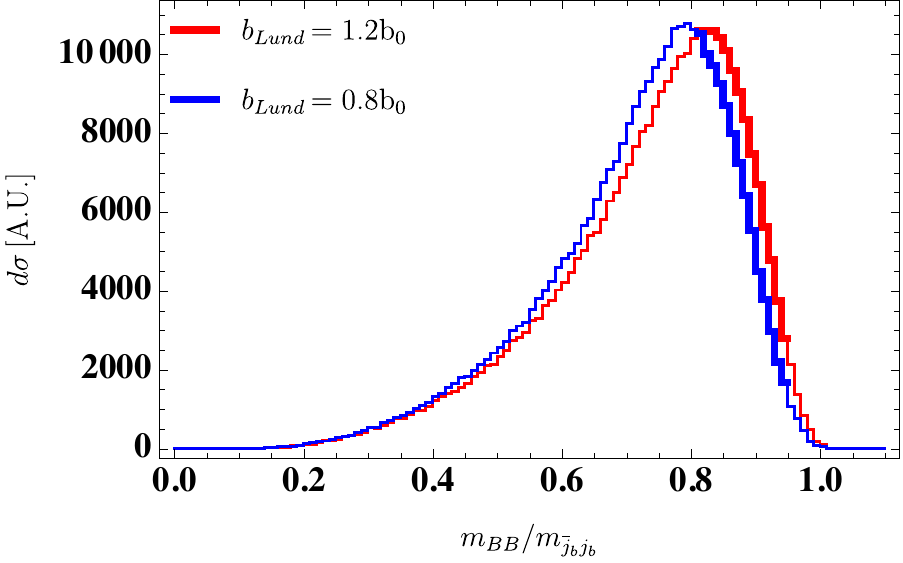}
\caption{\label{fig:Example-spectra-calibration-chi_B} Example spectra for $\rho(r)$, 
$\chi_{B}(E_{j_b}+E_{\bar{j}_b})$, $E_{B}/E_{\ell}$, $\Delta R(BB)-\Delta R(j_{\bar{b}}j_{b})$, $p_{T,B}/p_{T,j_{b}}$, and $m_{B\bar B}/m_{j_{b}\bar{j}_{b}}$ for $m_{t}=174$~GeV.}
\end{figure}

For a simple but effective characterization of the impact of the Monte Carlo
parameters, we start by computing the average values of the
calibration-observable spectra, also referred in the following as
the first Mellin moment, {\it i.e.,} $\mathcal{M}^{(\mathcal{O})}_{1}$.
The general expression of the $k$-th Mellin moment of a distribution with counts $\mathcal{O}_{j}$ and bins centered in $b_{j}$ reads:
\begin{equation}
\mathcal{M}^{(\mathcal{O})}_{k}=\sum_{j}b_{j}\mathcal{O}_{j}^{k}\,\label{mellindef}
\end{equation}
Note that $k$ is not necessarily an integer.
Since we will deal 
with 
only the first moment 
in our study, we name it with a simpler notation $\mellinO$.
The spectra of the calibration observables and all our results in the
following sections are obtained with samples obeying the following cuts on final-state jets and leptons: 
\begin{equation}
p_{T,j}>30~\gev, |\eta_{j}|<2.4\,,\quad p_{T,\ell}>20~\gev, |\eta_{\ell}|<2.4 \,.
\end{equation}
In order to avoid sensitivities
to extreme kinematic configurations and the unknown unknowns about
their theory description, we compute the average
values in a range around the peak of the spectra.
This range usually corresponds to the full width at half-maximum (FWHM)
range, but in some cases we shall employ slightly different choices.
For each observable we fix this range from the distribution that
we obtain for one value of $m_{t}$, and we use it throughout the analysis.
Our choices for the observables are reported in Table \ref{tab:Sensitivity-Calibration-Obs}.
Example spectra in observables that are potentially interesting
for the calibration of the Monte Carlo parameters are shown in Figure
\ref{fig:Example-spectra-calibration-chi_B}, in which the thick part
of the histograms corresponds to the FWHM range.

\subsection{Dependence on the top-quark mass}

A preliminary step for the Monte Carlo tuning is 
the assessment of the dependence of the calibration observables on
the top-quark mass. Ideally, 
a calibration observable should have no dependence
on $m_t$, so that it could be used to constrain
the Monte Carlo parameters without any concern about a potential bias in the $m_{t}$ measurement. In
practice, all quantities have {\it some} sensitivity to $m_{t}$, so
the only viable approach is using 
 observables with minimal sensitivities to the top-quark mass.
In this section we study
the dependence on $m_t$ of the calibration observables, so that one can restrict the analysis to a specific set of quantities, clearly aimed at
calibration.

We determine the dependence on $m_{t}$ of the first 
Mellin moments, obtained from 21 $m_{t}$ values in-between 163 GeV and
183 GeV, by fitting a straight line. We then compute the
sensitivities defined in eq.~(\ref{eq:DeltaGenericDef}) and
repeat this procedure for the several 
 Monte Carlo settings needed to explore the dependence on
the parameters in Table~\ref{tab:varied}.
The Mellin moments and the straight-line fits for some illustrative
parametrizations are presented
in Figure~\ref{fig:Example-Mellin-of-calibration}. 
Once we have collected the values of the sensitivity to $m_{t}$ for all the variations of the Monte Carlo parameters, we calculate the 
mean value and standard deviation for each observable.
We highlight the fact that the standard deviation, arising from the
discrepancies
in the straight-line fits to data with different  parametrizations,
is to be read as a measure of the sensitivity of our result to the
Monte Carlo setting employed in the computation.

The results are reported
in Table~\ref{tab:Sensitivity-Calibration-Obs}; a few comments are
in order. First of all we remark that for our purposes it is sufficient to present
the broad picture of the sensitivity of the observables to showering and hadronization
parameters, and therefore we can content ourself with one or two digits of accuracy in the
determination of sensitivity parameters. This accuracy will be more than sufficient to draw
informative conclusions. Concerning the actual values of the sensitivity, we see that they
tend to be $\Delta_{m_{t}}^{(\mathcal{M_{O}})}\ll 1 $ for most observables,
with few notable exceptions, the most apparent being the jet mass, which has
$\Delta_{m_{t}}^{(\mathcal{M_{O}})}\simeq 0.23$.
This is the only observable not constructed to be dimensionless, hence it
is not surprising that it exhibits the largest sensitivity to the dimensional
$m_{t}$. The smallest dependence on $m_{t}$ is found
for the purely angular distributions $\Delta\phi(B\bar B)$
and $\Delta\phi(j_{b}j_{\bar{b}})$, as well as
their $\Delta R$ equivalents.
In addition, we remark that the sample of variables
$\chi_{B}$ tends to have slightly larger sensitivity to $m_{t}$.
The choice of different denominators in $\chi_{B}$
can lead to a milder dependence on $m_{t}$ in the
numerator, as can be seen by comparing the results for
$\chi_{B}\left(\left|p_{T,j_{b}}\right|+\left|p_{T,\bar{j}_{b}}\right|\right)$
with respect to $\chi_{B}\left(E_{j_{b}}+E_{\bar{j}_{b}}\right)$. 

\subsection{Constraining power of the calibration observables
  \label{subsec:Constraining-power}}

\begin{table*}
\begin{centering}
\begin{tabular}{|c|c|c|c|c|c|c|c|c|c|}
\hline 
\multirow{2}{*}{$\mathcal{O}$} & \multirow{2}{*}{Range} & \multirow{2}{*}{$\Delta_{m_{t}}^{(\mathcal{M_{O}})}$} & \multicolumn{7}{c|}{$\Delta_{\theta}^{(\mathcal{M_{O}})}$}\tabularnewline
\cline{4-10} 
 &  &  & $\alpha_{s,FSR}$ & $m_{b}$ & $p_{T,\text{min}}$ & $a$ & $b$ & $r_{B}$ & recoil\tabularnewline
\hline 
\hline 
$\rho(r)$ & 0-0.04 & -0.007(7) & 0.78(1) & 0.204(4) & -0.1286(8) & 0.029(3) & -0.043(4) & 0.056(7) & 0.020(1)\tabularnewline
\hline 
$p_{T,B}/p_{T,j_{b}}$ & 0.6-0.998 & -0.053(1) & -0.220(3) & -0.1397(8) & 0.0353(5) & -0.0187(4) & 0.0451(6) & -0.0518(9) & -0.0108(3)\tabularnewline
\hline 
$E_{B}/E_{j_{b}}$ & 0.6-0.998 & -0.049(1) & -0.220(3) & -0.1381(8) & 0.0360(5) & -0.0186(4) & 0.0447(6) & -0.052(1) & -0.0107(3)\tabularnewline
\hline 
$E_{B}/E_{\ell}$ & 0.05-1.5 & -0.155(7) & -0.156(3) & -0.053(3) & 0.0149(7) & -0.007(2) & 0.016(2) & -0.016(10) & -0.0087(7)\tabularnewline
\hline 
$E_{B}/(E_{\ell}+E_{\bar{\ell}})$ & 0.05-1.0 & 0.021(5) & -0.231(2) & -0.082(4) & 0.0228(4) & -0.011(2) & 0.026(2) & -0.028(6) & -0.0113(3)\tabularnewline
\hline 
$m(j_{\bar{b}})$/GeV & 8-20 & 0.229(3) & 0.218(1) & 0.022(1) & -0.0219(7) & 0.000(1) & -0.001(1) & 0.001(3) & 0.0050(3)\tabularnewline
\hline 
\hline 
$\chi_{B}(\sqrt{s_{\text{min},bb}})$ & 0.075-0.875 & -0.177(4) & -0.262(4) & -0.086(1) & 0.0255(3) & -0.0105(10) & 0.027(1) & -0.031(3) & -0.0137(2)\tabularnewline
\hline 
$\chi_{B}\left(E_{j_{b}}+E_{\bar{j}_{b}}\right)$ & 0.175-1.375 & -0.109(2) & -0.357(4) & -0.134(1) & 0.0373(3) & -0.016(1) & 0.040(1) & -0.045(4) & -0.0175(3)\tabularnewline
\hline 
$\chi_{B}(m_{j_{b}j_{\bar{b}}})$ & 0.175-1.375 & -0.089(3) & -0.252(3) & -0.080(1) & 0.0248(3) & -0.010(1) & 0.024(1) & -0.028(5) & -0.0126(2)\tabularnewline
\hline 
$\chi_{B}\left(\left|p_{T,j_{b}}\right|+\left|p_{T,\bar{j}_{b}}\right|\right)$ & 0.46-1.38 & -0.15(2) & -0.47(1) & -0.189(10) & 0.054(3) & -0.023(10) & 0.06(1) & -0.07(4) & -0.022(2)\tabularnewline
\hline 
\hline 
$m_{BB}/m{}_{j_{b}j_{\bar{b}}}$ & 0.8-0.95 & -0.0191(8) & -0.0623(7) & -0.0464(5) & 0.0146(2) & -0.0093(3) & 0.0180(4) & -0.0212(9) & -0.00296(10)\tabularnewline
\hline 
$\Delta\phi(j_{b}j_{\bar{b}})$ & 0.28-3. & -0.210(7) & 0.027(3) & 0.001(2) & -0.0014(5) & -0.000(3) & -0.000(1) & -0.003(9) & 0.0003(5)\tabularnewline
\hline 
$\Delta R(j_{b}j_{\bar{b}})$ & 1.4-3.3 & -0.071(3) & 0.010(1) & 0.0005(10) & -0.0004(2) & -0.000(1) & 0.0004(9) & 0.001(3) & 0.0001(2)\tabularnewline
\hline 
$\Delta\phi(BB)$ & 0.28-3. & -0.207(7) & 0.026(2) & 0.001(1) & -0.0008(4) & 0.000(4) & 0.000(2) & -0.000(8) & 0.0002(5)\tabularnewline
\hline 
$\Delta R(BB)$ & 1.4-3.3 & -0.070(3) & 0.009(1) & 0.000(1) & -0.0003(2) & -0.0003(10) & 0.0002(9) & -0.000(4) & 0.0001(2)\tabularnewline
\hline 
$|\Delta\phi(BB)-\Delta\phi(j_{b}j_{\bar{b}})|$ & 0-0.0488 & 0.06(1) & 0.734(6) & 0.099(5) & -0.088(2) & 0.006(5) & -0.004(5) & 0.01(2) & 0.026(2)\tabularnewline
\hline 
$|\Delta R(BB)-\Delta R(j_{b}j_{\bar{b}})|$ & 0-0.0992 & 0.10(1) & 0.920(3) & 0.079(5) & -0.075(1) & -0.000(4) & 0.005(4) & -0.00(2) & 0.0418(8)\tabularnewline
\hline 
\end{tabular}
\par\end{centering}
\caption{\label{tab:Sensitivity-Calibration-Obs}Sensitivity of the calibration
  observables to the top quark mass and to the parameters
  of PYTHIA parton
  shower and hadronization. The quantities $\Delta_{m_{t}}^{(\mathcal{M_{O}})}$
  and $\Delta_{\theta}^{(\mathcal{M_{O}})}$ are defined in the text.
  $\theta=\{\alpha_{s,FSR},\,m_{b},\,p_{T,\text{min}},\,a,\,b,\,r_{B},\,recoil\}$
 denotes a generic PYTHIA parameter.}
\end{table*}

\begin{figure}
\centering{}\includegraphics[width=0.49\linewidth]{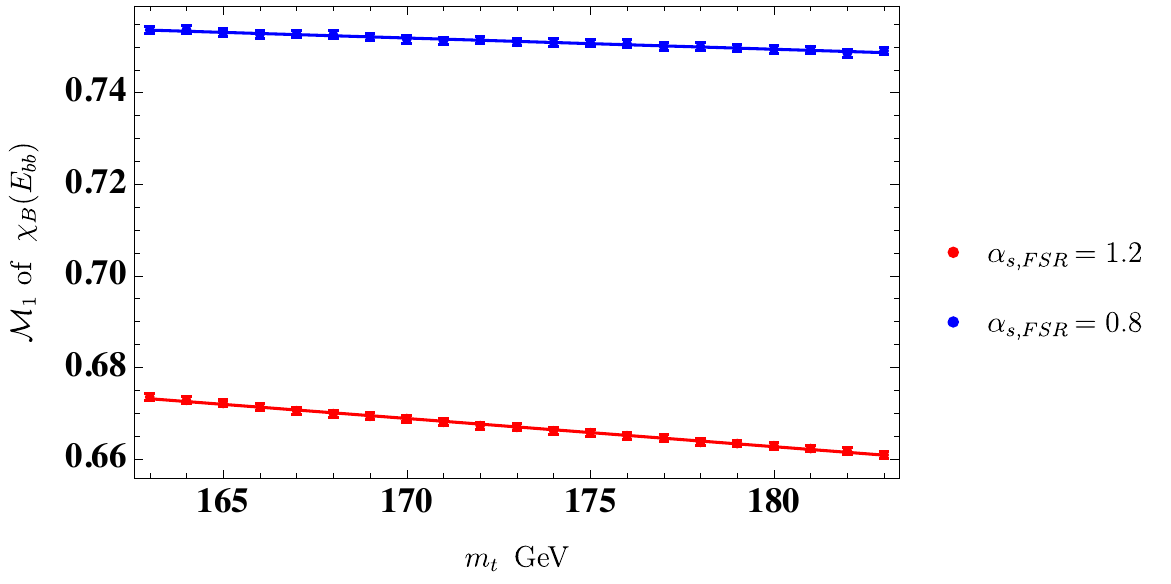}\includegraphics[width=0.49\linewidth]{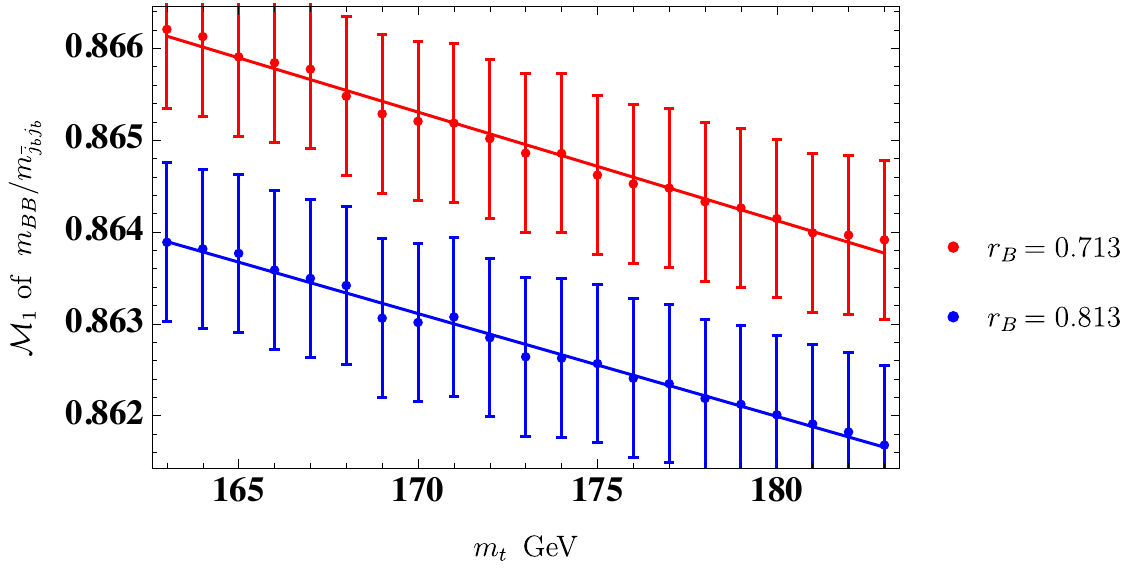}\caption{\label{fig:Example-Mellin-of-calibration}Dependence on $m_{t}$ of
the average values $\langle\chi_{B}(E_{B}+E_{B})\rangle$ and $\langle m_{B\bar B}/m_{j_{b}\bar{j}_{b}}\rangle$.
The error bars are due to the limited statistics of our calculation; the
straight lines through the points are the fits employed to compute
$\DeltaMO$.}
\end{figure}

In order to quantify the dependence of the first Mellin moments on
the Monte Carlo tuning, we evaluate the difference
between the moments obtained for two values of each 
parameter at fixed $m_t$ and compute the sensitivity according to eq.~(\ref{eq:DeltaGenericDef}) for the $m_{t}$ value at hand. Considering several $m_t$ values between 163 GeV and 183 GeV, we calculate 
the average value and standard
deviation for the sensitivity $\Delta^{(\mathcal{M_{O}})}_{\theta}$ to each
PYTHIA parameter in Table~\ref{tab:Sensitivity-Calibration-Obs}. The obtained estimate for the sensitivity essentially represents the distance between the lines shown in Figure~\ref{fig:Example-Mellin-of-calibration} for different values of the same Monte Carlo parameter. 

From Table~\ref{tab:Sensitivity-Calibration-Obs},
we can derive what precision in the measurement of the Mellin moments
$\delta \mellinO/\mellinO$
should be reached to notice a given variation $\delta\theta/\theta$
of the Monte Carlo parameter. From the definition of the logarithmic derivative eq.~(\ref{eq:DeltaGenericDef})
that we have tabulated, it follows:
\begin{equation}
  \frac{\delta \mellinO}{\mellinO}=\Delta_{\theta}^{(\mellinO)}\frac{\delta\theta}{\theta},
  \label{eq:constrain_eq}
\end{equation}
hence, observables with large $\Delta_{\theta}^{(\mellinO)}$, are the best diagnostics of
the accuracy of the tuning or, equivalently, the best observables for an in-situ calibration in the $t\bar{t}$ environment.

From a quick glance to the table, we find that a few interesting patterns emerge:
the radial energy distribution $\rho(r)$ is among the most sensitive
observables to both hadronization and shower parameters and
tends to have opposite-sign sensitivity with respect to the other
quantities. 
This tendency can be easily justified: for instance,
when increasing $\alpha_{S,FSR}$, it is natural to expect more radiation,
even at larger angles, so that $\langle\rho(r)\rangle$ grows, while $p_{T,B}$
or any possible denominator of $\chi_{B}$ decreases;
a similar but reversed argument holds for $p_{T,\text{min}}$. 
In spite of this apparently different dependence of $\rho(r)$,
we warn the readers that the actual usefulness of the combination of several observables to obtain constraints from data on the parameters has to be evaluated globally, by looking at the dependence of all observables on all parameters.
We defer to Section~\ref{sec:combined-constraints} a full commentary on the information that we can obtain on the Monte Carlo parameters combining several calibration observables.

Angular variables like $\Delta\phi(B\bar B)$ and $\Delta R$, as well as
their analogs at $b$-jet level, tend to exhibit very small dependence
on all parameters. However, the difference between hadron- and jet-level
observables exhibits significant dependence on most parameters, as well as
some similarity with the results for $\rho(r)$. 

The hadron-to-jet ratios $p_{T,B}/p_{T,j_{b}}$, $E_{B}/E_{j_{b}}$, and $m_{B\bar B}/m_{j_{b}j_{\bar{b}}}$
exhibit substantial sensitivity to the Monte Carlo parameters and are
expected to probe similar aspects of the PYTHIA description of the events. Due
to the little dependence of  $\Delta\phi$ and $\Delta R$ 
on both $m_{t}$ and shower/hadronization parameters,
we can expect the sensitivity
of the mass ratio to be closely related to that of the transverse-momentum
ratio.

The ratios that involve lepton energies usually show very little sensitivity to the Monte Carlo parameters and $m_t$, and therefore they have little use for the calibration.

The variables $\chi_{B}$ show some of the largest sensitivity to
the Monte Carlo parameters. We remark that for the denominator of $\chi_B$, the variants most sensitive to Monte Carlo parameters are those mostly dependent on $m_t$ as well. 
Usually, the increase
of the dependence on the Monte Carlo parameters is smaller
than the one on $m_{t}$. Nevertheless, care must be taken when one uses
the definitions of $\chi_{B}$ more sensitive to $m_{t}$ in the calibration
and in the $m_{t}$ determination.

The jet mass tends to have small dependence on the parameters, 
at most comparable to that exhibited by other observables.
Putting this together with its large sensitivity
to $m_{t}$, it is clearly a non-optimal candidate for a tuning or a measurement of $m_t$ 
with an in-situ Monte Carlo calibration.

\subsection{Combined constraining power\label{sec:combined-constraints}}
In order to go beyond the qualitative analysis of the preceding paragraphs, we have devised a procedure, based on the $\Delta_{\theta}^{(\mellinO)}$ values, to quantify the constraints on the
Monte Carlo parameters which can be obtained
from the combined use of several calibration observables. 
This procedure returns an estimate of the relative error on the determination of the parameters from the first Mellin moments, but can be straightforwardly applicable as well to other input observable quantities such as the higher Mellin moments or other properties of the spectrum of the calibration observables. The procedure goes as follows.
We denote our parameters generically as a vector $\theta$, and collectively, we
adopt the following notation: 
\[
\theta=\{\alpha_{s,FSR},\,m_{b},\,p_{T,\text{min}},\,a,\,b,\,r_{B},\,recoil\}\,.
\]
We put the observables as well in a vector that we denote as $O=\lbrace {\mathcal{O}_{i}} \rbrace$ for the $\mathcal{O}_{i} $ listed in the first column of Table~\ref{tab:Sensitivity-Calibration-Obs}. In this notation the sensitivity of each observable to the several parameters can be expressed in matrix notation as:
\begin{equation}
  \frac{\delta\mathcal{M_{O}}_{i}}{\mathcal{M_{O}}_{i}}=\left(\Delta_{\theta}^{(\mathcal{M_{O}})}\right)_{ij}
  \frac{\delta\theta_{j}}{\theta_{j}},\label{eq:matrix_constrain}
\end{equation}
where Table~\ref{tab:Sensitivity-Calibration-Obs} gives the entries
$\left(\Delta_{\theta}^{(\mathcal{M_{O}})}\right)_{ij}$. This matrix contains all the necessary information to estimate how strong a constraint on the parameters $\theta$ can be obtained from measurements of the observables and their Mellin moments. 
In order to derive
the constraining power of our set of calibration observables, we would like to invert the matrix $\DeltaMO$, as to express the sensitivity of the parameters as functions of the observables, which would take the relation of 
\begin{equation}
 \frac{\delta\theta_{j}}{\theta_{j}} =\left(\tilde{\Delta}_{\theta}^{(\mathcal{M_{O}})}\right)_{ij}  \frac{\delta\mathcal{M_{O}}_{i}}{\mathcal{M_{O}}_{i}}
 \,,\label{eq:parameters_constrain}
\end{equation}
with $\DeltaMOtilde$ satisfying $$\DeltaMOtilde\cdot\DeltaMO=\mathds{1}\,.$$
Being aware that $\DeltaMO$ is,  in general, a rectangular matrix,
we may not be able to invert it to find $\DeltaMOtilde$.  
However, we can utilize the pseudo-inverse \cite{Penrose_1955,Dresden_1920} procedure to define $\tilde{\Delta}_{\theta}^{(\mathcal{M_{O}})}$: one defines linear combinations of observables and linear combinations of parameters such that each of the new observables depends only on one of the new parameters. To find these linear combinations, one can use a singular-value decomposition of the matrix $\DeltaMO$. The pseudo-inverse matrix is then the matrix obtained by acting on a diagonal matrix with values reciprocal to the singular values of $\DeltaMO$, with the inverse of the transformation that defines the new observables and new parameters.  Further details on this procedure are given in Appendix~\ref{appA}.
 
The strength of the constraints that can be obtained on the parameters $\theta$ is given by the general transformation of the covariance matrix. From eq.~(\ref{eq:parameters_constrain}) it follows that 
\begin{equation}
\cov_{\frac{d\theta}{\theta}} = \DeltaMOtilde \cdot \cov_{\frac{dO}{O}} \cdot \left.\DeltaMOtilde\right.^{t}\,. \label{covTheta}
\end{equation}
Assuming for simplicity that the measurements of the Mellin moments are all uncorrelated and that an overall precision of 1\% in their measurement is achievable,
we find that the constraining power implied by Table~\ref{tab:Sensitivity-Calibration-Obs} would be in general very modest. In fact, the standard deviations read off the diagonal elements of $\cov_{\frac{d\theta}{\theta}}$ are all around or above 100\% relative uncertainty. The origin of this fact can easily be tracked back to the fact that the matrix $\DeltaMO$ has several small singular values: for Table~\ref{tab:Sensitivity-Calibration-Obs} the singular values read {1.7, 0.26, 0.048, 0.0075, 0.005, 0.0033, 0.0014}. This means that the information contained in the FWHM Mellin moments of the several observables which
we consider 
allows us to constrain at most
one or maybe two parameters (that would be $\alpha_{s,FSR}$ and $m_{b}$). In fact a singular value of order one means that from a measurement of the observables with precision 1\% we can extract one parameter (or combination of Monte Carlo parameters) with the same accuracy, that is 1\%. Similarly a singular value of order $10^{-3}$ implies that even in presence of a measurement of the observables with $10^{-4}$ precision we would obtain a constrain on the relevant (combination of) Monte Carlo parameters at a mere $10^{-1}$ level, as suggested by eq.~(\ref{eq:constrain_eq}). Given the singular values obtained, we remark that, even imagining measurements of the calibration observables with precision better than 1\%, it is clear that an analysis of very inclusive quantities such as the Mellin moments is not capable of yielding any useful constraint on the Monte Carlo parameters.

\begin{figure}[htbp]
\begin{center}
\includegraphics[width=0.99\linewidth]{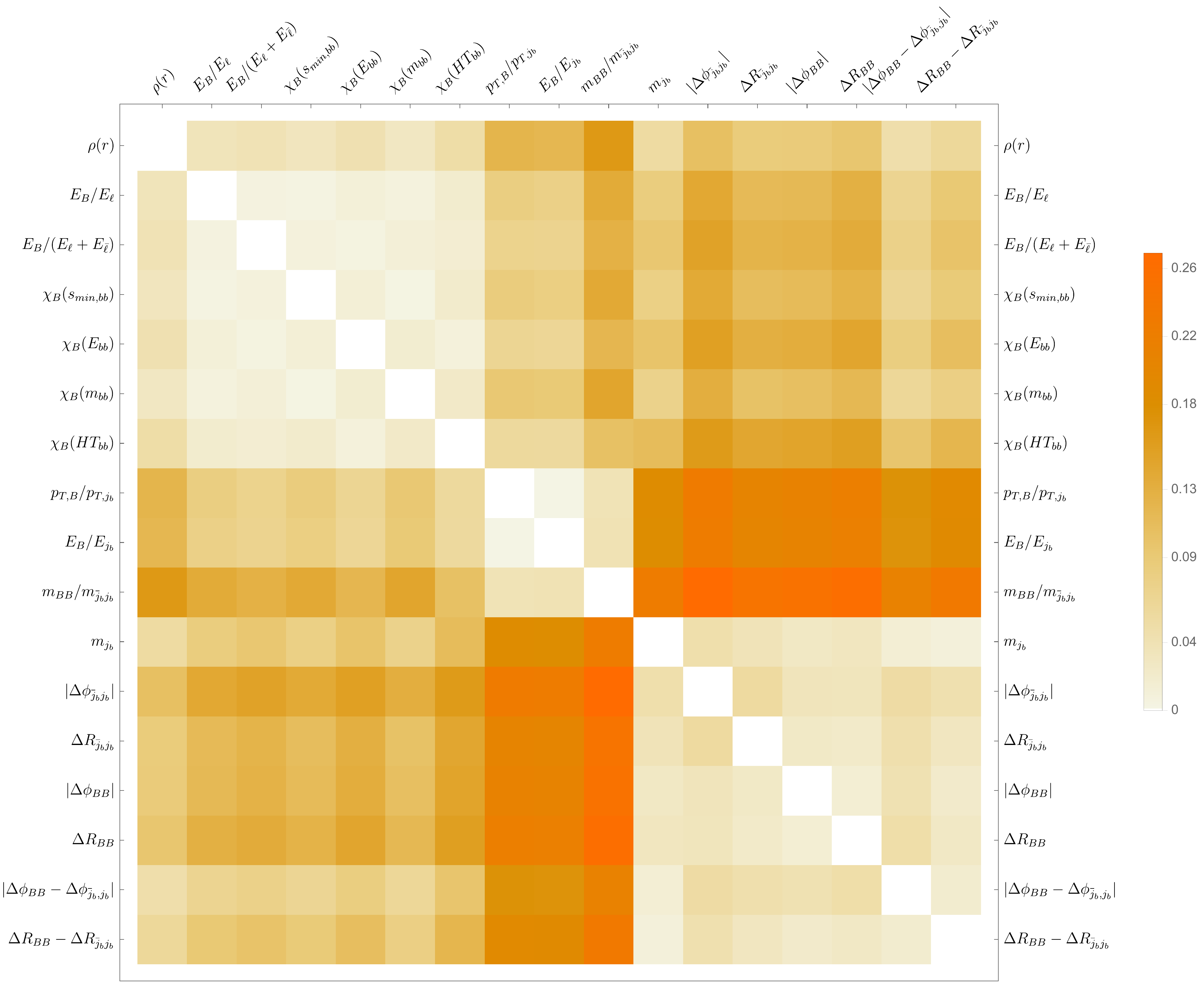}
\caption{Angular distance between the directions in parameter space pointed by the rows of Table~\ref{tab:Sensitivity-Calibration-Obs}.  \label{fig:anglesmellin1}}
\end{center}
\end{figure}

A more graphical way to picture why the Mellin moments do not contain enough information to constrain all the relevant Monte Carlo parameters is to look at the angles between the directions pointed by the rows of the matrix $\DeltaMO$. If two rows point in significantly different directions it means that two observables are sensitive to different combinations of the Monte Carlo parameters, hence their changes upon variations of the parameters are not much correlated. However, it turns out that for Table~\ref{tab:Sensitivity-Calibration-Obs} all the rows point in a very similar direction. In Figure~\ref{fig:anglesmellin1} we show the values of $1-|\cos\alpha_{ij}|$, where $\alpha_{ij}$ is the angle between the direction of the row $i$ and the row $j$. From the figure it is apparent that several Mellin moments point essentially in the same direction ({\it e.g.,} $m_{BB}/m{}_{j_{b}j_{\bar{b}}}$, $p_{T,B}/p_{T,j_{b}}$, and $E_{B}/E_{j_{b}}$ are well visible in the center part of the plot); the two furthest apart directions are separated by $1-|\cos\alpha|\simeq 0.25$. Overall we can define three main directions which might be (somewhat arbitrarily) labeled by $\rho(r)$, for the observables in the upper left corner of Figure~\ref{fig:anglesmellin1},  $p_{T,B}/p_{T,j_{b}}$ for the observables in the middle part, and $|\Delta R(BB)-\Delta R(j_{b}j_{\bar{b}})|$  for the observables in the lower-right part.

Adding several higher Mellin moments to the analysis, we find very little change in the results of the analysis. We have tested values of $k$ in the definition of the $k$-th Mellin moment eq.~(\ref{mellindef}) from 0.25 to 4 to explore the sensitivity of Mellin moments to the Monte Carlo parameters. We find that, although they have different magnitudes, all Mellin moments are sensitive to the same combinations of Monte Carlo parameters as in the first moment. We ascribe this similarity to the fact that we are considering FWHM ranges, which limits the possibility to be sensitive to different physical regimes, hence different Monte Carlo
parameters. In the following section, seeking for an improvement of the prospected constraints, we explore the constraints that can be obtained if one considers the full range of the calibration observables.

%
%

\subsection{Differential constraining power \label{subsec:differential-constraining-power}}

More stringent constraints on the Monte Carlo parameters can be obtained
exploring more in detail the full shape of the distribution of the
calibration observables. A common practice in this case is to study
several Mellin moments of the full distribution as a mean to probe the full shape of the observable at hand.
However, in most cases one cannot extract the highest Mellin moments from the data
very accurately, since they would be sensitive even to small errors
in the distributions.\footnote{More generally, the use of the moments in the analyses is often motivated
by theory arguments, namely the fact that convolution integrals, such as
those entering in the DGLAP equations, are turned into
ordinary products in $N$-space. In our study, however, these properties are not compelling, so Mellin moments are really just a way to describe the shape of the spectra of our observables.}

As an alternative, and probably more transparent, manner to study the shapes of the several observables, we use directly the bin counts of a subset of the calibration observables in suitably chosen ranges.\footnote{This allows us to access different regimes of full phase space, in which observables may have different sensitivities to the Monte Carlo parameters.} 
The sensitivity of the bin counts to changes of the Monte Carlo parameters will be examined comparing bin-by-bin the spectra obtained from different Monte Carlo settings.

\begin{figure}
\begin{centering}
  \includegraphics[width=0.49\linewidth]{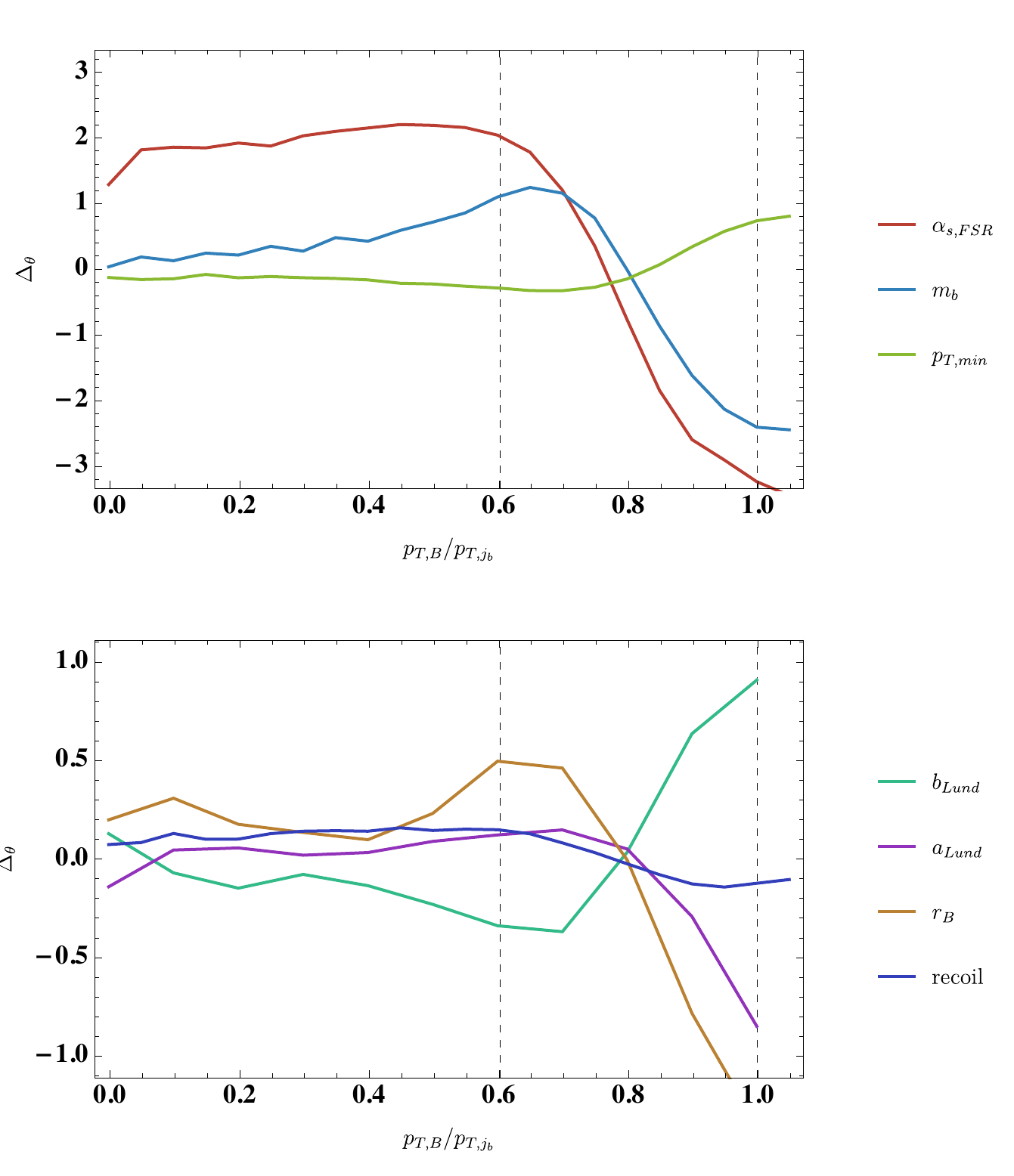}
  \includegraphics[width=0.49\linewidth]{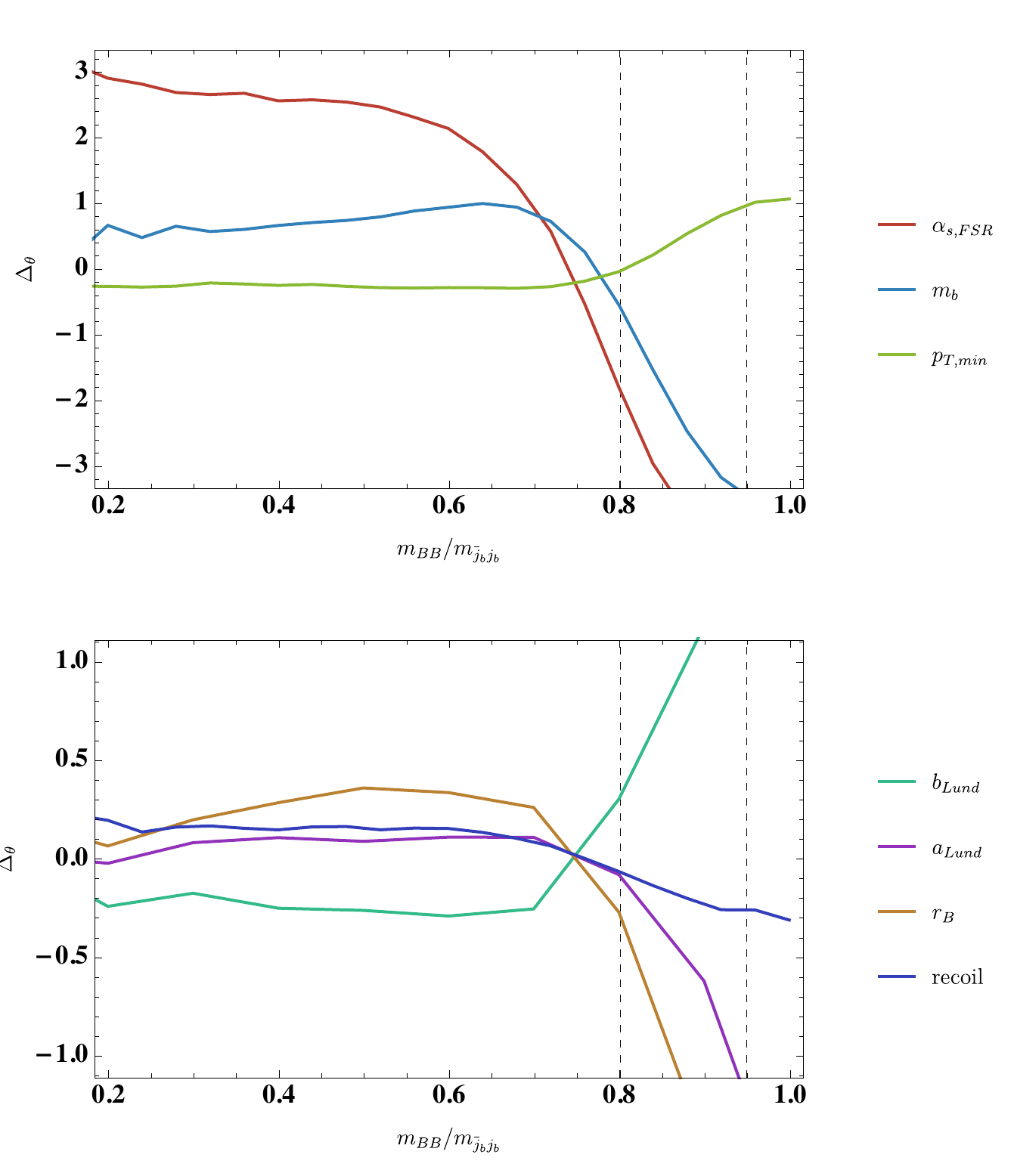}
\par\end{centering}
\begin{centering}
  \includegraphics[width=0.49\linewidth]{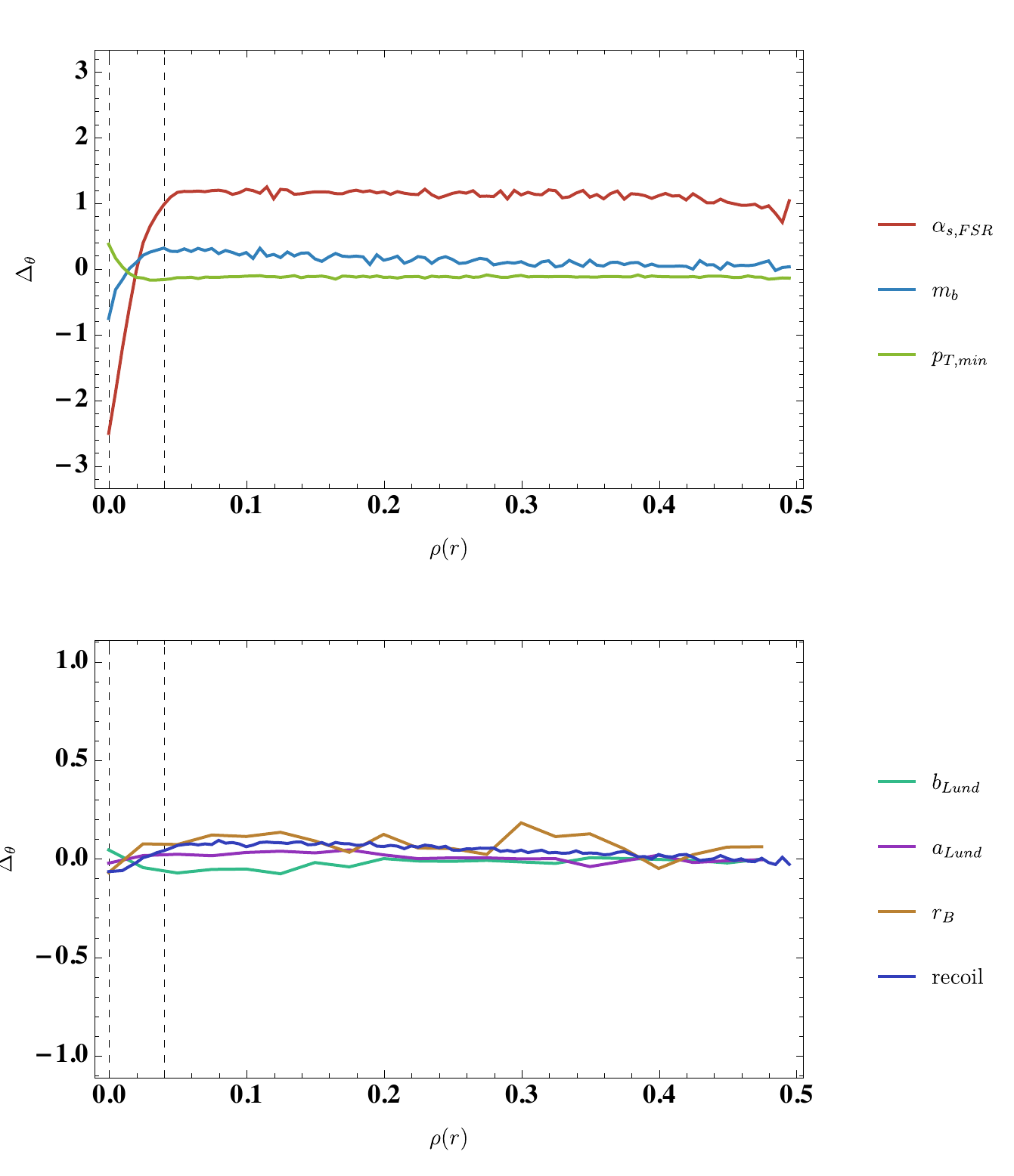}
  \includegraphics[width=0.49\linewidth]{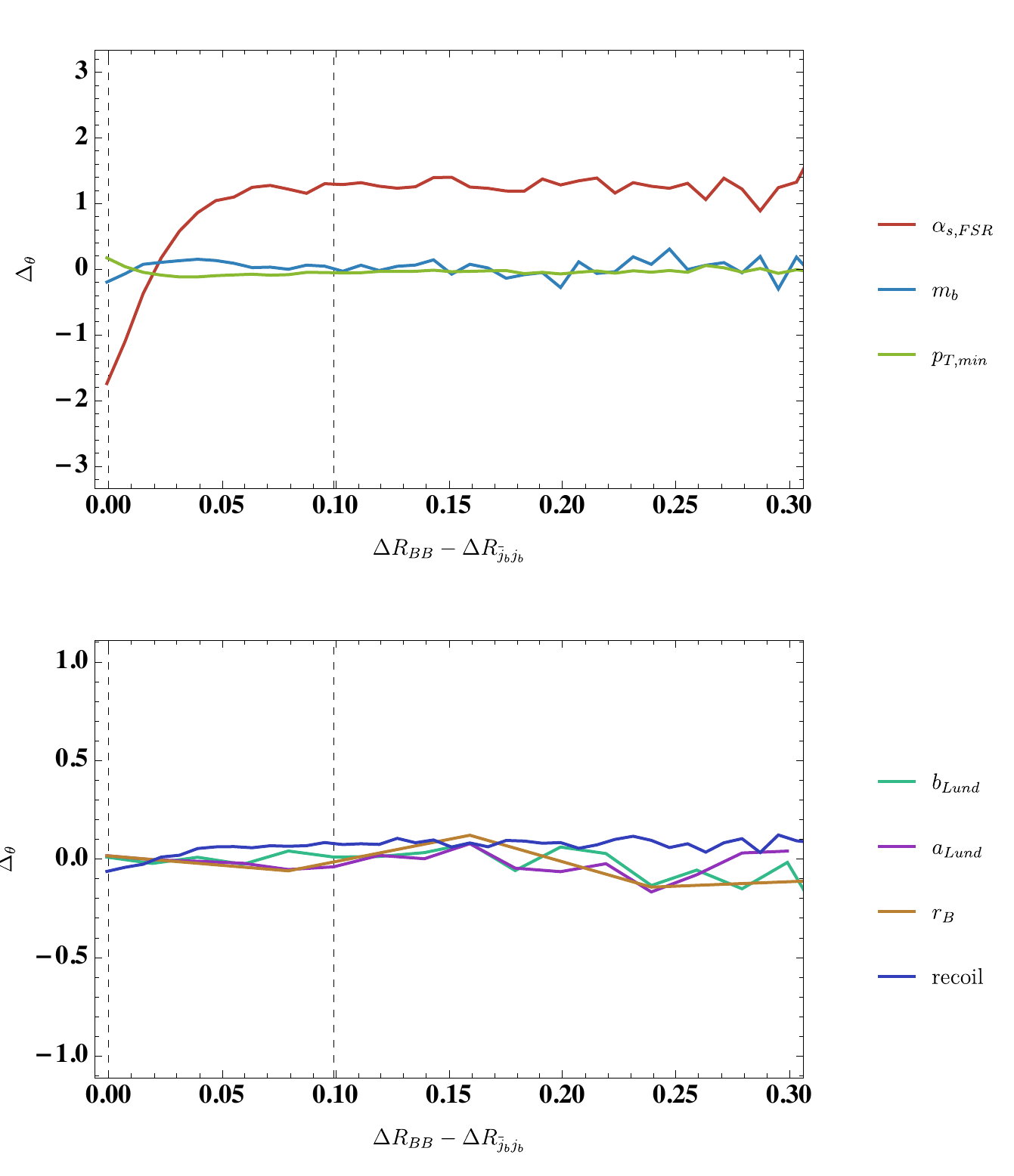}
\par\end{centering}
\caption{\label{fig:Differential-sensitivity-plots}Differential sensitivity
to the Monte Carlo parameters of the distributions of $p_{T,B}/p_{T,j_{b}}$, $m_{B\bar B}/m_{j_{b}\bar{j}_{b}}$, $\rho(r)$, and $\Delta R(BB)-\Delta R(j_{\bar{b}}j_{b})$. The two dashed vertical lines denote
the FWHM ranges.}
\end{figure}

In order to keep the problem at a minimum of complexity,
we start by identifying the most interesting physical quantities and
studying the dependence of each bin count on the Monte Carlo parameters.
Results for a few representative observables are presented
in Figure~\ref{fig:Differential-sensitivity-plots}, where
the vertical dashed lines limit the FWHM ranges used in the previous section to compute the first Mellin moments of the distributions. These lines also serve as guidance to get a feeling of which parts of the distribution belong to the tails and which to the bulk. Clearly in some tails there is a good deal of sensitivity to the Monte Carlo parameters, but it is likely that data
will be scarce in those regions.

From these results, we learn that sensitivities 
to $\alpha_{S,FSR}$ and $m_{b}$ exceeding 2 can be achieved,
{\it e.g.,} in $p_{T,B}/p_{T,j_{b}}$, whereas it is overall quite hard to find regions of the spectra
with a similar dependence on $p_{T,min}$. The dependence on $a$, $b$,
$r_B$ and $recoil$ is in general much milder.

Learning from the experience of the previous section, we can generalize the analysis by considering each bin as an independent observable and attempting to use its theoretical dependence on the Monte Carlo setting to constrain these. Similarly to the previous section, we can compute a matrix of sensitivity $\DeltaBinO$ with $(b_{j,\mathcal{O}})$ being the $j$-th bin of the distribution of the observable $\mathcal{O}$.

The observables that we use in this analysis are 
$\rho(r)$, $\chi_{B}(E_{j_b}+E_{\bar j_b})$,
$E_{B}/E_{\ell}$, 
$\Delta R(B\bar B)-\Delta R(j_bj_{\bar b})$, 
 $p_{T,B}/p_{T,j_{b}}$, 
 and $m_{B\bar B}/m_{j_{b}\bar{j}_{b}}$. 
 They are the observables that in the previous section have shown the greatest sensitivity in absolute sense and the most distinct dependence on linearly independent combinations of Monte Carlo parameters. In place of the
 $\chi_{B}(E_{j_b}+E_{\bar j_b})$ variable we could have used other options for the denominator that appears in $\chi_{B}$, but we find it convenient employing  $\chi_{B}(E_{j_b}+E_{\bar j_b})$, because it is bound to be in the range [0,2], facilitating the identification of an optimal bin-size for the calibration. This fact is quite important, because the study of distributions rather than integrated quantities inflates the expected statistical errors on the measurements. Using an observable that is limited to a range by definition (such as 
 $\chi_{B}(E_{j_b}+E_{\bar {j_b}})$, $p_{T,B}/p_{T,j_{b}}$, and $m_{B\bar B}/m_{j_{b}\bar{j}_{b}}$), we can more easily define bins with sufficiently large expected number of events. In order not to deal with too small expected bin counts, we have divided each spectrum in about 10 bins, the details of the binning are reported in Table~\ref{fullrange} for completeness.

\begin{table}
\begin{centering}
\begin{tabular}{|c|c|c|}
\hline 
{$\mathcal{O}$} & 
{Range} &
{$N_{bins}$}  \tabularnewline
\hline 
\hline 
$\rho(r)$ & 0.-0.4 & 16 \tabularnewline
\hline 
$p_{T,B}/p_{T,j_{b}}$ & 0.-0.99 & 11 \tabularnewline
\hline 
$E_{B}/E_{\ell}$ & 0.05-4.55 & 9 \tabularnewline
\hline 
$\chi_{B}\left(E_{j_{b}}+E_{\bar{j}_{b}}\right)$ & 0.-2. & 10\tabularnewline
\hline 
$m_{BB}/m{}_{j_{b}j_{\bar{b}}}$ & 0.-0.998 & 11 \tabularnewline
\hline 
$|\Delta R(BB)-\Delta R(j_{b}j_{\bar{b}})|$ & 0.-0.288 & 9 \tabularnewline
\hline 
\end{tabular}
\par\end{centering}
\caption{Ranges and binning used for the calculation of the sensitivity of the bin counts of the full spectrum to changes of the Monte Carlo parameters.\label{fullrange}}

\end{table}

We remark that by using a single observable and studying the whole spectrum the best results are obtained from the distribution of  $p_{T,B}/p_{T,j_{b}}$. In this case the sensitivity matrix has singular values {7.0, 1.8, 0.28, 0.11, 0.11, 0.037, 0.018} from which we see a substantial improvement with respect to the case of Mellin moments in the FWHM range of the previous section. The largest angular distance in this case is $1-|\cos\alpha_{ij}|\simeq 0.45$, which clearly supports the observed sensitivity to more independent combinations of Monte Carlo parameters than in the inclusive analysis of Mellin moment. In spite of this, there are still several small singular values, which indicates that it is not possible to look only
at this distribution to constrain all the parameters we set out to constrain.  For completeness we report the sensitivity matrix for the $p_{T,B}/p_{T,j_{b}}$ distribution in Table~\ref{tab:Sensitivity-of-bins-pTbratio} in Appendix~\ref{appA}.

In order to obtain meaningful constraints on all Monte Carlo parameters of interest, it seems necessary to gather more information using a larger set of input observables. Performing such a global analysis on the six observables mentioned above, we find that the singular values of the sensitivity matrix are {15.0, 4.2, 0.75, 0.42, 0.27, 0.16, 0.13}. Given the lowest singular value of order $10^{-1}$,  we expect that it is possible
to put meaningful bounds on all the parameters. For instance input observables measured at $10^{-2}$ precision should give $O(10^{-1})$ even on the most loosely constrained parameter. 

We use eq.~(\ref{covTheta}) to determine the covariance matrix expected for the determination of the Monte Carlo parameters from the study of the full spectrum of the six observables.  If we assume that each bin of these distributions is an independent observable, uncorrelated with all the rest, and measured with 1\% precision the expected relative uncertainty ranges from 70\% for the recoil to about 4\% for $\alpha_{s,FSR}$.  
Such uncertainties for all the parameters as well as
the correlation matrix are given in Table~\ref{Correlation}. 
The reported uncertainties are for 1\% precision on the input observables,
treated as independent  and uncorrelated. Had we considered a different precision for the input observables, for instance a 0.1\% precision
(reachable at the HL-LHC if one considers purely statistical uncertainties),
we would have obtained linearly rescaled uncertainties (that is 10 times smaller than what are reported in Table~\ref{Correlation}), as eq.~(\ref{covTheta})
dictates. These results clearly indicate that a global analysis of several spectra of calibration observables has a potential to constrain all the
PYTHIA parameters explored here.

\begin{figure}[htbp]
\begin{center}
\includegraphics[width=0.30\linewidth]{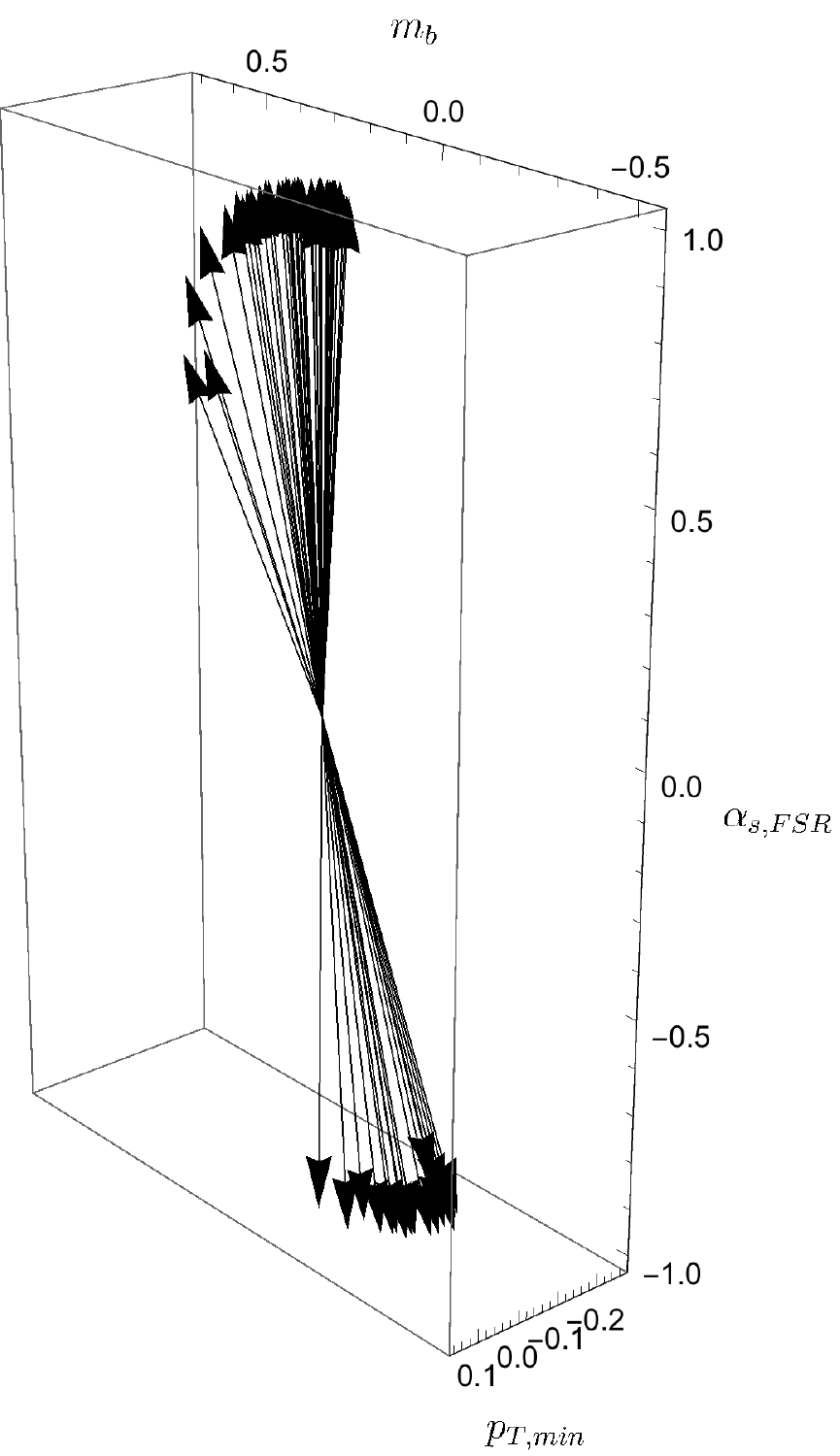}
\includegraphics[width=0.45\linewidth]{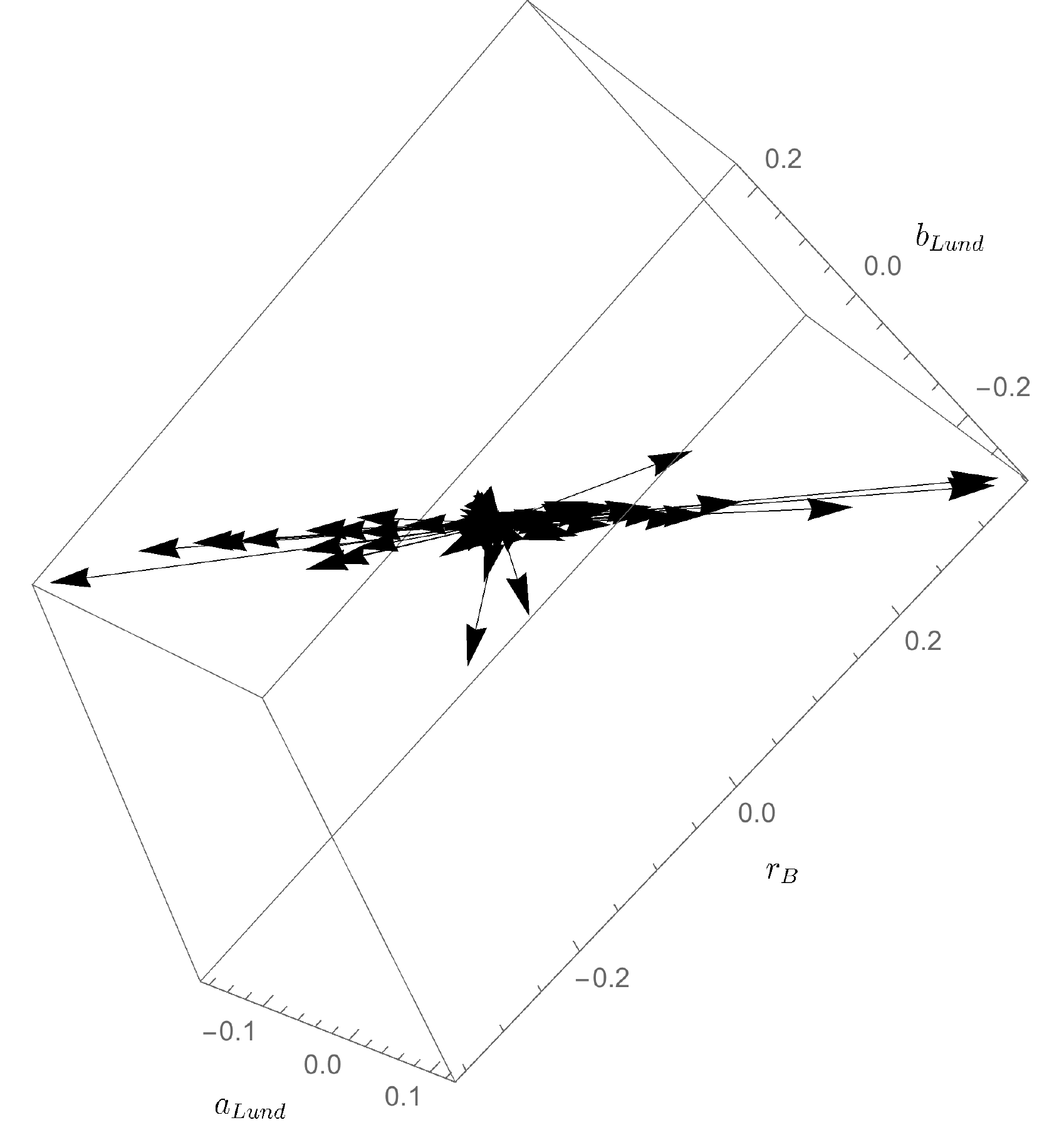}\\
\includegraphics[width=0.45\linewidth]{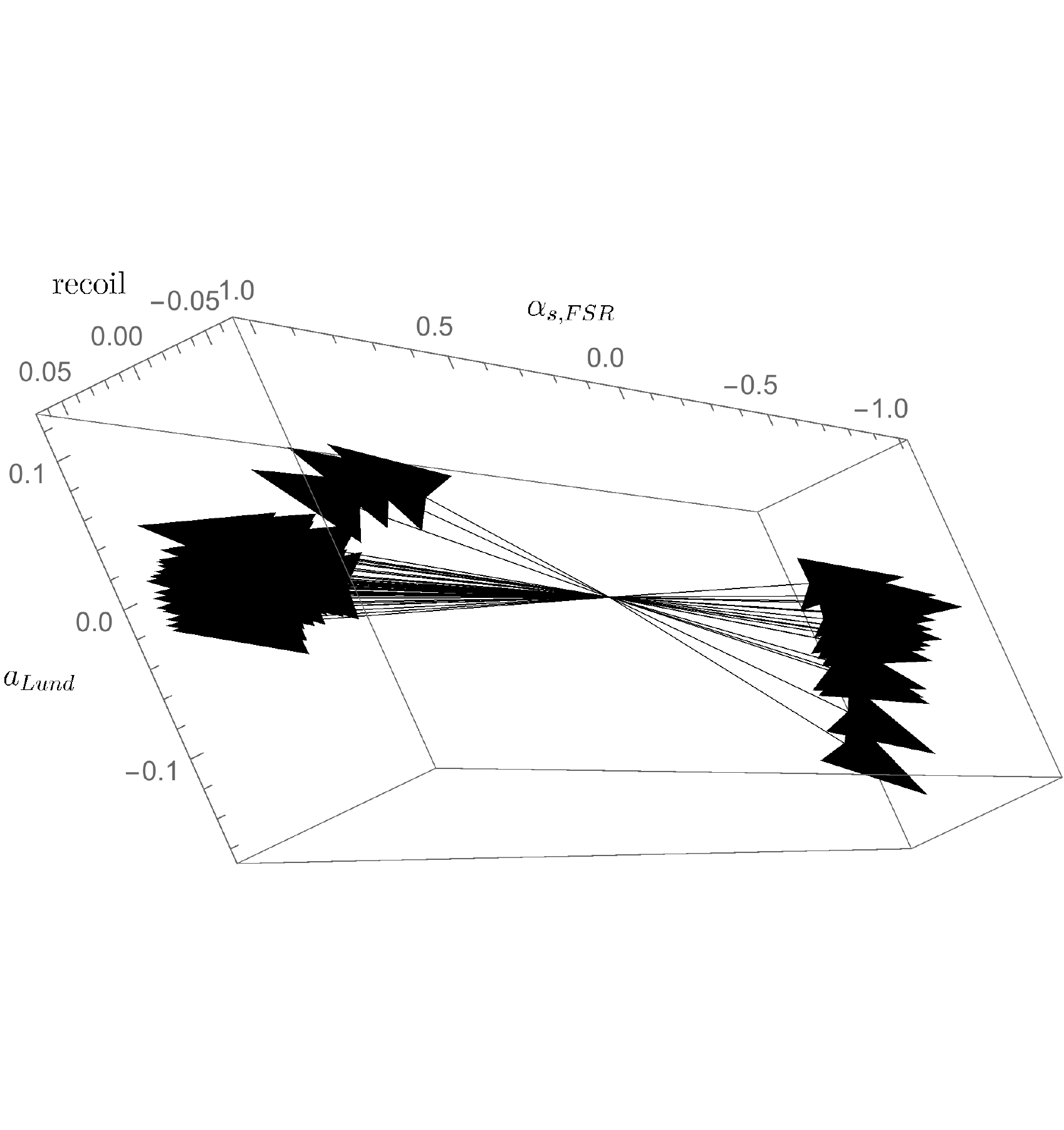}\\
\caption{Directions in the $\lbrace \alpha_{s,FSR},\,m_{b},\,p_{T,\text{min}} \rbrace$  (top left), $\lbrace a,\,b,\,r_{B} \rbrace$ (top right) and $\lbrace \alpha_{s,FSR},a,recoil \rbrace$ (bottom) space pointed by the combination of parameters upon which the considered 66 bins of the distributions depend. \label{directions}}
\end{center}
\end{figure}

The improvement of the achievable constraint can be tracked to the angular distance between the rows of the sensitivity matrix that is larger than in the case of single observable analysis: for the six observables at hand we get $1-|\cos\alpha_{ij}|\lesssim0.55$. Still, most of the information is extracted from bins in the distributions that depend on very similar combinations of the Monte Carlo parameters. This can be seen in Figure~\ref{directions}, where we display the direction of the gradient of the 66 bin counts used in our global analysis in various subspaces of our 7-dimensional parameter space. The shower parameters tend to appear in overall different combinations, although $p_{T,\textrm{min}}$ has usually small impact on the calibration observables, hence the gradient vectors in the top left panel have small components in the $p_{T,\textrm{min}}$ direction. The hadronization parameters,  presented in the top right panel, tend to appear in combination $b-a-r_{B}$ except for a few observables, which generates the high degree of correlation in the matrix in Table~\ref{Correlation}. Finally in the bottom panel we find the dependence on the recoil switch parameter which shows significant correlation with $\alpha_{s,FSR}$.

We stress the fact that we have not gone through an intense optimization of the choice of the observables and ranges of the spectra, but nevertheless
we expect these results to be illustrative of the achievable constraints on
PYTHIA 8.

\begin{table}[htp]
\begin{center}
\begin{tabular}{ccccccc}
$\alpha_{s,FSR}$ & $m_{b}$ & $p_{T,\text{min}}$ & $a$ & $b$ & $r_{B}$ & recoil \\ \hline
0.045 & 0.14 & 0.35 & 0.5 & 0.48 & 0.21 & 0.73 \\
\end{tabular}
$$
\left(
\begin{array}{ccccccc}
 1. & -0.13 & 0.48 & -0.37 & -0.24 & 0.38 & -0.85 \\
 -0.13 & 1. & 0.01 & 0.62 & 0.81 & -0.46 & -0.06 \\
 0.48 & 0.01 & 1. & -0.09 & -0.13 & 0.53 & -0.08 \\
 -0.37 & 0.62 & -0.09 & 1. & 0.8 & -0.47 & 0.31 \\
 -0.24 & 0.81 & -0.13 & 0.8 & 1. & -0.23 & 0.15 \\
 0.38 & -0.46 & 0.53 & -0.47 & -0.23 & 1. & 0. \\
 -0.85 & -0.06 & -0.08 & 0.31 & 0.15 & 0. & 1. \\
\end{array}
\right)$$
\caption{Top: expected relative uncertainties on the PYTHIA 8 parameters.
  assuming input observables measured with 1\% accuracy.
  Bottom: correlation matrix for the constraint on these Monte Carlo parameters.\label{Correlation}}
\end{center}
\end{table}%

\section{Results on observables sensitive to $m_{t}$ \label{sec:topmass}}

\begin{figure}[htbp]
\begin{center}
\includegraphics[width=0.49\linewidth]{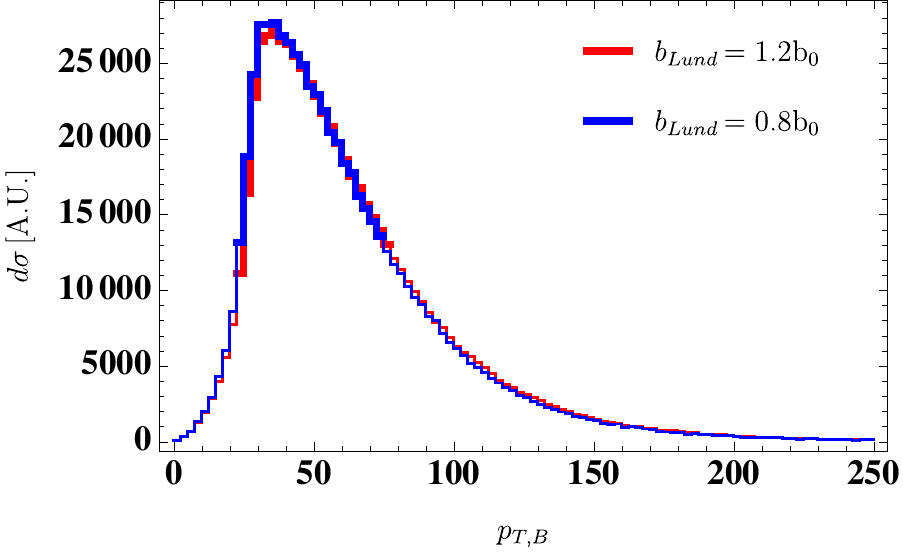}\includegraphics[width=0.49\linewidth]{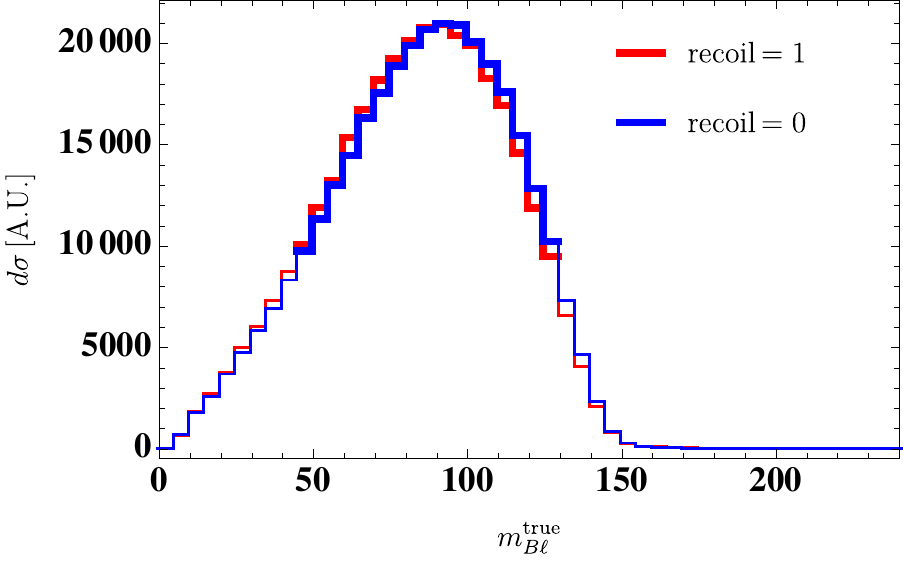}\\
\includegraphics[width=0.49\linewidth]{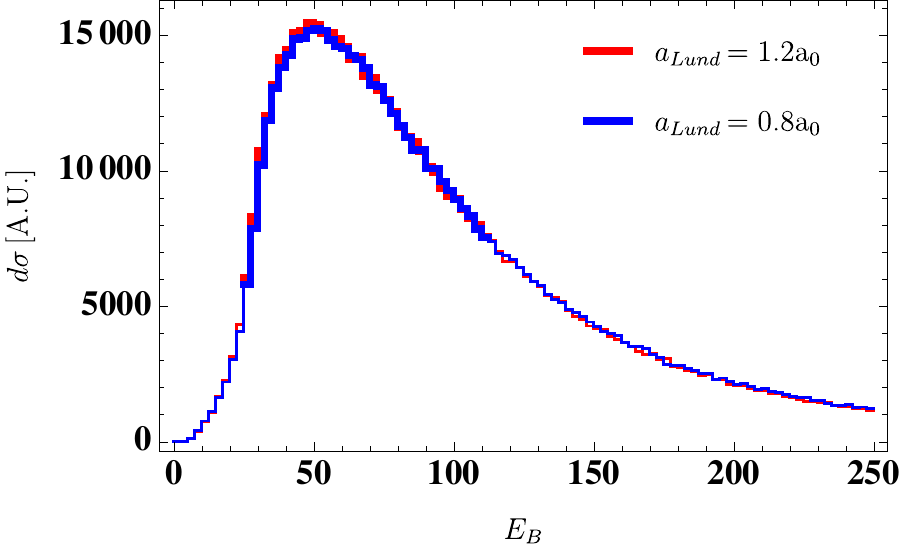}\includegraphics[width=0.49\linewidth]{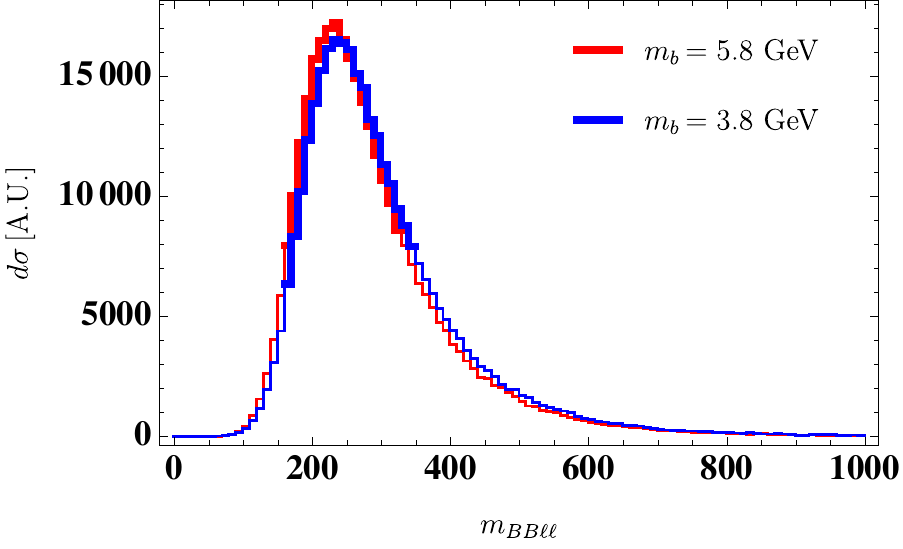}\\
\includegraphics[width=0.49\linewidth]{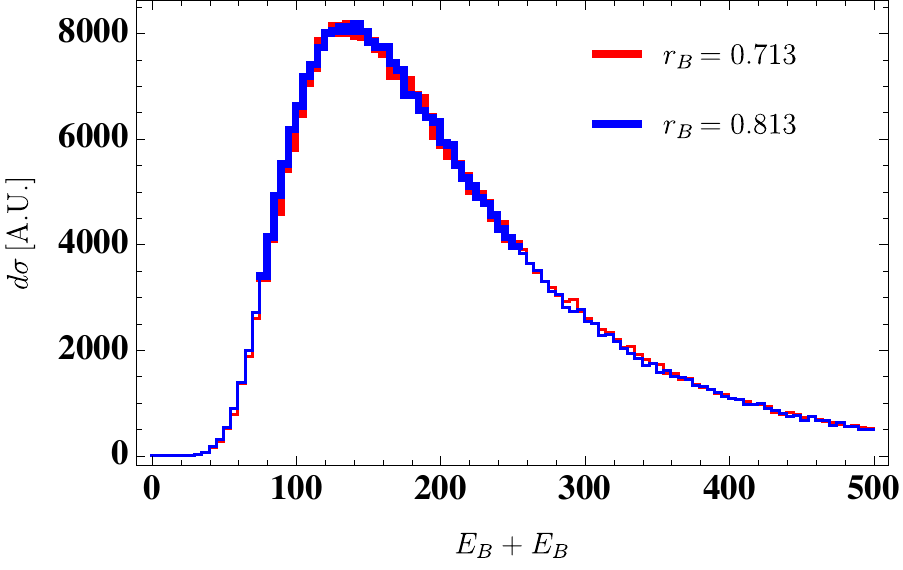}\includegraphics[width=0.49\linewidth]{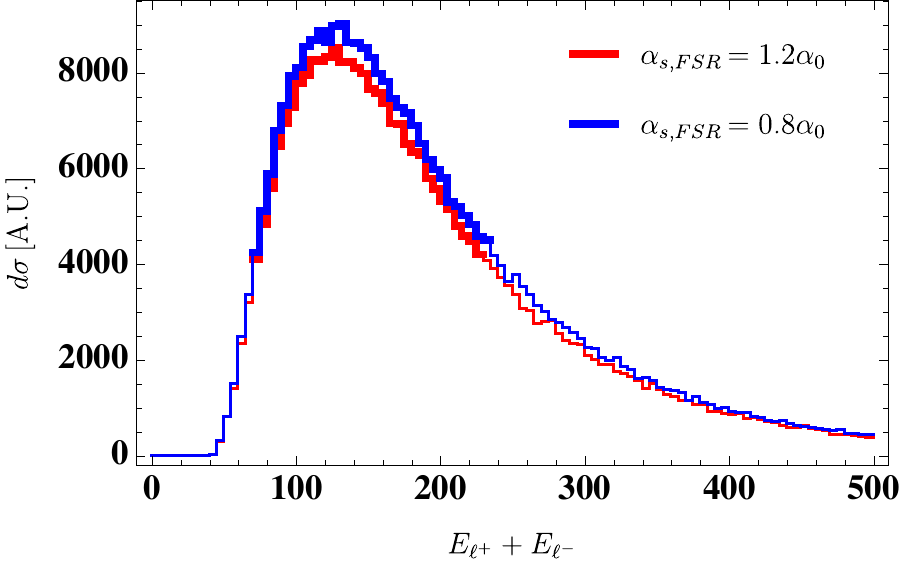}
\caption{\label{fig:ExampleSpectraMass}Example spectra of $p_{T,B}$, $E_{B}$, $E_{B}+E_{\bar{B}}$, $m_{B\ell}^{\textrm{true}}$, $m_{BB\ell\ell}$, and $E_{\ell^{+}}+E_{\ell^{-}}$  for $m_{t}=174\,\gev$. }
\label{default}
\end{center}
\end{figure}

We are now in position to report our results on observables sensitive to $m_t$, following a similar path to the preceding section on calibration observables.
We quantify the sensitivity of a mass-sensitive observable $\mathcal{O}$ 
to a given Monte Carlo parameter $\theta$ or to $m_{t}$, using a logarithmic derivative defined in eq.~(\ref{eq:DeltaGenericDef}). 
The numerical derivatives necessary to compute $\Delta_{\theta}^{(m_{t})}$ are obtained from straight-line fits and comparison with Monte Carlo predictions
varying the parameters in the same ranges as discussed for the purpose 
of the calibration variables.
Examples of spectra for different Monte Carlo parameter settings are shown in Figure~\ref{fig:ExampleSpectraMass}. 

For mass-sensitive observables we would like to have a pattern of sensitivity that is opposite to the one wished 
for calibrations observables, that is to say large sensitivity to $m_{t}$ and small dependence on $\theta$.
In fact, in order to reach a given accuracy $\delta m/m$ on 
$m_t$, we need to have under control each $\theta$
better than a relative precision
\[
\frac{\delta\theta}{\theta}=\frac{\delta m}{m}\frac{1}{\Delta_{\theta}^{(m_{t})}}\,.
\]
Therefore, the observables with smaller $\Delta_{\theta}^{(m_{t})}$ can be useful even
when shower and hadronization parameters $\theta$ are not 
known well.
As a consequence, for the observables to be used for the $m_t$ measurement we wish to have small $\Delta_{\theta}^{(m_{t})}$. %

Due to possible different sensitivity to $m_{t}$ which different
ranges of the distributions might
exhibit, we will carry out two analyses. One is more inclusive and is based on the dependence of the Mellin moment of each observable on $m_{t}$ and the Monte Carlo parameters; the other one 
is based on the shape analysis of certain spectral features. 
These shape analyses are motivated by known properties of kinematic endpoints of distributions in {\it e.g.,} $m_{B\ell}$ and $m_{T2}$, and of peaks of distributions in {\it e.g.,} $E_B$, since the corresponding quantities constructed with a $b$-jet instead of a $B$-hadron are less affected by dynamical details.
  Hereafter, we shall first discuss the Mellin moments and
  then the shape investigation.

\subsection{Mellin-moment observables \label{sec:Mellin-mt}}
Similarly to the analysis  carried out for the calibration observables in
Section~\ref{subsec:Constraining-power},
we look at the first Mellin moment of the distributions of mass-sensitive observables, in order to characterize the dependence on $m_{t}$ and on the parameters $\theta$.
We obtain each Mellin moment,
restricting the calculation to a range that roughly corresponds to the Full-Width Half-Maximum area. The exact choices that we adopt in our analyses
are reported in Table~\ref{tab:Sensitivity-Mellin-Mass}, together with the sensitivity values 
to each PYTHIA variables and to the top-quark mass.
In the first rows of Table~\ref{tab:Sensitivity-Mellin-Mass}, we present results for observables that can be reconstructed 
with just one hadron, followed by those using also leptons.
Then we present observables that would require the tagging of two $B$-hadrons and finally quantities that employ only leptons. 

A further comment is in order concerning the numbers in
Table~\ref{tab:Sensitivity-Mellin-Mass}. 
In the previous sections we have highlighted how a number of variations of some observables can be devised if we change strategy {\it i})  to pair together leptons and $B$-hadrons {\it ii}) to deal with the hadrons other than the $B$ in each jet {\it iii}) to deal with ISR, in the computation  of $m_{B\ell}$ and $m_{T2}$. It turns out that the several variants of these observables tend to give rather similar results when it comes to both sensitivity to $m_t$ and to Monte Carlo parameters. For this reason we show only representative results for each type of observables.
A similar table for HERWIG 6 parameters is given in Table~\ref{tab:Sensitivity-Mellin-Mass-1}, but
it is limited to the most representative observables among those
of the bigger set analyzed for PYTHIA.


\begin{figure}[htbp]
\begin{center}
\includegraphics[width=0.49\linewidth]{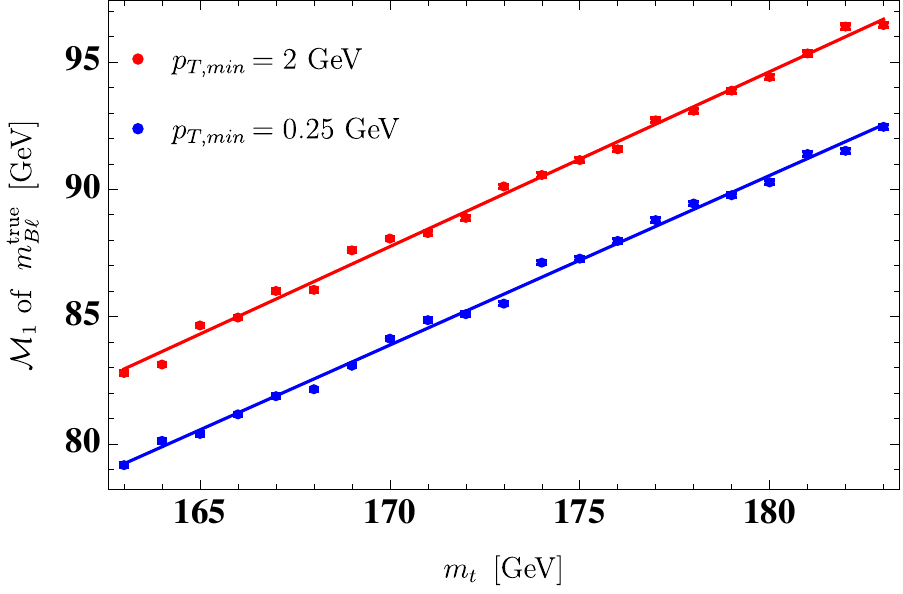}\includegraphics[width=0.49\linewidth]{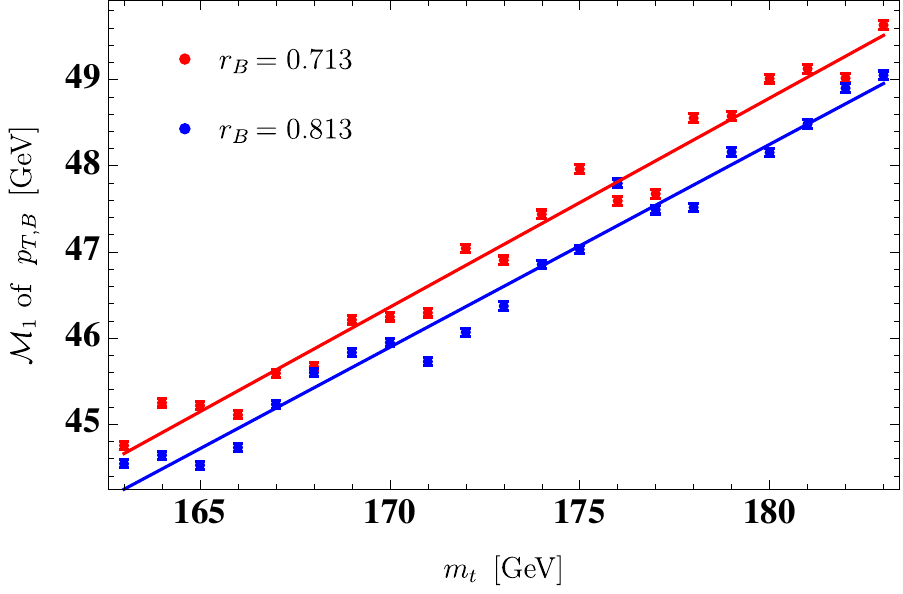}\\
\includegraphics[width=0.49\linewidth]{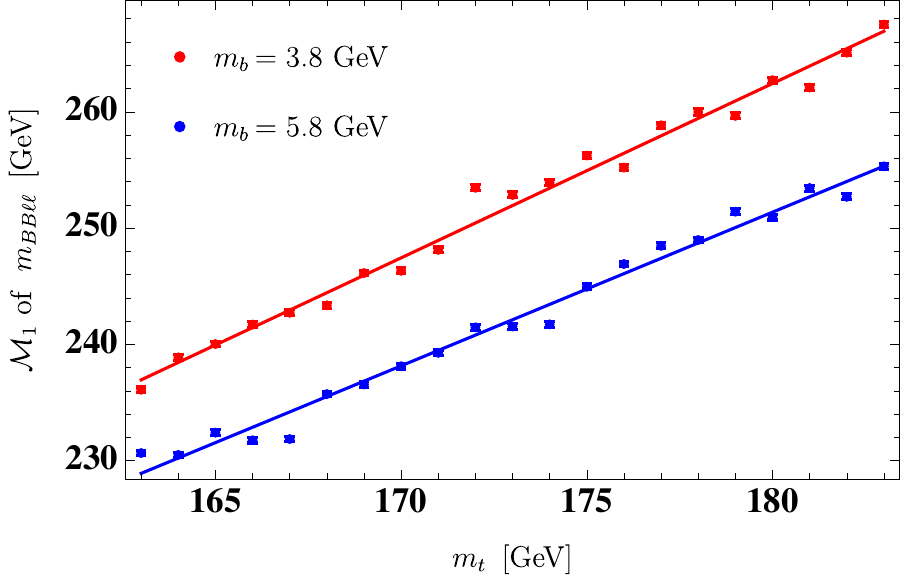}\includegraphics[width=0.49\linewidth]{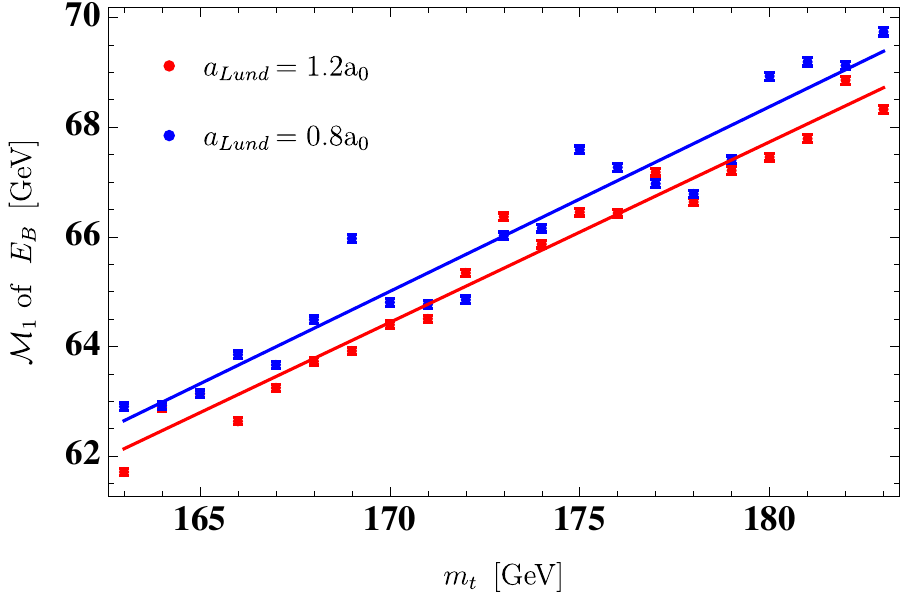}\\
\includegraphics[width=0.49\linewidth]{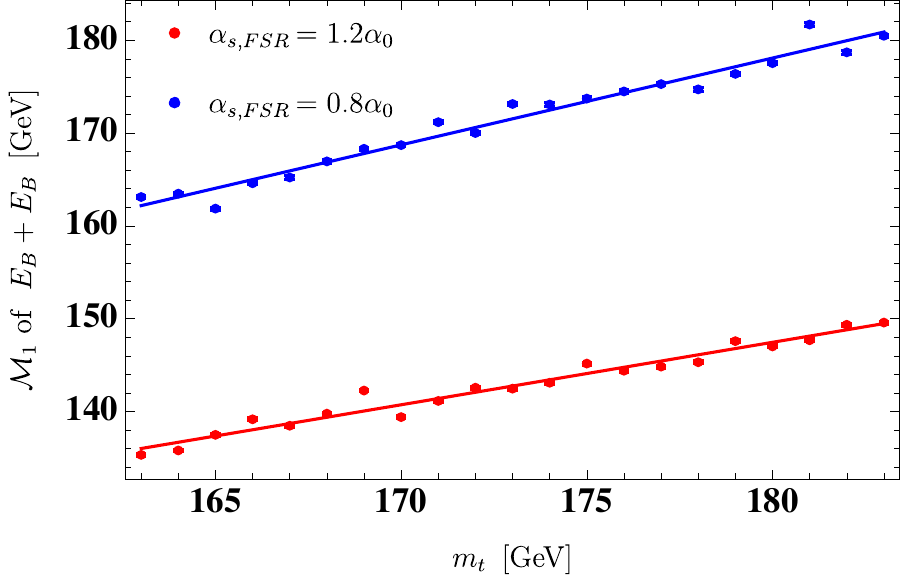}\includegraphics[width=0.49\linewidth]{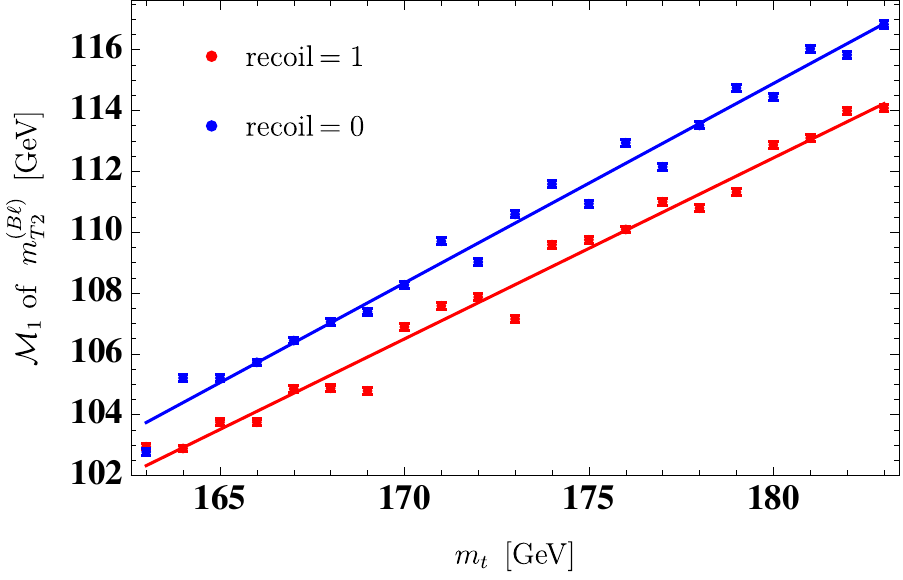}
\caption{\label{fig:ExampleSpectraMass1}Examples of $m_{t}$ dependence of the first Mellin moment  of $p_{T,B}$, $E_{B}$, $E_{B}+E_{\bar{B}}$, $m_{B\ell}^{\textrm{true}}$, $m_{BB\ell\ell}$, and $m_{T2}^{(B\ell)}$  for different Monte Carlo parameters settings. }
\label{default1}
\end{center}
\end{figure}

\begin{table*}
\begin{centering}
\begin{tabular}{|c|c|c|c|c|c|c|c|c|c|}
\hline 
\multirow{2}{*}{$\mathcal{O}$} & \multirow{2}{*}{Range} & \multirow{2}{*}{$\DeltaMO$} & \multicolumn{7}{c|}{$\Delta_{\theta}^{(m_{t})}$}\tabularnewline
\cline{4-10} 
 &  &  & $\alpha_{s,FSR}$ & $m_{b}$ & $p_{T,\text{min}}$ & $a$ &  $b$ & $r_{B}$  & recoil\tabularnewline
\hline 
\hline 
$E_{B}$ & 28-110 & 0.92(5) & -0.52(2) & -0.21(3) & 0.057(4) & -0.02(2) & 0.06(2) & -0.10(5) & -0.022(5)\tabularnewline
\hline 
$p_{T,B}$ & 24-72 & 0.92(3) & -0.54(2) & -0.21(2) & 0.056(4) & -0.03(2) & 0.07(1) & -0.09(4) & -0.023(2)\tabularnewline
 
\hline 
\hline 
$m_{B\ell,\text{true}}$ & 47-125 & 1.30(2) & -0.241(8) & -0.072(6) & 0.022(2) & -0.007(5) & 0.023(6) & -0.02(2) & -0.008(2)\tabularnewline
\hline 
$m_{B\ell^{+},\text{min}}$ & 30-115 & 1.16(2) & -0.282(5) & -0.078(7) & 0.024(2) & -0.011(7) & 0.021(7) & -0.04(2) & -0.010(1)\tabularnewline
\hline 
\hline 
$E_{B}+E_{B}$ & 83-244 & 0.92(4) & -0.50(2) & -0.21(2) & 0.056(6) & -0.02(2) & 0.07(3) & -0.08(6) & -0.020(4)\tabularnewline
\hline
$m_{BB\ell\ell}$ & 172-329 & 0.96(2) & -0.25(1) & -0.10(1) & 0.028(3) & -0.01(1) & 0.026(7) & -0.03(3) & -0.008(2)\tabularnewline
\hline 
$m_{T2,B\ell,\text{true}}^{\text{(mET)}}$  & 73-148 & 0.95(3) & -0.27(1) & -0.09(1) & 0.029(3) & -0.009(9) & 0.03(1) & -0.03(4) & -0.010(3)\tabularnewline
\hline 
$m_{T2,B\ell,\text{min}}^{\text{(mET)}}$  & 73-148 & 0.95(3) & -0.27(1) & -0.09(1) & 0.029(3) & -0.009(9) & 0.03(1) & -0.03(4) & -0.010(3)\tabularnewline
\hline 
\hline 
$m_{T2}^{(\ell\nu)}$ & 0.5-80 & -0.118(7) & -0.03(2) & 0.00(2) & 0.002(8) & 0.00(2) & -0.01(2) & 0.00(7) & 0.004(5)\tabularnewline
\hline 
$m_{\ell\ell}$ & 37.5-145 & 0.40(5) & -0.03(5) & -0.01(4) & 0.00(1) & 0.01(5) & 0.01(4) & 0.0(1) & 0.00(1)\tabularnewline
\hline 
$E_{\ell}+E_{\ell}$ & 75-230 & 0.54(5) & -0.03(3) & 0.00(3) & 0.003(9) & 0.01(3) & -0.00(2) & 0.06(9) & 0.003(8)\tabularnewline
\hline 
$E_{\ell}$ & 23-100 & 0.48(4) & -0.02(5) & 0.00(5) & 0.004(9) & 0.01(4) & -0.01(4) & -0.06(9) & 0.003(8)\tabularnewline
\hline 
\end{tabular}
\par\end{centering}
\caption{\label{tab:Sensitivity-Mellin-Mass} Columns under $\Delta_{\theta}^{(m_{t})}$ show the sensitivity  to shower and hadronization PYTHIA parameters of the top quark mass extracted from the first Mellin moment of mass-sensitive observables. The  column $\DeltaMO$ reports the sensitivity of the 
first Mellin moment to the top quark mass. The range extremal values reported in the second column are all in GeV. These ranges correspond to the FWHM for $m_{t}=174 \gev$, for other masses we have used the corresponding FWHM.}
\end{table*}

\begin{table*}
\begin{centering}
\begin{tabular}[t]{|c|c|c|c|c|c|c|c|c|c|c|}
\hline 
\multirow{2}{*}{$\mathcal{O}$} & \multirow{2}{*}{$\DeltaMO$} & \multicolumn{9}{c|}{$\Delta_{\theta}^{(m_{t})}$}\tabularnewline
\cline{3-11} 
 &  & PSPLT & QCDLAM & CLPOW & CLSMR(2) & CLMAX & RMASS(5) & RMASS(13) & VGCUT & VQCUT\tabularnewline
\hline 
\hline 
$m_{B\ell,\text{true}}$ & 0.52 & 0.036(4) & -0.008(2) & -0.007(5) & 0.002(3) & -0.007(4) & 0.058(1) & 0.06(5) & 0.003(1) & -0.003(3)\tabularnewline
\hline 
$p_{T,B}$ & 0.47 & 0.072(1) & -0.03(9) & -0.02(7) & 0.0035(5) & -0.03(5) & 0.11(9) & 0.12(5) & 0.0066(2) & -0.006(5)\tabularnewline
\hline 
$E_{B}$ & 0.43 & 0.069(7) & -0.026(7) & -0.017(5) & 0.0038(9) & -0.01(2) & 0.12(1) & 0.12(2) & 0.006(2) & -0.007(5)\tabularnewline
\hline 
\hline 
$E_{\ell}$ & 0.13 & 0.0005(5) & -0.04(3) & 0.04(2) & -0.0002(2) & -0.004(4) & 0.008(3) & 0.008(2) & -0.002(5) & 0.008(2)\tabularnewline
\hline 
\end{tabular}

\par\end{centering}
\caption{\label{tab:Sensitivity-Mellin-Mass-1}
  As in Table~\ref{tab:Sensitivity-Mellin-Mass}, but in terms of the
  HERWIG 6 shower and hadronization parameters. Ranges are those reported in Table~\ref{tab:Sensitivity-Mellin-Mass} for all values of $m_{t}$.}
\end{table*}

Several interesting observations can be made on these tables.
As for the PYTHIA study we find that the uncertainty on $\alpha_{S,FSR}$ has
the largest impact on the $m_t$
measurement: a relative error on $\alpha_{s,FSR}$ around 1\%
seems necessary to achieve a
0.5\%-level top quark mass determination. This level of precision in 
the knowledge of  $\alpha_{s,FSR}$ is clearly signaling the overall necessity to describe radiation very accurately to
attempt a precision measurement of the top quark mass. One could argue that the intrinsic error of a calculation based on a leading order matrix element attached to a leading-logarithmic parton shower, such as PYTHIA,
makes it meaningless to talk about fixing one of its main parameters to such a precision. In fact, one can interpret our results as a starting point 
to conclude that $i$) a calculation with matrix elements computed
beyond leading order matched to a parton shower might be used to (hopefully)
get a smaller sensitivity to the $\alpha_{s,FSR}$ parameter and $ii$) a more accurate description of the parton shower beyond leading logarithms
would probably be necessary as well to warrant  a top-quark mass
measurement from Monte Carlo simulations
with sub-percent precision. In any case, for the scope of this paper, we
will treat the findings on $\alpha_{s,FSR}$ at face value, hence we conclude
that this parameter should be known with a precision from 1\% to few \%,
for the sake of employing any observables which does not rely only
on lepton momenta.

Concerning the other Monte Carlo variables, we observe that the
$m_t$-dependent observables are much less sensitive to variations of the other
parameters. The second most relevant one is the bottom-quark mass $m_{b}$,
which, in order not to spoil
the $m_t$ measurement, should be known with precision between
few \% and 10\% from observables not made of just lepton momenta.
Even smaller sensitivity values are obtained for 
$a$, $b$, $r_{B}$, $recoil$, and $p_{T,\text{min}}$:
determining them to a 10\% accuracy should be enough to
warrant a 0.5\% precision on $m_t$.

Purely hadronic quantities such as $p_{T,B}$, $E_{B}$,  and $E_{B}+E_{\bar B}$  
yield quite similar values of 
$\Delta^{(m_{t})}_{\theta}$. Therefore, the uncertainties
due to hadronization and showering for a mass measurement based on any of these observables are expected to be essentially the same.
Variables like $m_{B\ell}$ are in general half as sensitive as purely
hadronic observables, hence they can tolerate a worse knowledge of the 
Monte Carlo parameters, even by a factor of 2.
To better put these results in context, we can compare them to the sensitivity $\Delta^{(m_{t})}_{\theta}$ of other observables defined in terms of the charged leptons, such as the single-lepton energy $E_{\ell}$ and the invariant mass of the two leptons $m_{\ell\ell}$. In general, we observe a reduction by a factor of 10 or greater in $\Delta^{(m_{t})}_{\theta}$, in such a way that
their parameter dependence  in most cases becomes
statistically compatible with zero. We can therefore argue
that the mild sensitivity of 
observables like $m_{B\ell}$ is a genuine consequence of the
small dependence of the lepton variables on the details of shower and
hadronization.

As for HERWIG 6, our investigation is limited to the $B\ell$ ``true''
invariant mass, and the energy and transverse momentum of $B$-hadrons, and
the energy of charged leptons in $W$ decay.
For all observables and parameters, the $\Delta$ values run between
values statistically compatible with zero and ${\cal O}(10^{-2})$, the highest
dependence being exhibited by $m_{B\ell}$, $E_B$, and $p_{T,B}$ on
$b$-quark and gluon effective masses, RMASS(5) and RMASS(13), and
on the cluster-spectrum variable PSPLT(2).
This implies that, aiming at a precision of 0.5\% on $m_t$,
it would be roughly sufficient to be aware of 
showering and hadronization parameters with about 10\% accuracy.
Before concluding this subsection, some comments on the
tuning of the strong coupling constant are in order.
Unlike PYTHIA, where different couplings for initial- and
final-state radiation are employed and the value of the coupling
at the $Z$ mass is possibly fitted, in HERWIG the user-defined
quantity QCDLAM is a modified version of the $\overline{\rm MS}$
parameter $\Lambda_{\rm QCD}$, according to the CMW prescription 
\cite{Catani:1990rr}. It is easy to prove, using the
formulas in \cite{Catani:1990rr} that a relative uncertainty of
10\% on QCDLAM implies a precision about 1-2\% on
$\alpha_S(m_Z)$ in the CMW scheme.

\subsection{Shape Analyses}

The Mellin-moment analysis above can be applied to all observables,
since they all have a rise-and-fall shape and it appears reasonable
to correlate it with $m_t$. 
However, for some observables, kinematic
features such as peaks and endpoints are likely to yield more precise and more robust extractions of $m_{t}$ than
the Mellin moments. 
These features
might possibly protect one against uncertainties on 
showering and hadronization parameters. For
example, at leading order and in the narrow-width approximation,
the $m_{B\ell}$ endpoint arises from on-shell conditions
of $W$ bosons and top quarks. For perfectly on-shell kinematics,
regardless of the parton shower, such an endpoint should
not be surpassed in any event, because it is completely dictated by the phase space of the top quark decay, and not by the underlying dynamics.
Barring the soft colored quanta exchanged
to neutralize the color of the top quark with the one of the rest
of the event, the endpoint should be respected by all events,
regardless of how the hadrons were formed.
Motivated by {this possible reduction of sensitivity to the Monte Carlo parameters}, we
extend our previous analysis
to the kinematic endpoints of the $B\ell$ invariant mass and  of
the transverse mass $m_{T2}$ distributions, and to 
the peak position of the $B$-hadron
energy spectrum. Endpoints and peaks will be determined
by fitting their spectra with suitable functions.

\subsubsection{Definition of the dedicated shape analyses}

In order to extract the kinematic endpoints and energy-peak position, we  analyze the part of the spectrum around them, employing fitting {\it ansatze}
  which capture the necessary features. 
For the kinematic endpoints, we take a simple second-order polynomial 
\bea
y(x)= c_1 (x-x_m)^2 + c_2 (x-x_m), \label{eq:parabola}
\eea
where $x_m$ is identified as the corresponding endpoint and $c_{1,2}$ are
other fit parameters, irrelevant to our study.
For determining the peak position in a $B$-hadron
spectrum, we adopt the following ansatz, inspired by the measurement of
the $b$-jet energy peak in Ref.~\cite{Agashe:2012bn}:
\bea
y(x) \sim \exp \left[-p\left(\frac{x^*}{x}+\frac{x}{x^*} \right) \right]\,,
\eea
where $x^*$ is identified as the peak position and $p$ is another
fitting parameter.

As for the endpoint of $m_{B\ell}$, we use the data in the
range above $m_{B\ell}=127$ GeV, which  corresponds to a fraction $\epsilon_{\text{tail}}\simeq 0.05$ of the total cross-section for $m_{B\ell,\text{true}}$
and $m_t=173$~GeV.
In terms of the number of events $N_{ev}$ expected in each experiment at the LHC we have
\[
N_{\text{ev}} \sim 5 \times 10^{3}\cdot \frac{\mathcal{L}}{1\,\text{ab}^{-1}}\cdot \frac{\epsilon_{\text{tagging}}}{10^{-4}}\cdot\frac{\epsilon_{\text{tail}}}{0.05}\cdot {\rm BR} (t\bar{t})
\]
where ${\rm BR}(t\bar{t})$ is the branching ratio of $t\bar{t}$ into a given decay mode, and $\epsilon_{\text{tagging}}$ is the tagging efficiency for an exclusive final state with a $B$-hadron.
This estimate implies that the foreseen LHC dataset will yield  statistical uncertainties ($\gtrsim 1.5\%$) somewhat above the level of the systematics
at which we are aiming ($\sim 0.5\%$).
There are several ways to get around the potential loss of overall precision due to the statistical uncertainty.
First of all, 
if a tagging efficiency better than $10^{-4}$ could be achieved, for instance by not requiring exclusive $B$-hadron decays for tags, then the resulting statistical errors might become comparable to the systematic ones.
Secondly,
other experimental considerations
may motivate to employ a broader range than the very end of the tail:
for instance, in the CMS analysis \cite{CMS-PAS-TOP-11-027}
the $m_{b\ell}$ region used in the fit 
corresponds to $\epsilon_{\text{tail}}\sim 0.1$.
One possible issue that one can raise with this trial is that the $m_{B\ell}$-endpoint analysis might {\it in}directly reintroduce sensitivities to Monte Carlo parameters, through including more bulk of the distribution than endpoint region for the fit,
resulting in an overall sensitivity to the Monte Carlo parameters similar to the Mellin moment
analysis of the above Section \ref{sec:Mellin-mt}.
We have explicitly checked the sensitivities with smaller statistics, {\it e.g.,} $\epsilon_{\text{tail}}\sim 0.02$, and found that there are no
relevant effects. Nevertheless, we leave to the experimental collaborations
the task of looking for an optimal balance; similar considerations,
including the fit-range choice, apply to the
other kinematic endpoints (see also Table~\ref{tab:Sensitivity-Mass-Obs}).

For the study of the  $E_B$-peak position it is much easier to secure  enough statistics. Of course, the exact range of energies in which we analyze data affects the sensitivity to shower and hadronization parameters. In this case as well,
we leave the optimal choice of range to the experimental collaborations. In our analysis, we simply take $[35, 85]$ GeV for illustration, which corresponds roughly to the full width at three quarters of the maximum.

Finally, we make a further comment on notational conventions.
Given the several options of reconstruction, subsystem and pairings,
hereafter, for the sake of a more effective notation,
the variants on which we focus will be labeled with roman letters in the math (min, true, ISR, mET). To avoid notational clutter with unnecessary indexes,
we shall denote the endpoint of a mass distribution by a breve accent, {\it i.e.,} $\breve{m}$, which should be evocative of our parabola fit near the endpoint.

\subsubsection{Results of the dedicated shape analyses}
The results on the sensitivity of our observables on the top-quark mass
and of $m_t$ on the Monte Carlo parameters
are reported in Table~\ref{tab:Sensitivity-Mass-Obs}.

First of all we would like to comment on the form
used to present these results: in fact,
whenever the sensitivity of an observable to a given parameter is very small,
it becomes numerically
challenging to determine it.
In the following we will be interested in identifying parameters that
have remarkable effects on the observables; therefore, in presence of
a small sensitivity, we will content ourselves with stating that 
it is below the few per-mil level, $\Delta_{\theta}^{(m_{t})}<0.003$.
In fact, for an observable to be used to measure $m_t$, this level
of sensitivity is clearly safe, as it implies that knowing just
the order of magnitude of a given $\theta$ will be enough to warrant
a $10^{-3}$ precision on $m_t$.
\begin{table*}
\begin{centering}
\begin{tabular}{|c|c|c|c|c|c|c|c|c|c|}
\hline 
\multirow{2}{*}{$\mathcal{O}$} & \multirow{2}{*}{Range}  & \multirow{2}{*}{$\Delta_{m_{t}}^{(\mathcal{O})}$}  & \multicolumn{7}{c|}{$\Delta_{\theta}^{(m_{t})}$}\tabularnewline
\cline{4-10} 
 &  &  & $\alpha_{s,FSR}$ & $m_b$ & $p_{T,\text{min}}$ & $a$ & $b$ & $r_{B}$ & recoil \tabularnewline
\hline 
\hline 
$E_{B,\text{peak}}$ & 35-85 & 0.8(1) & -0.74(9) & -0.26(4) & 0.05(1) & -0.04(2) & 0.08(3) & -0.07(9) &  -0.031(7) \tabularnewline
\hline 
$\breve{m}_{B\ell,\text{true}}$ & 127-150 &1.26(1)& 0.017(6) & 0.003(9) & -0.006(2) & -0.008(2) &  0.008(7) & -0.016(6) &  -0.00042(9) \tabularnewline
\hline 
$\breve{m}_{B\ell,\text{min}}$  & 127-150 & 1.28(1) & -0.023(3) & -0.022(2) & 0.006(3) & -0.008(3) & 0.008(3) & -0.02(1) & -0.0001(6)  \tabularnewline
\hline 
\hline 
$\breve{m}_{T2,B\ell,\text{true}}^{\text{(mET)}}$ & 150-170 & 0.98(2) & -0.01(2) & -0.023(3) & 0.007(1) & -0.006(3) & 0.010(4) & -0.011(9) & -0.0002(8) \tabularnewline
\hline 
$\breve{m}_{T2,B\ell,\text{min}}^{\text{(mET)}}$  & 150-170 & 0.97(2) & -0.02(1) & -0.021(5) & 0.006(2) & -0.006(3) & 0.009(4) & -0.01(1) & -0.0001(8) \tabularnewline
\hline 
$\breve{m}_{T2,B\ell,\text{min,}\perp}^{\text{(mET)}}$  & 138-170 & 0.89(2) & -0.071(5) & -0.046(7) & 0.012(2) & -0.011(7) & 0.010(8) & -0.01(2) & -0.002(1) \tabularnewline
\hline 
\hline
$\breve{m}_{T2,B}^{\text{(mET)}}$  & 142-170 & 0.95(3) & -0.089(6) & -0.064(6) & 0.018(1) & -0.017(4) & 0.031(4) & -0.04(2) & -0.0028(8) \tabularnewline
\hline
$\breve{m}_{T2,B,\perp}^{\text{(mET)}}$  & 126-170 & 0.94(4) & -0.07(1) & -0.04(1) & 0.011(3) & -0.009(9) & 0.02(1) & -0.03(4) & -0.001(2) \tabularnewline
\hline 
\end{tabular}
\par\end{centering}
\begin{centering}

\par\end{centering}

\caption{\label{tab:Sensitivity-Mass-Obs}
Sensitivity to shower and hadronization parameters of the top quark mass extracted from a shape analysis of special feature in the spectra of $E_{B}$, $m_{B\ell}$, $m_{T2}$. The range extremal values reported in the second column are all in GeV. }
\end{table*}

For the energy-peak fit, we find results rather
close to those from the Mellin-moment analysis: regarding 
hadronization and showering uncertainties,
determinations of $m_t$ from a moment or from energy peaks 
are on the same footing.
In fact, top-quark decays $t\to W^{+}B+{\rm hadrons}$ are genuinely
multi-body processes, hence the $B$-energy peak does not fully enjoy the process-independent ``invariance'' property of the peak in
$b$-jet energy spectra~\cite{Agashe:2012bn,Agashe:2012fs}, which, instead is typical for a two-body process such as 
$t\to j_{b}W^{+}$ with $j_{b}=B+\text{hadrons}$. 

For the $B\ell$ invariant mass, the endpoint analysis
shows a very significant reduction of shower and hadronization
sensitivity: in fact, 
the upper-bound on $m_{B\ell}$ is roughly the same as  
for an ideal two-step two-body on-shell decay, {\it i.e.,} $B \rightarrow b$ limit where no shower or hadronization effect comes into play. 
However, we remind that in practice the endpoint is extracted by fitting a certain amount of bulk region with a suitable template, {\it e.g.,} a parabola in Eq.~(\ref{eq:parabola}). 
As discussed before, the endpoint measurement is indirectly affected by the Monte Carlo parameters via fit parameters $c_1$ and $c_2$ in Eq.~(\ref{eq:parabola}) that are responsible for the detailed shape near the endpoint.
In the chosen range, the dependence of $m_{B\ell}$ on hadronization and showering parameters is milder by an order of magnitude with respect to the 
Mellin moment analysis.
In some cases, the sensitivity is even below the statistical uncertainty: for example, the binary choice of the conservation of momentum in the shower, which has a $\Delta_{\text{recoil}}\sim0.02$ in the $m_{B\ell}$ Mellin
analysis, becomes completely negligible for the endpoint strategy. 
As for $p_{T,\text{min}}$, $a$, $b$, and $r_{B}$, the corresponding
$\Delta_\theta^{(m_t)}$ are ${\cal O}(10^{-3})$, 
hence they are borderline to be irrelevant;
$\alpha_{S,FSR}$ and $m_b$ carry small impacts,
although not completely negligible.
It should be remarked that these results rely on using only portion
of the spectrum in the parabola fit. 

As for $m_{T2}$, all variants perform in a very similar way, with 
sensitivity being (mostly) as reasonably small as $\Delta\sim
  {\cal O}(10^{-3})$.
Very little difference appears between the ``min'' and ``true''
options, and therefore, it seems that the pairing or combinatorial issue 
does not make a worrisome impact on the mass measurement.
The mET and ISR prescriptions show a comparable dependence on the Monte Carlo parameters, for both $B$- and the $B\ell$-subsystems
(see also Appendix~\ref{appV} for a complete list).
The mET ones exhibit
the smaller sensitivity to some parameters, while the ISR ones 
are less dependent on other variables.
Therefore, both
prescriptions should be considered equally good to determine $m_t$.
Again, these results hold for the analysis with relatively narrow fitting ranges towards the tails, yielding reliable parabola fits of the endpoints, 
even with no reference to the details of the reconstruction.
As for $m_{T2}$ in the $B$-subsystem,
given the little difference between the ``min'' and ``true'' cases 
in the $B\ell$-subsystem, the use of the $B$-subsystem options loses
some of its initial motivations.
However, it exhibits a dependence on the
Monte Carlo parameters very close to the lowest of $m_{T2}$, hence
it can be a useful validation of the $B\ell$ results.

One notable feature has to do with the ``perp'' correction applied to the variables denoted by the $\perp$ subscript.
In this subsystem, the $\perp$-correction leads to
a slight reduction of the sensitivity to the Monte Carlo parameters, in particular, to $\alpha_{S,FSR}$ and $m_B$ for the mET prescription. 
By performing the $\perp$-correction, any process-dependent effect in association with initial-state radiation, which is somehow
convoluted with the Monte Carlo parameters of interest, is detached, so
that the resulting sensitivity gets reduced. 
By contrast, such a reduction does not arise for the ISR-reconstruction
scheme, since the $\perp$-correction itself carries over its own sensitivity to the parameters through $A +\bar{A}$.
The $\perp$-variants in the $B\ell$ subsystem are somewhat more involved: 
Here the inclusion of leptons in constructing the variables already mitigates the sensitivity, even before applying the $\perp$-correction.
It turns out that the net effect resulting from the $\perp$-correction is not friendly in the $B\ell$ subsystem. 
Nevertheless, this class of variables are worth investigating, in the sense of probing sensitivities to the Monte Carlo parameters with any production-dependent effects decoupled.

Overall, we find a conclusion similar to the Mellin analysis:
ISR in $t\bar{t}$ production, pairing leptons and $B$-hadrons,
as well as the treatment
of light-flavored hadrons in the $b$-jet,
are all largely orthogonal from the viewpoint
of the sensitivity to the Monte Carlo hadronization parameters. 

\section{Conclusions \label{sec:conclusions}}

The accurate
measurement of the top-quark mass is one of the major goals for the physics
program of the Large Hadron Collider. The level of precision that
is demanded for $m_t$ to be useful in the context of physics
of the Standard Model and in tests of new physics is around 500 MeV,
well below the percent level. 

Carrying out a measurement
at such a level of accuracy is challenging and
has generated a number of proposals. All
strategies rely to some extent on our ability to predict
some observables by means of a QCD calculation or a Monte
Carlo simulation, {\it e.g.,} a total or fiducial rate of top
quark production, or the shapes of some differential distributions:
all such methods exhibit an associated theoretical error.
Presently, most precise determinations of $m_t$ use quantities
that are believed to be less affected by theoretical errors, such
as the peak of invariant mass obtained with kinematic (template)
methods, which
essentially tries to capture the peak in $b$-jet$+W$ invariant mass
in top-quark decay. In this case the
most important systematic uncertainty is the jet energy scale, 
namely the correction necessary to measure the true jet energy from
the measured detector response. These corrections
have been the subject
of intense studies and could be improved in future, but probably only
modestly. Therefore, it seems that it may not be possible
to improve significantly the measurement of the top-quark mass from
these ``standard'' methods. 

Motivated by the above drawbacks, a number of ``alternative'' strategies
have been proposed and will
acquire even more importance in the future. 
In fact, many of the alternative techniques pursue
different strategies than the ``standard'' methods, and 
hence they will allow to cross-check the available and existing results,
eventually giving rise to a global determination of $m_t$,
accounting for measurements performed by using 
inherently different strategies.
Of course, the most useful measurements in this combination
will be those 
that are not affected significantly by the jet energy correction, which
presently dominates the uncertainty on $m_t$.
Among the proposals,
there are measurements that utilize only leptons from semi-leptonic
and di-leptonic final states of $t\bar{t}$ decays and others that
employ exclusive and semi-exclusive hadronic final states. In both
cases there is no reference to jets in the formulation of the observables
that are used to extract $m_{t}$ from the comparison of data and
calculations, and therefore the jet energy correction has a
very minor effect in these measurements. These observables are, however,
not free from uncertainties: in fact, when dealing with hadrons,
one clearly encounters the difficulty to carry out
theoretical predictions for such final states. For leptonic final
states as well, there are residual theory errors because the distributions
that one uses to extract $m_{t}$ are  usually less sensitive to $m_{t}$,
but they still carry some non-negligible
sensitivity to the theoretical modeling, \emph{e.g.,} to missing orders
in perturbation theory or to the description of the production mechanism
of the top quarks at the LHC.

The goal of our work was to quantify the uncertainty on the top-quark
mass extraction due to our limited ability to compute observables
for exclusive hadronic final states. We have considered a number of
possible observables that can be defined on exclusive hadronic final
states, both drawing from the literature and introducing some new ones,
potentially useful to explore the achievable precision
on $m_{t}$ from the study of hadrons in $t\bar{t}$ samples. We
have concentrated our efforts on Monte Carlo event generators and
the dependence of the predictions on their tunable parameters.
As discussed throughout the paper, Monte Carlo codes are often
used to derive predictions on exclusive hadronic final states at colliders
and are often preferred over theoretical approaches, such as
resummed calculations in the fragmentation function formalism, 
which, as discussed in the Introduction,
tend have a difficulty \emph{i}) to give predictions for generic
observables, that can instead be obtained from Monte Carlo programs;
\emph{ii}) to give precise-enough predictions at least by using
the currently available perturbative computations, which are
typically NLO+NLL, and \emph{iii}) to yield results at hadron
level without relying on phenomenological
non-perturbative models, which are often too simple to
describe the data and eventually meet the LHC precision goals.
Therefore we believe that a study of 
exclusive hadronic final states is best carried out employing Monte Carlo
event generators. 

We have simulated $t\bar t$ events by means of the 
PYTHIA~8 and HERWIG~6 programs, which differ considerably
in the description of both emissions of soft and collinear
partons in the so-called ``parton shower'' and of the 
formation of hadrons. Looking at both approaches,
we are able to evaluate the peculiarities of the codes and 
identify the common issues which are to be addressed in the top-quark
mass measurement strategies.

For both codes, we have evaluated the dependence
of several observables on $m_{t}$ and on
the most relevant Monte Carlo parameters.
To simplify the understanding of the results,
we have introduced a sensitivity measure $\Delta$, defined
in eq.(\ref{eq:DeltaGenericDef}), which allows to look in a uniform
way to the vast number of results that we have collected in our tables.
The sensitivity has been computed for the mean value of each
quantity, calculated in a specific range, which corresponds
approximately to the bulk of the distribution of the
observable.

The observables belong roughly to the following
classes: purely hadronic observables, \emph{e.g.} the $p_{T}$ of
the $B$-hadron, mixed leptonic-hadronic observables, \emph{e.g.}
the invariant mass $m_{B\ell}$, and purely leptonic observables,
\emph{e.g.} the lepton energy $E_{\ell}$. As could have been anticipated,
the description of the hadronic final states is very important
for purely hadronic observables and slightly less important for the mixed
leptonic-hadronic ones. Leptonic quantities usually have so
small dependence on the Monte Carlo parameters,
that we observe a dependence compatible with zero within our statistical
uncertainties arising from limited simulated data sample. 

More quantitatively, we find that, in order to warrant a 0.5 GeV precision
on the $m_t$ extraction from hadronic and mixed
leptonic-hadronic observables,
most parameters must be known with a precision around 10\%
in both PYTHIA and HERWIG: it is clear that
such an accuracy cannot be attained without inputs from data.
We will discuss shortly how the interplay between data, 
parameters and the measurement of the top quark mass might unfold.
Before doing that, we highlight that for both PYTHIA and HERWIG it
seems necessary to have an even more precise knowledge of the strong
coupling constant used in the parton shower.
For PYTHIA
which uses a parameter named
$\alpha_{s,FSR}$, corresponding to $\alpha_{s}(m_Z)$ with a one-loop
evolution,
we find that hadronic and mixed leptonic-hadronic observables
demand the knowledge of $\alpha_{s,FSR}$
with a precision from 1\% up to a
few \%.
The necessity to fix $\alpha_{s}$ so precisely 
clearly signals the need for improved theoretical predictions,
which could be ultimately achieved by
using beyond-LO hard matrix elements as well as 
a more accurate description of the parton shower,
{\it e.g.}, in the NLL approximation. In any case,
even within our study with the present parton shower accuracy,
the message is clear: the
description of radiation is of utmost importance for the measurement
of $m_t$ if one uses any observable that involves hadrons. 

The striking precision with which it is necessary to know the Monte
Carlo parameters motivated two further explorations in our analysis:
\emph{i}) the study of more robust observables, possibly less sensitive to
the Monte Carlo parameters that necessarily act as nuisances in the
$m_t$ determination; \emph{ii}) the investigation of a strategy
to derive directly from data the value of the Monte Carlo parameters
to be used for the mass extraction. 

For the former,
we have studied the use of kinematic endpoints of certain masses
that are known to carry information on the top quark mass.
To this end, we investigated the endpoints
of $m_{B\ell}$ and of several variants of the $m_{T2}$ variable which can be constructed from the four observable final state particles ({\it i.e.,} two $B$ hadrons and two leptons) of the fully leptonic $t\bar{t}$ decay. 
A crucial observation drawn from our scrutiny is that studying the regions very close to the endpoint enables us to make predictions sufficiently independent from Monte Carlo parameters. 
The price to pay for this is to reduce the number of events 
to be analyzed, hence inflating the statistical uncertainties to a level
jeopardizing the entire strategy.
In other words, the resulting statistical uncertainty can be even beyond the level of theoretical uncertainty that we aim to achieve.
Furthermore, it should be remarked that,
when one concentrates on the region of the spectrum so close
to the endpoint, the current limit in the Monte Carlo description,
\emph{e.g.,} ``off-shell''
production of top quarks, not-calculated perturbative higher orders,
which were assumed sub-dominant and hence neglected in our study, might
start to play an important role.
To alleviate this issue one would be forced to consider a larger portion
of the spectrum, hence reintroducing a larger dependence on the Monte
Carlo parameters. Seeking a balance between statistical
and Monte Carlo uncertainties parameters will hence
be a delicate task for the experimental collaborations. 

Clearly, if one had a good knowledge of the Monte Carlo parameters,
all the above-mentioned conclusions on the uncertainties in the top quark
mass determination could be improved.
For this reason we have studied what kind of constraints
can be obtained from data if one tries
to use a number of measured distributions in $t\bar t$ events in order
to adjust the parameters in the Monte Carlo event generators.
We have restricted a majority of our study to PYTHIA,
expecting that the general conclusions and observations
obtained with PYTHIA will be valid for HERWIG as well. 

The effort of reproducing measured distributions varying the parameters
might be considered as a ``tuning''
similar to the general-purpose tunings 
which exist for Monte Carlo
generators. However, we remark that for the mass measurements outlined
above it is necessary to ``tune'' $\alpha_{s}$ to about 1\% precision
and other parameters around 10\% precision. 
Therefore, one should take this tuning in part as an attempt
to capture physics effects
that go beyond the approximation inherent in the use of
the parton shower approach.
In other words, in the fitting procedure, one incorporates
in the parameters all the effects due to the incomplete
description of perturbative and non-perturbative physics
in the event generator.
Therefore, our tuning is quite likely 
to be specific to the top-quark events, 
while in general it may not be valid for other processes.
Therefore, even if PYTHIA is eventually tuned to top-quark data, 
it will be very unlikely for such fits 
to be ``universal'', {\it i.e.,} they are not
expected to be reliable outside the {\it very} same data
sample accounted for in the parameter adjustment.
For this reason we would like better to call this procedure a ``calibration''
of the Monte Carlo variables, specific to $t\bar t$ processes.
Such a calibration is ideally performed on the same data as what
will be used to measure the top-quark mass, hence we might call this strategy a $m_t$ determination, in conjunction with an ``in-situ'' Monte Carlo
calibration.

As for PYTHIA, we have studied seven parameters
which affect the showering and hadronization modeling.
We find that, in order to constrain them from $t\bar t$ data,
one has to examine the detailed shape of several observables, as the
study of more inclusive quantities (\emph{e.g.,} averages in a range)
may not be enough.
We further observe
that it is necessary to carry out a global analysis with several quantities,
as no single-observable exploration
can constrain all the necessary parameters. 

In summary, we find that using simple features, {\it e.g.,} the mean of the spectra, to extract the top-quark mass requires good knowledge
on the Monte Carlo parameters.
Such parameters
can be obtained from the very same data sample used to determine $m_t$,
but looking at observables which have sufficiently large
dependence on the parameters and very little sensitivity to
$m_t$.
This ``in-situ'' calibration of the Monte Carlo event generator might warrant a sufficiently small theoretical error to obtain
$m_t$ from exclusive hadronic final states, but it may be
necessary to carry out a detailed shape analysis of
several observables at once. 
Despite correlations among the calibration quantities,
it seems that the required precision can ultimately be achieved
if one is able to measure the observable shapes with about 1\% accuracy,
which may be challenging. 
Furthermore, our results might be altered by experimental systematics and correlations among measurements that we have not evaluated (\emph{e.g.,} a correlation in the calibration observables due to common experimental systematics). 
For this reason, one might look at endpoints of some distributions which are known to be sensitive to the top quark mass, \emph{e.g.,} the endpoint of
$m_{B\ell}$. 
For such observables, we have demonstrated that the effect of the Monte Carlo parameters can be suppressed
at the cost of larger statistical uncertainties.
It remains as a future work
to pursue an optimal balance between endpoint analyses versus studies
of the bulk of the distributions,
in view of the performances that can be achieved
through the ``in-situ'' Monte Carlo calibration.

The present work can also be extended along several lines.
Although in the introduction we discussed the role that it plays
in the uncertainty in the top-mass determination,
color reconnection was not explored in our investigation.
Nevertheless, studying the impact of color-reconnection models and
parameters, along the lines of \cite{spyros}, on Mellin
moments, peaks and endpoints of $B$-hadron observables is certainly
mandatory for the sake of completeness.
Furthermore, even if $t\bar t$-production processes are mostly
high transverse-momentum events, investigating the effect on the
top mass of the modeling of the underlying event and of the
implementation of multiple scatterings is definitely compelling.

Regarding Monte Carlo generators, the HERWIG 6 code, which we used
through our analysis, is outdated and not supported anymore.
Therefore, it will be much useful updating our
investigation by employing the latest object-oriented HERWIG 7.1
code. In fact, as debated in the introduction, its description of
bottom-quark fragmentation is not optimal, but, thanks to the
implementation of the new dipole-shower model and a number of
novel features for the purpose of the treatment of heavy quarks, 
the comparison with $B$-hadron data at $e^+e^-$ machines has much
improved with respect to previous versions \cite{Bellm:2017idv}.
Analyzing the dependence of calibration and $m_t$-dependent observables on 
the parameters of HERWIG 7.1 will hence be very instructive.

Finally, an obvious extension of our study consists of using
NLO+shower codes, such as the latest version of
aMC@NLO \cite{mcnlo}, SHERPA \cite{sherpa} and especially POWHEG
\cite{powheg}, which accounts for NLO corrections to top decays,
width and non-resonant effects.
We plan to explore our $B$-hadron observables by using aMC@NLO, POWHEG
and SHERPA, in order to
check their sensitivity to non-perturbative parameters and whether it
is different from the case of standard LO event generators.
In fact, on the one hand, one should expect that, thanks to the improved
description of the perturbative part of the $t\bar t$ event simulation,
the dependence on non-perturbative parameters
should become milder.
On the other hand, however, a recent analysis on the determination of the
top mass \cite{powtop} using NLO+shower codes
has displayed substantial discrepancies according to whether
the latest POWHEG version is interfaced to HERWIG 7 or PYTHIA 8
for parton cascades, hadronization and underlying event.
Exploring the parameter dependence of calibration and 
top-mass observables yielded by POWHEG interfaced to HERWIG 7 or
PYTHIA 8 should possibly help to shed light on the discrepancy in the mass
extraction pointed out in \cite{powtop}.
This is in progress as well.

\section*{Acknowledgements}
We thank the LHC Top Working Group and especially Michelangelo Mangano for discussions and inputs while this work was in progress.
RF thanks CERN Theory Division for hospitality during the completion of this work. RF thanks Simone Amoroso, Marco Ciuchini, Kyle Cranmer, Enrico Rinaldi,  Giuseppe Salamanna, and Peter Skands for discussions.
DK thanks Jared Evans, Konstantin Matchev, Steven Mrenna, Myeonghun Park, and Stefan Prestel for their constructive comments and discussions. 
The work of RF was supported by Programma per Giovani Ricercatori ``Rita Levi Montalcini'' granted by Ministero dell'Istruzione, dell'Università e della Ricerca (MIUR).
DK is supported by the Korean Research Foundation (KRF) through the CERN-Korea Fellowship program.
Figures \ref{fig:Schematic-view-of-event} and \ref{fig:Three-kinematical-configurations} have been drawn with JaxoDraw~\cite{Binosi_2009}. The steering of  simulation jobs on multiple CPU cores has been managed by GNU Parallel~\cite{Tange2011a}.

\appendix
\section{Determination of the sensitivity matrix\label{appA}}

In the regime in which the observables are linear functions of the
parameters
\[
O_{i}=c_{i}+D_{ij}\theta_{j}
\]
we define $\tilde{O}_{i}=O_{i}-c_{i}$, such that 
\begin{equation}
\tilde{O}=D\theta\label{eq:linearsystem}
\end{equation}
where $D$ is a $N_{O}\times N_{\theta}$ matrix. The matrix can be
said to contain the derivatives of the observables w.r.t. the parameters

\[
\frac{d\tilde{O}_{i}}{d\theta_{j}}=\frac{dO_{i}}{d\theta_{j}}=D_{ij}\,.
\]
Since $d\tilde{O}=dO$ we will use them interchangeably at our convenience. 

If we want to derive the parameters that correspond to a certain measurement
of the observables, we should invert $D$, which may not be rigorously
possible when $D$ is not a square matrix. In place of the inverse,
we use its pseudo-inverse~\cite{Penrose_1955,Dresden_1920}, that
we denote as $D^{-1}$, and which can be computed by a singular-value
decomposition of $D$. The singular-value decomposition allows to write $D$ as the product of three matrices $D=\Omega^{-1}\hat{D}P$ which can be understood as a change of basis in the vector parameter space operated by $P$, a change of basis in the observable space operated by $\Omega$, and a matrix
$\hat{D}$ which gives the linear dependence of the observables on the
parameters in the new basis. The matrix $\hat{D}$ is diagonal, hence the matrix made of its reciprocal diagonal values can be used to define the pseudo-inverse in such a way that $\hat{D}D=\mathds{1}$. With this definition we can write
\[
\theta=D^{-1}\tilde{O}=\left(P^{-1}\hat{D}^{-1}\Omega\right)\tilde{O}\,.
\]

If we desire to work entirely with relative uncertainties, we can restart
from eq.~(\ref{eq:linearsystem}) and make a change of variables by
rescaling the parameters and the observables by suitable factors that
turn error propagation into an expression where only relative errors
appear. We define $\bar{\theta}$ rescaling values so that each parameter
is a dimensionless fudge-factor
\[
f_{\theta_{i}}\equiv\frac{\theta_{i}}{\bar{\theta}_{i}}=\Lambda_{\theta}\theta=\left(\begin{array}{ccc}
\frac{1}{\bar{\theta}_{1}}\\
 & \ddots\\
 &  & \frac{1}{\bar{\theta}_{N}}
\end{array}\right)_{ij}\theta_{j}
\]
Using this redefinition in eq.~(\ref{eq:linearsystem}) we get 

\[
\tilde{O}=D\theta=D\Lambda_{\theta}^{-1}\,f_{\theta_{i}}\,.
\]
Similarly we can define a convenient value to rescale $\tilde{O}$
and make them dimensionless 
\[
f_{\tilde{O}_{i}}=\frac{\tilde{O}_{i}}{\bar{O}_{i}}=\left(\Lambda_{O}\tilde{O}\right)_{i},
\]
so that eq.~(\ref{eq:linearsystem}) now reads 
\[
\Lambda_{O}^{-1}f_{\tilde{O}}=D\Lambda_{\theta}^{-1}\,f_{\theta_{i}}\,
\]
and ultimately 
\[
f_{\tilde{O}}=\Lambda_{O}D\Lambda_{\theta}^{-1}\,f_{\theta_{i}}\,.
\]
The last line involves dimensionless fudge-factors $f$ and the sensitivity
matrix $\Delta$ whose elements are
\[
\Delta_{ij}\equiv\frac{d\tilde{O}_{i}}{d\theta_{j}}\frac{\bar{\theta_{j}}}{\bar{O_{i}}}=\frac{dO_{i}}{d\theta_{j}}\frac{\bar{\theta_{j}}}{\bar{O_{i}}}\,.
\]
Once we know the linear mapping between the dimensionless fudge factors, we can use eq.~(\ref{covTheta}) to find out the covariance matrix of one as a function of the covariance of the others.

As an example, Table~\ref{tab:Sensitivity-of-bins-pTbratio}
presents the sensitivity matrix for the $p_{T,B}/p_{T,j_b}$
ratio in terms of the PYTHIA parameters, in each bin.

\begin{table*}
\begin{centering}
\begin{tabular}{|c|c|c|c|c|c|c|c|}
\hline 
$p_{T,B}/p_{T,j_{b}}$ range & $\alpha_{s,FSR}$ & $m_{b}$ & $p_{T,\text{min}}$ & $a$ & $b$ & $r_{B}$ & recoil\tabularnewline
\hline 
\hline 
{[}0,0.09) & 1.76(8) & 0.12(10) & -0.05(4) & 0.00(9) & -0.1(1) & -0.1(3) & 0.09(2)\tabularnewline
\hline 
{[}0.09,0.18) & 1.90(3) & 0.12(3) & -0.08(1) & 0.01(6) & -0.04(6) & -0.0(1) & 0.094(9)\tabularnewline
\hline 
{[}0.18,0.27) & 1.95(4) & 0.18(4) & -0.095(6) & 0.02(3) & -0.08(4) & 0.1(1) & 0.114(7)\tabularnewline
\hline 
{[}0.27,0.36) & 2.06(3) & 0.27(2) & -0.11(1) & 0.01(3) & -0.07(3) & 0.08(9) & 0.128(8)\tabularnewline
\hline 
{[}0.36,0.45) & 2.18(2) & 0.40(2) & -0.141(4) & 0.04(2) & -0.11(2) & 0.15(7) & 0.142(5)\tabularnewline
\hline 
{[}0.45,0.54) & 2.26(1) & 0.61(1) & -0.186(5) & 0.04(2) & -0.16(1) & 0.11(7) & 0.155(4)\tabularnewline
\hline 
{[}0.54,0.63) & 2.22(1) & 0.90(1) & -0.246(4) & 0.07(1) & -0.24(1) & 0.27(4) & 0.150(4)\tabularnewline
\hline 
{[}0.63,0.72) & 1.81(1) & 1.18(1) & -0.305(3) & 0.11(1) & -0.36(1) & 0.39(4) & 0.121(3)\tabularnewline
\hline 
{[}0.72,0.81) & 0.55(3) & 0.78(2) & -0.256(2) & 0.130(8) & -0.32(1) & 0.38(3) & 0.036(2)\tabularnewline
\hline 
{[}0.81,0.90) & -1.38(4) & -0.60(2) & 0.015(3) & 0.028(8) & 0.08(1) & -0.06(2) & -0.063(3)\tabularnewline
\hline 
{[}0.90,0.99) & -2.65(5) & -1.82(2) & 0.449(10) & -0.300(8) & 0.656(7) & -0.77(2) & -0.133(4)\tabularnewline
\hline 
\end{tabular}
\par\end{centering}
\caption{\label{tab:Sensitivity-of-bins-pTbratio}Sensitivity of the bin counting
of the $p_{T,B}/p_{T,j_{b}}$ distribution to variations of the PYTHIA 8
parameters. }
\end{table*}

\section{Sensitivities and
  special features of $m_{T2}$\label{appV}}

For completeness, we report the sensitivities of $m_t$ 
to shower and hadronization parameters, once it is
extracted from a shape analysis of special features of the 
$m_{T2}$ variable, obtained according to other reconstruction schemes
with respect to those presented in Table~\ref{tab:Sensitivity-Mass-Obs}.
The corresponding numbers for $\Delta_{m_t}^{({\cal O})}$ and
  $\Delta_\theta^{(m_t)}$ are quoted in Table~\ref{tab:Sensitivity-Mass-Obs-spec}.

\begin{table*}
\begin{centering}
\begin{tabular}{|c|c|c|c|c|c|c|c|c|c|}
\hline 
\multirow{2}{*}{$\mathcal{O}$} & \multirow{2}{*}{Range}  & \multirow{2}{*}{$\Delta_{m_{t}}^{(\mathcal{O})}$}  & \multicolumn{7}{c|}{$\Delta_{\theta}^{(m_{t})}$}\tabularnewline
\cline{4-10} 
 &  &  & $\alpha_{s,FSR}$ & $m_b$ & $p_{T,\text{min}}$ & $a$ & $b$ & $r_{B}$ & recoil \tabularnewline
\hline 
\hline  
$\breve{m}_{T2,B\ell,\text{true}}^{\text{(ISR)}}$ & 152-170 & 1.01(2) & 0.008(4) & -0.005(8) & 0.000(5) & -0.008(2) & 0.012(7) & -0.01(2) & 0.000(2) \tabularnewline
\hline 
$\breve{m}_{T2,B\ell,\text{min}}^{\text{(ISR)}}$  & 152-170 & 1.01(1) & 0.004(6) & -0.02(2) & 0.005(6) & -0.007(2) & 0.013(6) & -0.01(1) & -0.0004(5) \tabularnewline
\hline 
$\breve{m}_{T2,B\ell,\text{min,}\perp}^{\text{(ISR)}}$  & 140-170 & 0.92(3) & -0.079(5) & -0.054(8) & 0.011(2) & -0.008(4) & 0.018(5) & -0.02(2) & -0.002(1) \tabularnewline
\hline 
\hline  
$\breve{m}_{T2,B\ell,\text{true}}^{\text{(true)}}$ & 152-170 & 1.01(2) & 0.008(4) & -0.005(8) & 0.000(5) & -0.008(2) & 0.012(7) & -0.01(2) & 0.000(1) \tabularnewline
\hline 
$\breve{m}_{T2,B\ell,\text{min}}^{\text{(true)}}$  & 152-170 & 1.01(1) & 0.004(6) & -0.02(2) & 0.005(6) & -0.007(2) & 0.013(6) & -0.01(1) & -0.0005(4) \tabularnewline
\hline 
$\breve{m}_{T2,B\ell,\text{min,}\perp}^{\text{(true)}}$  & 140-170 & 0.85(3) & -0.074(6) & -0.049(6) & 0.012(4) & -0.011(9) & 0.023(8) & -0.01(2) & -0.001(1) \tabularnewline
\hline 
\hline
$\breve{m}_{T2,B}^{\text{(ISR)}}$  & 147-170 & 1.03(1) & 0.007(5) & -0.002(4) & 0.0014(4) & -0.001(1) & 0.002(1) & -0.003(3) & 0.0001(3) \tabularnewline
\hline
$\breve{m}_{T2,B,\perp}^{\text{(ISR)}}$  & 128-170 & 0.98(2) & -0.071(6) & -0.068(4) & 0.013(2) & -0.012(3) & 0.021(4) & -0.03(1) & -0.0026(9) \tabularnewline
\hline 
\end{tabular}
\par\end{centering}
%
%
\caption{ 
\label{tab:Sensitivity-Mass-Obs-spec}
  Sensitivity $\Delta_\theta^{(m_t)}$ for an analysis extracting $m_t$
  by using special features of the $m_t$ spectra, as discussed in the text.
  The range extremal values reported in the second column are all in GeV. }
\end{table*}

\providecommand{\href}[2]{#2}\begingroup\raggedright\endgroup


\begin{thebibliography}{10}

\bibitem{degrassi}
G. Degrassi, S. Di Vita, J. Elias-Miro, J. R. Espinosa, G. F. Giudice, 
G. Isidori and A. Strumia, JHEP 1208 (2012) 098.

\bibitem{andreassen} 
  A.~Andreassen and M.~D.~Schwartz,
  \href{http://dx.doi.org/10.1007/JHEP10(2017)151}{{\em JHEP} {\bfseries 10} (2017) 151}, \href{http://arxiv.org/abs/1705.07135}{{\ttfamily arXiv:1705.07135 [hep-ph]}}.

\bibitem{corcella}
  G. Corcella,
PoS (TOP2015) (2016) 037.

\bibitem{hoang}
  M. Butenschoen, B. Dehnadi, A.H. Hoang, V. Mateu, M. Preisser
  and I.W. Stewart, Phys. Rev. Lett. {\bf 117} 232001 (2016).

\bibitem{hoang1}
  A.~H.~Hoang, S.~Mantry, A.~Pathak and I.~W.~Stewart,
  arXiv:1708.02586 [hep-ph].


\bibitem{CMS-Collaboration:2016nr}
{CMS Collaboration}, ``{Measurement of the mass of the top quark in decays with
  a $J/\psi$ meson in $pp$ collisions at 8 TeV},'' {\em ArXiv e-prints} (Aug.,
  2016) , \href{http://arxiv.org/abs/1608.03560}{{\ttfamily arXiv:1608.03560
  [hep-ex]}}.

\bibitem{Aad:2012bh}
{ATLAS Collaboration}, ``{Measurement of the top quark mass with the template
  method in the $ t \bar{t} \to$~leptons+jets channel using ATLAS data},''
  \href{http://dx.doi.org/10.1140/epjc/s10052-012-2046-6}{{\em European
  Physical Journal C} {\bfseries 72} (June, 2012) 2046},
  \href{http://arxiv.org/abs/1203.5755}{{\ttfamily arXiv:1203.5755 [hep-ex]}}.

\bibitem{Aad:2015sf}
{ATLAS Collaboration}, ``{Measurement of the top quark mass in the and channels
  using ATLAS data},''
  \href{http://dx.doi.org/10.1140/epjc/s10052-015-3544-0}{{\em European
  Physical Journal C} {\bfseries 75} (July, 2015) 330},
  \href{http://arxiv.org/abs/1503.05427}{{\ttfamily arXiv:1503.05427
  [hep-ex]}}.

\bibitem{Corcella:2001fp}
G.~{Corcella}, I.~G. {Knowles}, G.~{Marchesini}, S.~{Moretti}, K.~{Odagiri},
  P.~{Richardson}, M.~H. {Seymour}, and B.~R. {Webber}, ``{HERWIG 6: an event
  generator for hadron emission reactions with interfering gluons (including
  supersymmetric processes)},''
  \href{http://dx.doi.org/10.1088/1126-6708/2001/01/010}{{\em Journal of High
  Energy Physics} {\bfseries 1} (Jan., 2001) 010},
  \href{http://arxiv.org/abs/hep-ph/0011363}{{\ttfamily hep-ph/0011363}}.

\bibitem{Sjostrand:2014rr}
T.~{Sj{\"o}strand}, S.~{Ask}, J.~R. {Christiansen}, R.~{Corke}, N.~{Desai},
  P.~{Ilten}, S.~{Mrenna}, S.~{Prestel}, C.~O. {Rasmussen}, and P.~Z. {Skands},
  ``An introduction to pythia 8.2,''
  \href{http://arxiv.org/abs/1410.3012v1}{{\ttfamily arXiv:1410.3012v1
  [hep-ph]}}.

\bibitem{sigmaatl}
ATLAS Collaboration, Eur. Phys. J. C74 (2015) 3109.

\bibitem{mcnlo}
J. Alwall et al., JHEP 1407 (2014) 079.

\bibitem{powheg}
S. Alioli, P. Nason, C. Oleari and E. Re, JHEP 1006 (2010) 043.

\bibitem{rikk}
A.S. Papanastasiou, R. Frederix, S. Frixione, V. Hirschi and F. Maltoni,
Phys. Lett. B726 (2013) 223.

\bibitem{madspin}
P. Artoisenet, R. Frederix, O. Mattelaer and R. Rietkerk,
JHEP 1303 (2013) 015.

  \bibitem{pow}
J.M. Campbell, R.K. Ellis, P. Nason and E. Re, JHEP 1504 (2015) 114.


\bibitem{bb4l}
  T. Jezo, J.M. Lindert, P. Nason, C. Oleari and S. Pozzorini,
  Eur. Phys. J. C76 (2016) 691.
	
\bibitem{sherpa}
  T. Gleisberg, S. H\"oche, F. Krauss, M. Schonherr, S. Schumann,
  F. Siegert and J. Winter, JHEP 0902 (2009) 007.

\bibitem{sherpa1}
  S. H\"oche, N. Moretti, S. Pozzorini and F. Siegert,
  Eur. Phys. J. C77 (2017) 145.

\bibitem{sherpa2}
  S. H\"oche, S. Kuttimalai, Steffen Schumann, Frank Siegert,
  Eur. Phys. J. C75 (2015) 135.
  
 \bibitem{wave}
ATLAS and CDF and CMS and D0 Collaborations,  arXiv:1403.4427 [hep-ex].

 

\bibitem{Kharchilava:2000yk}
A.~{Kharchilava}, ``{Top mass determination in leptonic final states with
  {$J/\psi$}},'' \href{http://dx.doi.org/10.1016/S0370-2693(00)00120-9}{{\em
  Physics Letters B} {\bfseries 476} (Mar., 2000) 73--78},
  \href{http://arxiv.org/abs/hep-ph/9912320}{{\ttfamily hep-ph/9912320}}.

\bibitem{Hill:2005dq}
C.~S. {Hill}, J.~R. {Incandela}, and J.~M. {Lamb}, ``{Method for measurement of
  the top quark mass using the mean decay length of $b$ hadrons in $t\bar{t}$
  events},'' \href{http://dx.doi.org/10.1103/PhysRevD.71.054029}{{\em {Phys.
  Rev. D}} {\bfseries 71} no.~5, (Mar., 2005) 054029},
  \href{http://arxiv.org/abs/hep-ex/0501043}{{\ttfamily hep-ex/0501043}}.

\bibitem{CMS-PAS-TOP-12-030}
{CMS Collaboration}, ``{Measurement of the top quark mass using the $B$-hadron
  lifetime technique},'' {\em CMS-PAS-TOP-12-030} .


\bibitem{CMS-PAS-FTR-13-017}
{CMS Collaboration}, ``{Projected improvement of the accuracy of top-quark mass
  measurements at the upgraded LHC},'' Tech. Rep. CMS-PAS-FTR-13-017, CERN,
  Geneva, 2013.
\newblock
  \url{https://twiki.cern.ch/twiki/bin/view/CMSPublic/PhysicsResultsFTR13017}.

\bibitem{Aaltonen:2009zl}
{CDF Collaboration}, ``{Measurement of the top quark mass using the invariant
  mass of lepton pairs in soft muon $b$-tagged events},''
  \href{http://dx.doi.org/10.1103/PhysRevD.80.051104}{{\em {Phys. Rev. D}}
  {\bfseries 80} no.~5, (Sept., 2009) 051104},
  \href{http://arxiv.org/abs/0906.5371}{{\ttfamily arXiv:0906.5371 [hep-ex]}}.

\bibitem{CMS-Collaboration:2016eu}
{CMS Collaboration}, ``{Measurement of the top quark mass using charged
  particles in pp collisions at $\sqrt{s}=8$~TeV},'' {\em ArXiv e-prints} (Mar.,
  2016) , \href{http://arxiv.org/abs/1603.06536}{{\ttfamily arXiv:1603.06536
  [hep-ex]}}.

\bibitem{Abdallah:2011az}
J.~{Abdallah}, P.~{Abreu}, W.~{Adam}, P.~{Adzic}, T.~{Albrecht},
  R.~{Alemany-Fernandez}, T.~{Allmendinger}, P.~P. {Allport}, U.~{Amaldi},
  N.~{Amapane}, and et~al., ``{A study of the $b$-quark fragmentation function
  with the DELPHI detector at LEP I and an averaged distribution obtained at
  the $Z$ Pole},'' \href{http://dx.doi.org/10.1140/epjc/s10052-011-1557-x}{{\em
  European Physical Journal C} {\bfseries 71} (Feb., 2011) 1557},
  \href{http://arxiv.org/abs/1102.4748}{{\ttfamily arXiv:1102.4748 [hep-ex]}}.

\bibitem{Aleph:2001sy}
{Aleph}, {CDF}, {Delphi}, {L3}, {Opal}, and {SLD}, ``{Combined results on
  $b$-hadron production rates and decay properties},'' {\em ArXiv High Energy
  Physics - Experiment e-prints} (Dec., 2001) ,
  \href{http://arxiv.org/abs/hep-ex/0112028}{{\ttfamily hep-ex/0112028}}.

\bibitem{Abe:2002fc}
SLD, ``{Measurement of the $b$-quark fragmentation function in Z$^{0}$ decays},''
  \href{http://dx.doi.org/10.1103/PhysRevD.65.092006}{{\em Phys. Rev. D}
  {\bfseries 65} no.~9, (May, 2002) 092006},
  \href{http://arxiv.org/abs/hep-ex/0202031}{{\ttfamily hep-ex/0202031}}.

\bibitem{mele}
  B. Mele and P. Nason,  Nucl. Phys. B361 (1991) 626, Erratum:
  Nucl. Phys. B921 (2017) 841.

  \bibitem{cno}
    M. Cacciari, P. Nason and C. Oleari, JHEP 0604 (2006)

  \bibitem{ccm}
    M. Cacciari, G. Corcella and A.D. Mitov,  JHEP 0212 (2002) 015.

  \bibitem{corc}
    G. Corcella, Nucl. Phys. B705 (2005) 363; Erratum: Nucl. Phys. B713 (2005)
    609.

  \bibitem{kart}
    V.G. Kartvelishvili, A.K. Likodek
    and V.A. Petrov, Phys. Lett. 78B (1978) 615.

  \bibitem{peter}
     C. Peterson, D. Schlatter, I. Schmitt, P.M. Zerwas, 
     Phys. Rev. D27 (1983) 105.
     
  \bibitem{Fickinger:2016fk}
M.~{Fickinger}, S.~{Fleming}, C.~{Kim}, and E.~{Mereghetti}, ``{Effective field
  theory approach to heavy quark fragmentation},'' {\em ArXiv e-prints} (June,
  2016) , \href{http://arxiv.org/abs/1606.07737}{{\ttfamily arXiv:1606.07737
  [hep-ph]}}.

\bibitem{Andersson:1983iaa}
B.~Andersson, G.~Gustafson, G.~Ingelman, and T.~Sjostrand, ``{Parton
  Fragmentation and String Dynamics},''
\href{http://dx.doi.org/10.1016/0370-1573(83)90080-7}{{\em Phys. Rept.}
  {\bfseries 97} (1983) 31--145}.

\bibitem{Bowler:1981sb}
M.~G. Bowler, ``{$e^+e^-$ Production of Heavy Quarks in the String Model},''
\href{http://dx.doi.org/10.1007/BF01574001}{{\em Z. Phys.} {\bfseries C11}
  (1981) 169}.

\bibitem{cluster}
B.R. Webber, Nucl. Phys. B238 (1984) 492.
  
\bibitem{Biswas:2010vq}
S.~{Biswas}, K.~{Melnikov}, and M.~{Schulze}, ``{Next-to-leading order QCD
  effects and the top quark mass measurements at the LHC},''
  \href{http://dx.doi.org/10.1007/JHEP08(2010)048}{{\em JHEP} {\bfseries 8}
  (Aug., 2010) 48}, \href{http://arxiv.org/abs/1006.0910}{{\ttfamily
  arXiv:1006.0910 [hep-ph]}}.

\bibitem{Agashe:2016xq}
K.~{Agashe}, D.~{Kim}, R.~{Franceschini}, and M.~{Schulze}, ``Top quark mass
  determination from the energy peaks of b-jets and b-hadrons at NLO QCD,''
  \href{http://arxiv.org/abs/1603.03445v1}{{\ttfamily arXiv:1603.03445v1
  [hep-ph]}}.

\bibitem{Gieseke:2004sh}
S.~{Gieseke}, A.~{Ribon}, M.H.~{Seymour}, P.~{Stephens}, and B.~{Webber},
  ``{Herwig++ 1.0: an event generator for $e^{+}e^{-}$ annihilation},''
  \href{http://dx.doi.org/10.1088/1126-6708/2004/02/005}{{\em Journal of High
  Energy Physics} {\bfseries 2} (Feb., 2004) 005},
  \href{http://arxiv.org/abs/hep-ph/0311208}{{\ttfamily hep-ph/0311208}}.

\bibitem{Bellm:2015rm}
J.~{Bellm}, S.~{Gieseke}, D.~{Grellscheid}, S.~{Pl{\"a}tzer}, M.~{Rauch},
  C.~{Reuschle}, P.~{Richardson}, P.~{Schichtel}, M.~H. {Seymour},
  A.~{Si{\'o}dmok}, A.~{Wilcock}, N.~{Fischer}, M.~A. {Harrendorf}, G.~{Nail},
  A.~{Papaefstathiou}, and D.~{Rauch}, ``{Herwig 7.0 / Herwig++ 3.0 Release
  Note},'' {\em ArXiv e-prints} (Dec., 2015) ,
  \href{http://arxiv.org/abs/1512.01178}{{\ttfamily arXiv:1512.01178
  [hep-ph]}}.

\bibitem{Karneyeu:2014uq}
A.~{Karneyeu}, L.~{Mijovic}, S.~{Prestel}, and P.~Z. {Skands}, ``{MCPLOTS: a
  particle physics resource based on volunteer computing},''
  \href{http://dx.doi.org/10.1140/epjc/s10052-014-2714-9}{{\em European
  Physical Journal C} {\bfseries 74} (Feb., 2014) 2714},
  \href{http://arxiv.org/abs/1306.3436}{{\ttfamily arXiv:1306.3436 [hep-ph]}}.

\bibitem{Skands:2014fj}
P.~{Skands}, S.~{Carrazza}, and J.~{Rojo}, ``{Tuning PYTHIA 8.1: the Monash
  2013 Tune},'' {\em ArXiv e-prints} (Apr., 2014) ,
  \href{http://arxiv.org/abs/1404.5630}{{\ttfamily arXiv:1404.5630 [hep-ph]}}.

\bibitem{ATL-PHYS-PUB-2015-007}
``{A study of the sensitivity to the Pythia8 parton shower parameters of
  $t\bar{t}$ production measurements in $pp$ collisions at $\sqrt{s} = 7$ TeV
  with the ATLAS experiment at the LHC},'' Tech. Rep. ATL-PHYS-PUB-2015-007,
  CERN, Geneva, Mar, 2015.
\newblock \url{http://cds.cern.ch/record/2004362}.

\bibitem{spyros}
S. Argyropoulos and T. Sj\"ostrand, JHEP 1411 (2014) 043.
  
\bibitem{corc1}
G. Corcella, EPJ Web Conf. 80 (2014) 00019.


\bibitem{mescia}
  G. Corcella and F. Mescia, Eur.Phys.J. C65 (2010) 171;
  Erratum: Eur.Phys.J. C68 (2010) 687.

  \bibitem{CMS-PAS-TOP-14-001}
{CMS Collaboration}, ``{Measurement of the top-quark
  mass in $t\bar t$ events with lepton+jets final states in $pp$ collisions at
  $\sqrt{s}=8$~TeV},'' Tech. Rep. CMS-PAS-TOP-14-001, CERN, Geneva, 2014.





\bibitem{corsey}
G. Corcella and M.H. Seymour, Phys. Lett. B442 (1998) 417.

\bibitem{norb}
  E. Norrbin and T. Sj\"ostrand, Nucl. Phys. B603 (2001) 297.

  \bibitem{CMS-Collaboration:af}
{CMS Collaboration},
  ``{Investigations of the impact of the parton shower tuning in Pythia 8 in
  the modelling of $\mathrm{t\overline{t}}$ at $\sqrt{s}=8$ and 13 TeV},'' {\em
  {CERN Note}} no.~CMS-PAS-TOP-16-021, ,
  \href{http://arxiv.org/abs/CMS-PAS-TOP-16-021}{{\ttfamily
      CMS-PAS-TOP-16-021}}. \url{https://cds.cern.ch/record/2235192/}.

\bibitem{Bellm:2017aa}
J.~{Bellm}, S.~{Gieseke}, D.~{Grellscheid}, P.~{Kirchgae{\ss}er}, F.~{Loshaj},
  G.~{Nail}, A.~{Papaefstathiou}, S.~{Pl{\"a}tzer}, R.~{Podskubka}, M.~{Rauch},
  C.~{Reuschle}, P.~{Richardson}, P.~{Schichtel}, M.~H. {Seymour},
  A.~{Si{\'o}dmok}, and S.~{Webster}, ``{Herwig 7.1 Release Note},'' {\em ArXiv
  e-prints} (May, 2017) , \href{http://arxiv.org/abs/1705.06919}{{\ttfamily
  arXiv:1705.06919 [hep-ph]}}.

\bibitem{Bellm:2017idv}
  J.~Bellm, K.~Cormier, S.~Gieseke, S.~Plätzer, C.~Reuschle, P.~Richardson and S.~Webster,
  arXiv:1711.11570 [hep-ph].

  
\bibitem{Corcella:2005tg}
G.~{Corcella} and V.~{Drollinger}, ``{Bottom-quark fragmentation: Comparing
  results from tuned event generators and resummed calculations},''
  \href{http://dx.doi.org/10.1016/j.nuclphysb.2005.09.030}{{\em Nuclear Physics
  B} {\bfseries 730} (Dec., 2005) 82--102},
  \href{http://arxiv.org/abs/hep-ph/0508013}{{\ttfamily hep-ph/0508013}}.

\bibitem{Catani:1990rr}
S.~Catani, B.R.~Webber, and G.~Marchesini, ``{QCD coherent branching and
  semiinclusive processes at large $x$},''
\href{http://dx.doi.org/10.1016/0550-3213(91)90390-J}{{\em Nucl. Phys.}
  {\bfseries B349} (1991) 635--654}.


  

\bibitem{Lester:1999tx} 
  C.~G.~Lester and D.~J.~Summers,
  ``{Measuring masses of semiinvisibly decaying particles pair produced at hadron colliders},''
  \href{http://dx.doi.org/10.1016/S0370-2693(99)00945-4}{{\em Phys.\ Lett.\ B} {\bfseries 463}, 99 (1999)},
  \href{http://arxiv.org/abs/hep-ph/9906349}{{\ttfamily hep-ph/9906349}}.

\bibitem{Matchev:2009ad} 
  K.~T.~Matchev and M.~Park, ``{A General method for determining the masses of semi-invisibly decaying particles at hadron colliders},''
  \href{http://dx.doi.org/10.1103/PhysRevLett.107.061801} 
 { {\em Phys.\ Rev.\ Lett.}  {\bfseries 107}, 061801 (2011)},    
 \href{http://arxiv.org/abs/arXiv:0910.1584} {{\ttfamily arXiv:0910.1584}}.

\bibitem{CMS-PAS-TOP-11-027}
``Mass determination in the $t\bar t$ system with kinematic endpoints,'' Tech.
  Rep. CMS-PAS-TOP-11-027, CERN, Geneva, 2012.

\bibitem{CMS-PAS-TOP-15-008}
{CMS Collaboration}, ``{Measurement of the top quark
  mass in the dileptonic $t\bar t$ decay channel using the $M_{bl}$,
  $M_{T2}$, and MAOS
  $M_{bl\nu}$
  observables},'' Tech. Rep. CMS-PAS-TOP-15-008, CERN, Geneva, 2016.
\newblock \url{http://cds.cern.ch/record/2204924}.

\bibitem{Alioli:2013mz}
S.~{Alioli}, P.~{Fernandez}, J.~{Fuster}, A.~{Irles}, S.-O. {Moch}, P.~{Uwer},
  and M.~{Vos}, ``{A new observable to measure the top-quark mass at hadron
  colliders},'' {\em ArXiv e-prints} (Mar., 2013) ,
  \href{http://arxiv.org/abs/1303.6415}{{\ttfamily arXiv:1303.6415 [hep-ph]}}.

\bibitem{Burns:2008va} 
  M.~Burns, K.~Kong, K.~T.~Matchev and M.~Park, ``{Using Subsystem MT2 for Complete Mass Determinations in Decay Chains with Missing Energy at Hadron Colliders},''
  \href{http://dx.doi.org/10.1088/1126-6708/2009/03/143} {{\em Journal of High Energy Physics} {\bfseries 0903}, 143 (2009)},
  \href{http://arxiv.org/abs/arXiv:0810.5576} {{\ttfamily arXiv:0810.5576}}.

\bibitem{dEnterria:2014mq}
D.~{d'Enterria}, K.~J. {Eskola}, I.~{Helenius}, and H.~{Paukkunen},
  ``{Confronting current NLO parton fragmentation functions with inclusive
  charged-particle spectra at hadron colliders},''
  \href{http://dx.doi.org/10.1016/j.nuclphysb.2014.04.006}{{\em Nuclear Physics
  B} {\bfseries 883} (June, 2014) 615--628},
  \href{http://arxiv.org/abs/1311.1415}{{\ttfamily arXiv:1311.1415 [hep-ph]}}.

  \bibitem{tancrediTopLHCWG14}
{Tancredi Carli}, ``{Hadronisation and radiation systematics and
  $b$-fragmentation},'' {\em Top quark LHC Working group meeting May 2014}'',
  \url{https://indico.cern.ch/event/301787/contributions/690356/}.



\bibitem{ATLAS-Collaboration:2013tx}
{ATLAS Collaboration}, ``{Measurement of jet shapes in top pair events at
  $\sqrt{s} = 7$~TeV using the ATLAS detector},'' {\em ArXiv e-prints} (July,
  2013) , \href{http://arxiv.org/abs/1307.5749}{{\ttfamily arXiv:1307.5749
  [hep-ex]}}.

\bibitem{Konar:2011sf}
P.~{Konar}, K.~{Kong}, K.~T. {Matchev}, and M.~{Park}, ``{RECO level
  $\sqrt{s}_{min}$ and subsystem improved $\sqrt{s}_{min}$: global inclusive
  variables for measuring the new physics mass scale in events with missing
  energy at hadron colliders},''
  \href{http://dx.doi.org/10.1007/JHEP06(2011)041}{{\em Journal of High Energy
  Physics} {\bfseries 6} (June, 2011) 41},
  \href{http://arxiv.org/abs/1006.0653}{{\ttfamily arXiv:1006.0653 [hep-ph]}}.
  




\bibitem{Penrose_1955}
R.~Penrose and J.~A. Todd, ``A generalized inverse for matrices,''
  \href{http://dx.doi.org/10.1017/s0305004100030401}{{\em Mathematical
  Proceedings of the Cambridge Philosophical Society} {\bfseries 51} no.~03,
  (Jul, 1955) 406}. \url{http://dx.doi.org/10.1017/S0305004100030401}.

\bibitem{Dresden_1920}
A.~Dresden, ``The fourteenth western meeting of the american mathematical
  society,'' \href{http://dx.doi.org/10.1090/s0002-9904-1920-03322-7}{{\em
  Bulletin of the American Mathematical Society} {\bfseries 26} no.~9, (Jun,
  1920) 385--397}. \url{http://dx.doi.org/10.1090/S0002-9904-1920-03322-7}.

\bibitem{Agashe:2012bn} 
  K.~Agashe, R.~Franceschini and D.~Kim, ``Simple ``invariance'' of two-body decay kinematics,''
  \href{http://dx.doi.org/10.1103/PhysRevD.88.057701}{{\em Phys.\ Rev.\ D} {\bf 88}, no. 5, 057701 (2013)},
  \href{http://arxiv.org/abs/arXiv:1209.0772}{{\ttfamily arXiv:1209.0772}}.
  
\bibitem{Agashe:2012fs} 
  K.~Agashe, R.~Franceschini, D.~Kim and K.~Wardlow, ``Using Energy Peaks to Count Dark Matter Particles in Decays,''
  \href{http://dx.doi.org/10.1016/j.dark.2013.03.003}{{\em Phys.\ Dark Univ}  {\bf 2}, 72 (2013)}, 
  \href{http://arxiv.org/abs/arXiv:1212.5230}{{\ttfamily arXiv:1212.5230}}.

\bibitem{powtop}
  S. Ferrario Ravasio, T. Jezo, P. Nason and C. Oleari, arXiv:1801.03944 [hep-ph].

  
\bibitem{Binosi_2009}
D.~Binosi, J.~Collins, C.~Kaufhold, and L.~Theussl, ``Jaxodraw: A graphical
  user interface for drawing feynman diagrams. version 2.0 release notes,''
  \href{http://dx.doi.org/10.1016/j.cpc.2009.02.020}{{\em Computer Physics
  Communications} {\bfseries 180} no.~9, (Sep, 2009) 1709--1715}.
  \url{http://dx.doi.org/10.1016/j.cpc.2009.02.020}.

\bibitem{Tange2011a}
O.~Tange, ``Gnu parallel - the command-line power tool,''
  \href{http://dx.doi.org/http://dx.doi.org/10.5281/zenodo.16303}{{\em ;login:
  The USENIX Magazine} {\bfseries 36} no.~1, (Feb, 2011) 42--47}.
  \url{http://www.gnu.org/s/parallel}.



\end{thebibliography}
\end{document}